\documentclass[aps, final, superscriptaddress]{revtex4-1}

\draft % marks overfull lines with a black rule on the right
\usepackage{dcolumn}
\usepackage{graphicx}
\usepackage[caption=false]{subfig}
\usepackage{url}
\usepackage{hyperref}
\usepackage{amsmath}

\usepackage{dcolumn}% Align table columns on decimal point
\usepackage{bm}% bold math
\usepackage{mathptmx}
\usepackage{xcolor}

\begin{document}

%\preprint{}

%\title{} %Title of paper



%\title{Particle effects on Target Patterns in Phase Separating Mixtures}% Force line breaks with \\
\title{Impact of Particle Arrays on Phase Separation Composition Patterns}% Force line breaks with \\
%\thanks{}

\author{Supriyo Ghosh}
\email{supriyo.ghosh@lanl.gov}
%\altaffiliation[Corresponding author.}
\affiliation{Theoretical Division, Los Alamos National Laboratory, Los Alamos, NM 87545, USA}
%%%Lines break automatically or can be forced with \\
%%
\author{Arnab Mukherjee}
\affiliation{Center for Hierarchical Materials Design, Northwestern University, Evanston, IL 60208, USA}
%%arnab.mukherjee@northwestern.edu
%\author{T.~A.~Abinandanan}
%\affiliation{Materials Engineering Department, Indian Institute of Science, Bangalore 560012, India}
%%
\author{Raymundo Arroyave}
\affiliation{Materials Science \& Engineering Department, Texas A\&M University, College Station, TX 77843, USA}
\author{Jack F. Douglas}
\affiliation{Materials Science \& Engineering Division, National Institute of Standards and Technology, Gaithersburg, MD 20899, USA}

\date{\today}
%\keywords{Spinodal decomposition, particles, wetting, coarsening, phase-field}
\begin{abstract}
We examine the symmetry-breaking effect of fixed constellations of particles on the surface-directed spinodal decomposition of binary blends in the presence of particles whose surfaces have a preferential affinity for one of the components. Our phase-field simulations indicate that the phase separation morphology in the presence of particle arrays can be tuned to have a continuous, droplet, lamellar, or hybrid morphology depending on the interparticle spacing, blend composition, and time. In particular, when the interparticle spacing is large compared to the spinodal wavelength, a transient target pattern composed of alternate rings of preferred and non-preferred phases emerge at early times, tending to adopt the symmetry of the particle configuration. We reveal that such target patterns stabilize for certain characteristic length, time, and composition scales characteristic of the pure phase separating mixture. To illustrate the general range of phenomena exhibited by mixture-particle systems, we simulate the effects of single-particle, multi-particle, and cluster-particle systems having multiple geometrical configurations of the particle characteristic of pattern substrates on phase separation. Our simulations show that tailoring the particle configuration, or substrate pattern configuration, a relative fluid-particle composition should allow the desirable control of the phase separation morphology as in block copolymer materials, but where the scales accessible to this approach of organizing phase-separated fluids usually are significantly larger. Limited experiments confirm the trends observed in our simulations, which should provide some guidance in engineering patterned blend and other mixtures of technological interest.
\end{abstract}


\maketitle %\maketitle must follow title, authors, abstract and \pacs
\section{Introduction}\label{sec_introduction}
Phase separation \textit{via} spinodal decomposition in metallic or polymeric mixtures generates complex morphologies of practical interest~\cite{paul2012polymer,strobl2007polymer}. Micro- to nano-sized filler particles at finite loadings are often added in these mixtures to improve the processability and properties of the material~\cite{paul2012polymer,strobl2007polymer}. Several factors, such as the geometry, size, and the concentration of the particles, are found to affect the phase separation morphology. However, there is a limited understanding of how particles, or lithographically etched relief patterns on films on supporting substrates, can be used to engineer the phase separation morphology and, in turn, tailor the properties of the blend composites and thin films.

The potential technological relevance of the resulting morphologies studied in the present work can be realized by analogy to many applications undergoing development for block copolymer materials, where regular polymer composition modulations in space at the nanoscale are created by controlling chemical compositional variations using polymer molecules~\cite{epps2016block,mai2012block}. Depending on the chemical nature and sizes of the component polymers, block copolymer can phase separate into different morphologies, such as spherical, cylindrical, gyroid, and lamella~\cite{hu2014block,bang2009block,hamley2009block}, which have excited interest in a myriad of engineering applications. For example, spherical and cylindrical morphologies have potential applications in bit-patterned media~\cite{segalman2005block,cummins2016block}, membranes~\cite{phillip2010block}, nanowires~\cite{majewski2015block}, polarizers~\cite{kim2014block} etc., gyroids have applications for ion conduction channels~\cite{werner2018block}, whereas lamellar structures are potential candidates for nanolithography~\cite{bates2013block}, dielectric capacitors~\cite{samant2016block}, filtration, and electrolytes in energy storage devices~\cite{liu2019block,young2014block}. While patterns of these kinds on the scale of nanometers are often desirable, we envision that facile creation of patterns of tunable structures at somewhat larger scales should also be useful, motivating the present work exploring the use of organized patterns of particles to organize the morphology of the phase separating polymer species for the many of the same type of applications now considered for block copolymer materials.

It is well-understood that the phase separation morphology depends on the mixture composition~\cite{Puri,Krausch1994,PBinder,Benderly,Karim,Brown,Chakrabarti,Ma}. For a critical 50:50 volume mixture, the phase separation results in a bicontinuous morphology. For off-critical binary mixtures, the resultant morphology depends on the component phase fractions. It is often observed that the majority phase is continuous, and the minority phase forms isolated domains or droplets. When particles are added to a binary mixture, it introduces several new effects on bulk phase separation \textit{via} (a) particle configuration, (b) interparticle spacing, (c) particle-matrix interaction, (d) particle geometry, (e) particle mobility, and (f) particle concentration. Even the presence of \textit{immobile} particles in simulations and experiments on thin films and polymer blends has resulted in specific phase separation morphologies~\cite{Lee, FCBalazs, DCBalazs}. The presence of fixed particles that provide a symmetry breaking perturbation of the phase separation process serves as a ``template'' around which the phase separation morphology organizes to a greater or lesser degree of faithfulness to the original particle pattern. We mention several examples where this method has been applied experimentally in the case of polymer blends cast on chemically-patterned substrates~\cite{ermi1998,boltau1998,karim1998,nisato1999}. Our simulations below are for mixtures in two dimensions and can be expected to apply qualitatively to these thin films with surface patterns and perhaps more appropriately to thin blend films with lithographically arrayed patterns or organized particle arrays on surfaces. As a first approximation, we work with immobile particles with spherical geometry in the present study. The magnitude of the interaction between particles and component phases is tailored such that the particle favors one of the phases to surround it, leading to surface-directed spinodal decomposition (SDSD)~\cite{Puri,Zeng2008,Binder2010}.  

The microstructure evolution during SDSD of mixture A:B:C can be rationalized as follows. The matrix A:B phase separates to A-rich $\alpha$ and B-rich $\beta$ phases in the presence of C-rich $\gamma$ particles. At early times, the morphology of the phases evolves as alternate concentric $\alpha$ and $\beta$ rings around the $\gamma$ particles. This is referred to as a ``target'' pattern~\cite{Karim,Qui,ACBalazs,Millett2014,Park}. Such patterns have been observed experimentally in metallic~\cite{Karim,aichmayer2003}, polymer~\cite{Lee}, and metallic glass~\cite{Park} mixtures. At later times, after phase separation adjacent to particles, the \textit{transient} target morphology dissolves due to the coarsening process, and a continuous, transition, or droplet morphology prevails in the matrix.

The particle distribution and concentration can potentially be exploited for controlling SDSD morphologies. For instance, it is natural to ask how the target morphology, which is transient in nature, can be stabilized, given its utility in many industrial applications. One possibility that was often explored in measurements is to influence the SDSD morphology by dissolving the filler particles in the mixture. This is equivalent to the effects of randomly distributed particles on SDSD that are relatively well-explored~\cite{ghosh_pccp, FCBalazs, DCBalazs,amoabeng2017}. In these systems, no target pattern survived past early phase separation times, and the subsequent phase coarsening led to bicontinuous, transition, or droplet domains with particles pinned either in the continuous or droplet phase or along the phase boundary, depending primarily on the thermodynamic forces such as phase boundary energy. Since particle separation can not be controlled in these systems, hence the resultant morphologies can not be specified. In a wide range of practical applications, however, filler particles are distributed periodically or in the form of ordered clusters~\cite{Chang2012,Jiang}. These systems are characteristic of pattern substrates that could potentially influence SDSD for obtaining specified morphologies. Phase separation on such particle substrates remains unstudied. Also, the effects of a wide range of particle loadings on SDSD morphologies have not been systematically investigated before. 

Spinodal decomposition occurs through the interference of composition waves emanating from the bulk phases, introducing a length scale during phase separation~\cite{cahn,cahn1961spinodal}. During SDSD, the presence of particles at finite volume fractions introduces another length scale, i.e., the interparticle spacing ($\lambda$). The rationale behind this work is to explore different regimes of $\lambda$ compared to the spinodal length scale ($\lambda_{sp}$) that affect the formation and stability of target morphologies in critical and off-critical mixtures. We focus on two different particle systems: one in which the particles are dispersed periodically, and the other in which the particles are ordered within a cluster. Such patterned systems allow us to explore the length and time scales for which, for example, the transient target patterns can be stabilized, and thus provide guidelines for obtaining materials with a specified morphology.

Numerical simulations of phase separation in ternary (A:B:C) mixtures have taken several approaches, including Cahn-Hilliard-type~\cite{lamorgese2018triphase,copetti2000numerical,Huang,Eyre,Nauman,Lee,Abi}, molecular dynamics~\cite{singh2015phase,Mac}, Monte Carlo~\cite{tafa2001kinetics}, lattice Boltzmann~\cite{DCBalazs}, and others~\cite{Chen,VVBalazs,FCBalazs}. In this paper, we characterize the ternary mixture, which consists of particles (C) embedded in a matrix (A:B), through a Cahn-Hilliard-based phase-field model, as described in Sec.~\ref{sec_model}. The results on particle configurations on phase separating mixtures are elaborated in Sec.~\ref{sec_results}. The results are discussed in Sec.~\ref{sec_discussion}, and a summary of this work is given in Sec.~\ref{sec_conclusions}.
% Is phase separation around particles remain trapped in a metastable state, unlike phase separation in bulk?

\section{Computational Method}

Simulations of particle effects on SDSD have mostly taken binary Cahn-Hilliard-type approaches with a surface interaction term at the particle-matrix interfaces~\cite{Lee,Chakrabarti,Oono1988,ACBalazs,FCBalazs}. The implementation of such an extra set of boundary conditions on the particle surface to realize SDSD is quite arduous. In our work, we treat particles as the C-rich phase that coexists with the binary A:B matrix so that a study of the effects of general interface properties between particle and matrix phases such as curvature-driven coarsening (Ostwald ripening) becomes possible. In our model, we can assign distinct energy to each interface to tailor particle effects on SDSD, which avoids the implementation of any additional interaction terms at particle-matrix interfaces and allows us to straightforwardly extend our work to probe the generic mechanisms for real ternary systems in terms of the morphological patterns and the sequence of phase transformations.

Arguably, the treatment of the particles as a phase is appropriate when the particles are ``small'', comparable to the inherent scale of phase separation, and this is the situation that is probably of greatest interest in applications of our simulations. Our choice of modeling is also advantageous in that it obviates the need to specify the parameters describing the polymer-particle interaction, where instead, such interaction is described by an interaction familiar from the phase-field modeling of phase-separating mixtures. Besides, we should admit some idealizations in our modeling. In particular, our simulations are confined to two dimensions and do not include hydrodynamic interaction effects, which preclude the modeling of phase separation in its late stage~\cite{ACBalazs,clarke2002target}.

\subsection{Ternary Phase-Field Model}\label{sec_model}

The phase-field model is a diffuse-interface approach that accurately describes the physics of phase separation with a minimum of computational effort (for reviews, see Refs.~\cite{chen2002,moelans2008,phdthesis}). We have used a ternary phase-field model to study the dynamics of phase separation in a three-component three-phase setting that was developed and validated in Refs.~\cite{huang1995phase,Huang,Nauman,Ghosh}. In particular, this model was used in Refs.~\cite{huang1995phase,Huang} to study the interface effects during phase separation of ternary polymer mixtures. Here we present the main features of the model; see Appendix~\ref{sec_appendix} and Refs.~\cite{huang1995phase,Nauman,Ghosh} for further details. 

We consider that $i = 3$ components, A, B, and C, with their respective local volume fractions $c_i$ ($c_A$, $c_B$, and $c_C$) make up the system such that 
\begin{equation}\label{eq_sum1}
    c_A + c_B + c_C = 1.
\end{equation}
We use a dimensionless Cahn-Hilliard~\cite{cahn} based ternary free energy functional~\cite{Abi,saswata2020} that describes the evolution of a three-component, three-phase system (i.e., A-rich $\alpha$, B-rich $\beta$, and C-rich $\gamma$) following the leading order expansion,
\begin{equation}\label{eq_ch}
\mathcal{F} = N_V\, \int_{V} f\left(c_A, c_B, c_C\right) + \sum_{i}\kappa_i\left(\nabla c_i\right)^2 \, dV, 
\end{equation}
where $N_V$ is the number of molecules per unit volume $V$, $f(c_i)$ is the bulk homogeneous free energy, and $\kappa_i (\nabla c_i)^2$ is the gradient energy with $\kappa_i$ the coefficients. The form of $f(c_i)$ is approximated using the regular solution model~\cite{Porter} given by,
\begin{equation}\label{eq_bf}
f\left(c_A,c_B,c_C\right)=\frac{1}{2}\sum_{i\neq j}\chi_{ij}c_ic_j+\sum_ic_i \ln c_i,
\end{equation}
where $\chi_{ij}$ is pairwise interaction energy between the coexisting phases. Since $\sum c_i$ is conserved, the temporal evolution of the components follows the continuity equation:
\begin{equation}\label{eq_continuity}
\frac{\partial c_i}{\partial t}= - \nabla\cdot \mathbf{\bar{J}}_i.
\end{equation} 
The net flux of the components, $\mathbf{\bar{J}}_i$, is approximated using mobility $M$ and chemical potential $\mu$ (see Appendix~\ref{sec_appendix} for details),
\begin{eqnarray}\label{eq_flux}
\mathbf{\bar{J}}_A &=& -M_{AA} (\nabla\mu_A - \nabla\mu_C) + M_{AB} (\nabla\mu_B - \nabla\mu_C) \; \text{and} \nonumber \\
\mathbf{\bar{J}}_B &=& -M_{BB} (\nabla\mu_B - \nabla\mu_C) + M_{AB} (\nabla\mu_A - \nabla\mu_C).
\end{eqnarray}
The effective mobilities are expressed by
\begin{eqnarray}\label{eq_mobility}
M_{AA}&=&\left(1-c_A\right)^2M_A+c_A^2\left(M_B+M_C\right), \nonumber\\
M_{BB}&=&\left(1-c_B\right)^2M_B+c_B^2\left(M_A+M_C\right), \; \text{and} \nonumber\\
M_{AB}=M_{BA}&=&\left(1-c_A\right)c_BM_A+c_A\left(1-c_B\right)M_B-c_Ac_BM_C. 
\end{eqnarray} 
Applying the Euler-Lagrange variational derivative~\cite{riley2002,arfken1999} of Eq.~(\ref{eq_ch}), we obtain $\mu_i$ in Eq.~(\ref{eq_flux}). Finally, the following nonlinear equations of motion simulate the time ($t$) evolution of the components:
\begin{eqnarray}\label{eq_dcadt}
\frac{\partial c_A}{\partial t}= M_{AA}\left[\nabla^2 \left(\partial f/\partial c_A\right) -2(\kappa_A + \kappa_C)\nabla^4 c_A - 2\kappa_C\nabla^4 c_B\right] \nonumber\\
-M_{AB}\left[\nabla^2 \left(\partial f/\partial c_B\right) -2(\kappa_B + \kappa_C)\nabla^4 c_B-2\kappa_C\nabla^4 c_A\right], \; \text{and}
\end{eqnarray}
\begin{eqnarray}\label{eq_dcbdt}
\frac{\partial c_B}{\partial t}= M_{BB}\left[\nabla^2 \left(\partial f/\partial c_B\right)-2(\kappa_B + \kappa_C)\nabla^4 c_B - 2\kappa_C\nabla^4 c_A\right]\nonumber\\
-M_{AB}\left[\nabla^2 \left(\partial f/\partial c_A\right)-2(\kappa_A + \kappa_C)\nabla^4 c_A-2\kappa_C\nabla^4 c_B\right].
\end{eqnarray} 

\subsection{Simulation Details}\label{sec_parameters}
Phase-field simulations were carried out solving Eqs.~(\ref{eq_dcadt}) and (\ref{eq_dcbdt}) on a 512 $\times$ 512 lattice, using the semi-implicit Fourier spectral method~\cite{zhu1999coarsening}. A dimensionless mesh spacing of $\Delta x = \Delta y = 0.5$ and a discrete Euler time step of $\Delta t = 0.0025$ were used in all simulations. The periodic boundary condition was applied in all directions.

The particle effect on spinodal decomposition is governed mainly through the magnitude of interfacial ($\sigma_{ij}$) energy between the particle $j$ and matrix phases $i =$ A-rich $\alpha$ and B-rich $\beta$, and interparticle distance $\lambda$. The values of $\sigma_{ij}$ between coexisting phases were determined using the simulations of equilibrium composition profile (across a flat interface between phases) for the values of $\chi_{ij}$ and $\kappa_i$, following Refs.~\cite{Ghosh,Huang}.

The values of $\chi_{ij}$, $\kappa_i$, and $\sigma_{ij}$ used in the present work are given in Table.~\ref{tab_param}. For the present values of $\chi_{ij}$, A:B matrix phase separates spontaneously to product A-rich $\alpha$ and B-rich $\beta$ phases. Also, the values of interfacial energy are such that (i.e., $\sigma_{\alpha\gamma} < \sigma_{\beta\gamma}$), $\gamma$ particles always prefer the $A$-rich $\alpha$ to surround it. Such a selective preference of $\alpha$ about the particle is referred to as ``wetting.'' Note that the magnitude of the ``quench'' from the one-phase region to the two-phase region of the spinodal phase diagram is given by $\chi_{\text{AB}}/{\chi_c}$, where $\chi_c = 2$ is the critical value beyond of which A:B mixture spinodally decomposes~\cite{paul2012polymer,strobl2007polymer}. Since $\chi_{ij} \propto 1/T$, our simulations correspond to $T/T_c = 0.8$, where $T$ is the quench (or final) temperature and $T_c$ the critical temperature.

\begin{table}[h]
\centering
\caption{The $\chi_{ij}$, $\kappa_i$, and $\sigma_{ij}$ parameters used in simulations.\label{tab_param}}
\begin{tabular}{c | c | c | c | c | c | c | c | c}
\hline \hline
$\chi_{AB}$ & $\chi_{BC}$ & $\chi_{AC}$ & $\kappa_A$ & $\kappa_B$ & $\kappa_C$ & $\sigma_{\alpha\beta}$ & $\sigma_{\beta\gamma}$ & $\sigma_{\alpha\gamma}$ \\
\hline
2.5 & 5.0 & 3.5 & 2.0 & 6.0 & 6.0 & 0.23 & 1.22 & 0.76 \\
\hline \hline
\end{tabular}
\end{table}

Phase-field simulations began with a specified distribution of $\gamma$ particles with a finite interparticle distance $\lambda$ within the A:B matrix. The particles were spherical in shape and were present in square or rectangular arrays with finite loadings within the matrix. Depending on $\lambda$, the particle loading varied between 2~\% and 20~\% in our simulations. Dimensionless particle radii of $R = 8$ and $R = 16$ units were used in simulations. Particles were small enough for the persistence of the composition pattern around them. 

\begin{figure}[h]
\centering
\includegraphics[scale=0.5]{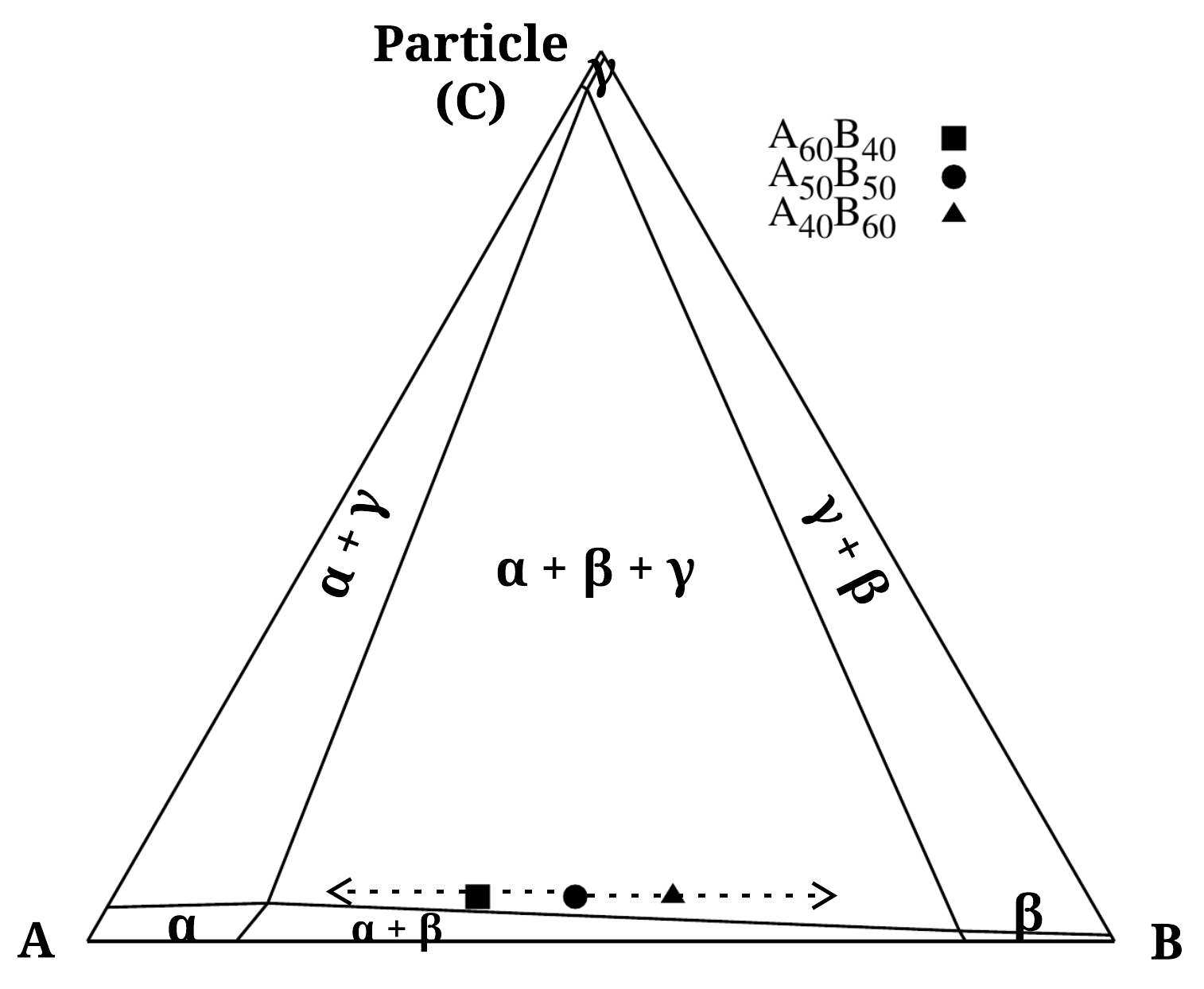}
\caption{An isothermal section of the ternary phase diagram is estimated using the $\chi_{ij}$ values in Table~\ref{tab_param} (Appendix~\ref{sec_appendix2}). The Gibbs triangle contains three single-phase (A-rich $\alpha$, B-rich $\beta$, C-rich $\gamma$) regions, three binary ($\alpha + \beta$, $\alpha + \gamma$, $\beta + \gamma$) spinodal regions, and one ternary ($\alpha + \beta + \gamma$) spinodal region. We work within the ternary spinodal region, with typical blend compositions simulated are given by ($c_A$, $c_B$, $c_C$) = (0.57, 0.38, 0.05), (0.475, 0.475, 0.05), and (0.38, 0.57, 0.05). These compositions correspond to binary mixtures of $A_{60}B_{40}$, $A_{50}B_{50}$, and $A_{40}B_{60}$, which are simulated in a matrix of pre-existing particles represented by the equilibrium composition of the $\gamma$ phase ($c_A$, $c_B$, $c_C$ = 0.035, 0.008, 0.957). In our ternary simulation setting, the binary mixtures phase separate to $\alpha$ and $\beta$ phases (instability directions) in the presence of immobile $\gamma$ particles.}\label{fig_ternary}
\end{figure}

For the set of $\chi_{ij}$ values in Table~\ref{tab_param}, we have calculated the equilibrium composition of the particle and mixture phases in the A:B:C ternary system (see Appendix~\ref{sec_appendix2}). The equilibrium compositions are presented in a typical ternary composition diagram in Fig.~\ref{fig_ternary}. In this work, we have embedded $\gamma$ particles in the matrix with a composition that is in phase equilibrium with a binary $\alpha$-$\beta$ mixture. This approach allows the particles to remain stable without appreciable composition changes during the course of our simulations. The particle composition or the equilibrium composition of the $\gamma$-phase is given by ($c_A$, $c_B$, $c_C$) = (0.035, 0.008, 0.957). We ignored the role of interface curvature in our calculations of equilibrium particle composition~\cite{Porter}; therefore, small but negligible differences in the particle profile (e.g., composition, radius) may occur during temporal evolution. However, such effects do not alter the general observations in terms of particle effects on the morphological patterns and the sequence of phase transformations.

The matrix composition needs to be chosen judiciously, i.e., above the $\alpha+\beta$ tie line within the three-phase region (Fig.~\ref{fig_ternary}) so that particles do not dissolve. Therefore, some amount of $c_C$ is required in the matrix. A high composition of $c_C$ would imply the coarsening of C-rich particles. Thus, the matrix composition should be selected just above the $\alpha+\beta$ two-phase region with a minimal amount of $c_C$ such that particles neither dissolve nor do they coarsen appreciably. Our simulated matrix compositions limit the coarsening of the particles and the composition changes in the particle below 2~\%. Particle coarsening can be restricted further by using a matrix composition that is even closer to the $\alpha+\beta$ tie line but within the three-phase region as long as the particle does not dissolve; however, it does not alter the general observations of particle effects on composition patterns in our work.

The matrix composition was chosen according to the ``effective binary'' mixture studied, critical or off-critical. Unless otherwise specified, the matrix composition was chosen as ($c_A$, $c_B$, $c_C$) = (0.475, 0.475, 0.05) for $A_{50}B_{50}$ critical mixture. This matrix composition phase separates with steady-state compositions of $\alpha$ and $\beta$ phases close to the estimated equilibrium values. Corresponding to the above critical matrix composition, two off-critical mixtures were also simulated, $A_{60}B_{40}$ and $A_{40}B_{60}$, with respective initial compositions of ($c_A$, $c_B$, $c_C$) = (0.57, 0.38, 0.05) and ($c_A$, $c_B$, $c_C$) = (0.38, 0.57, 0.05). A small, random noise of $\pm$ 0.005 was added to the matrix composition at the start of our simulations for the phase separation of A:B to begin.

Finally, we model a test case where C-rich particles are immobile in the phase-separating A:B matrix. Substituting $M_C = 0$ and using the matrix composition ($c_A$, $c_B$, $c_C$) in Eq.~\eqref{eq_mobility}, we obtain the effective mobilities for which the mobility of particles becomes zero. The scaled mobilities used in our simulations for $A_{50}B_{50}$ mixture are given by: $M_{AA} = M_{BB} = 1.0$ and $M_{AB} = 0.995$ for which the determinant of the mobility matrix ($M_{AA}M_{BB}-M_{AB}^2$) remains positive definite~\cite{Ghosh,Huang}.

%------------------------------------------------------
\section{Results and Analysis}\label{sec_results}

\subsection{Single-Particle Effects in a Critical Blend}
\subsubsection{General Remarks}
We begin with the simulations of single-particle effects on SDSD in a critical blend. Due to the selective preference, at first, a thin layer of A-rich $\alpha$ forms as a ring around the particle. This results in a depletion of A in the immediate vicinity of $\alpha$, leading to the formation of a $\beta$ ring around $\alpha$. Such alternate concentric rings of $\alpha$ (preferred phase) and $\beta$ (non-preferred phase) around the particle are referred to as the target pattern (Fig.~\ref{target1}). The target pattern is illustrated using the alternate concentration profiles of the rings in Fig.~\ref{comp1}. The number and thickness of the rings in such patterns depend on the magnitude of relative interfacial energy ($\sigma_{ij}$) between coexisting phases. Such target pattern formation during SDSD was nicely realized in experiments on solid metallic mixtures by Aichmayer et al.~\cite{aichmayer2003} and experiments and simulations on particle-filled polymer blends by Karim et al.~\cite{Karim} and Lee et al.~\cite{Lee}

Followed by phase separation at early times, the phase inversion process takes place within the target pattern around the particle surface. Phase inversion is a curvature-driven coarsening process. Due to the Gibbs-Thomson effect~\cite{Porter,voorhees1985}, solute concentration in the matrix adjacent to each target ring increases as the radius of curvature decreases. The resultant concentration gradient in the target pattern leads to solute diffusion in the direction of the target ring (of the same phase) having the largest radius of curvature away from the smallest. As a result, the inner $\alpha$ ring in the target pattern shrinks and eventually disappears from the particle surface, exposing the next $\beta$ ring to surround the particle (Figs.~\ref{target2}, \ref{comp2}). We are not aware of any \textit{in-situ} experimental measurement of \textit{transient} phase inversion in blends that alters the sequence of phases in the target pattern. However, in related measurements, long-time instability of target patterns around particles in chemically patterned substrates~\cite{Karim,Lee} and particle-induced composition changes affecting the sequence of morphologies during phase inversion in polymer blend films have been observed~\cite{amoabeng2017,domenech2017}.

At late times, surrounding the target pattern, phase separation takes place in the bulk region followed by phase coarsening in the matrix. This results in continuous $\alpha$ and noncontinuous $\beta$ in the form of ``background'' spinodal pattern (Fig.~\ref{target2}). However, for the simulated matrix composition that is symmetric both in A and B (i.e., $A_{50}B_{50}$), a bicontinuous morphology is quite common in experiments and simulations~\cite{Puri,Benderly,Karim,Brown,Chakrabarti,Ma}. The general tendency of symmetry-breaking in bulk (i.e., bi-continuity) is due to the presence of the particle and its selective preference for $\alpha$, noting that, symmetric blend either without particles or with particles but no preference for both $\alpha$ and $\beta$ always exhibits a bicontinuous microstructure typical of spinodal decomposition~\cite{ghosh_pccp}. Therefore, the formation of noncontinuous $\beta$ in a symmetric matrix is, by all means, a particle effect. It is the phase with preferential affinity to the particle forms a continuous structure as the non-preferred phase tends to become noncontinuous. We explore later in Sec.~\ref{sec_mp_systems} that by changing the volume fraction of the particle in multi-particle systems, we can control the apparent symmetry of the composition, namely, whether the final pattern becomes continuous or isolated.

\subsubsection{Size Effects}
As a reference, we performed preliminary calculations on the effects of particle size on target and bulk spinodal patterns. Although not shown here, the finite size of the filler particle does not affect the steady-state thickness of the wetting layer surrounding it. However, particle size somewhat affects the duration of the transient target pattern. This is illustrated with simulations of different particle radii in Fig.~\ref{fig_target2}. At $t = 1500$, the $\alpha$ ring (next to particle surface) in the target pattern disappears at the end of the phase inversion process in small-particle simulations (Fig.~\ref{target11}), while $\alpha$ is still undergoing phase inversion in large-particle simulations (Fig.~\ref{target22}) that eventually dissolves at a later time ($t = 2000$). However, the size effect is limited in the late stages ($t = 3000$) when the scale of phase separation exceeds the size of the filler particle (compare Figs.~\ref{target2}, \ref{target33}). The size effects mentioned above are likely universal in particle-filled blends~\cite{Tanaka}, irrespective of simulation settings such as particle size, particle shape, and their distribution.

\subsubsection{Domain Growth} 
\begin{figure}[h]
\centering
\subfloat[$R = 8, t = 200$]{\label{target1}\includegraphics[trim={20cm 15 20 15},clip,scale=0.05]{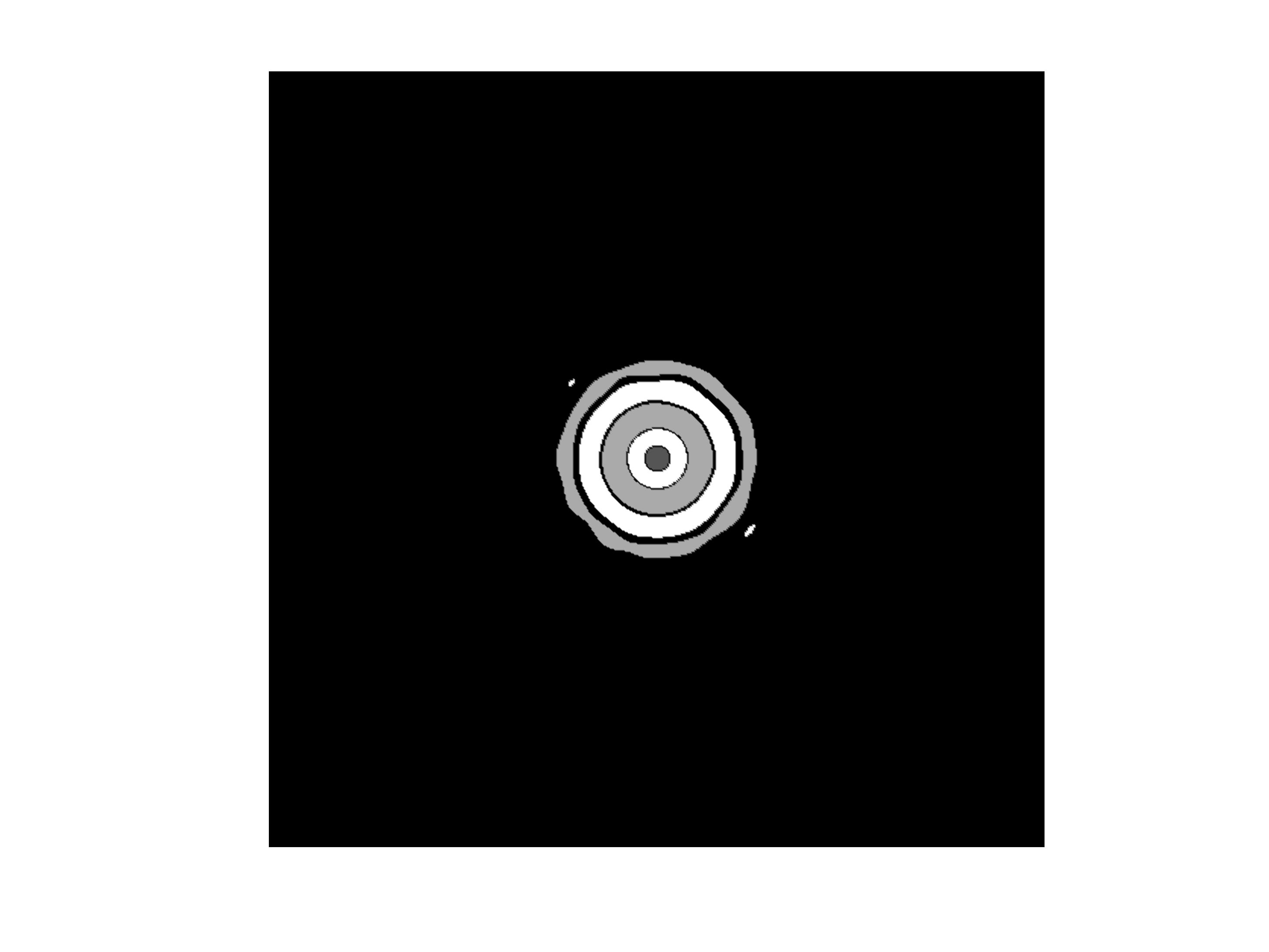}}\hspace{-8mm}
\subfloat[$t = 200$]{\label{comp1}\includegraphics[scale=0.42]{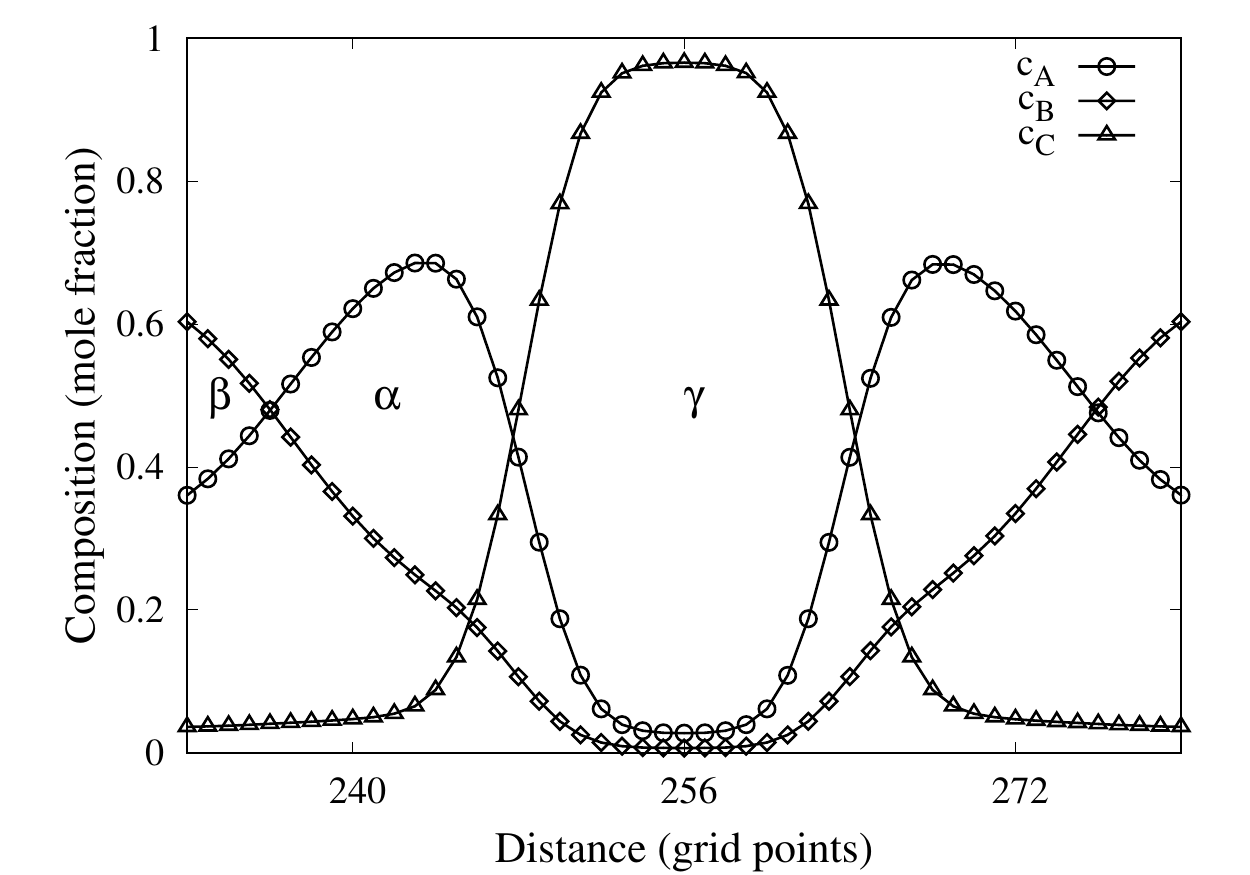}}\hspace{-1mm}
\subfloat[$R = 8, t = 3000$]{\label{target2}\includegraphics[trim={20cm 15 20 15},clip,scale=0.05]{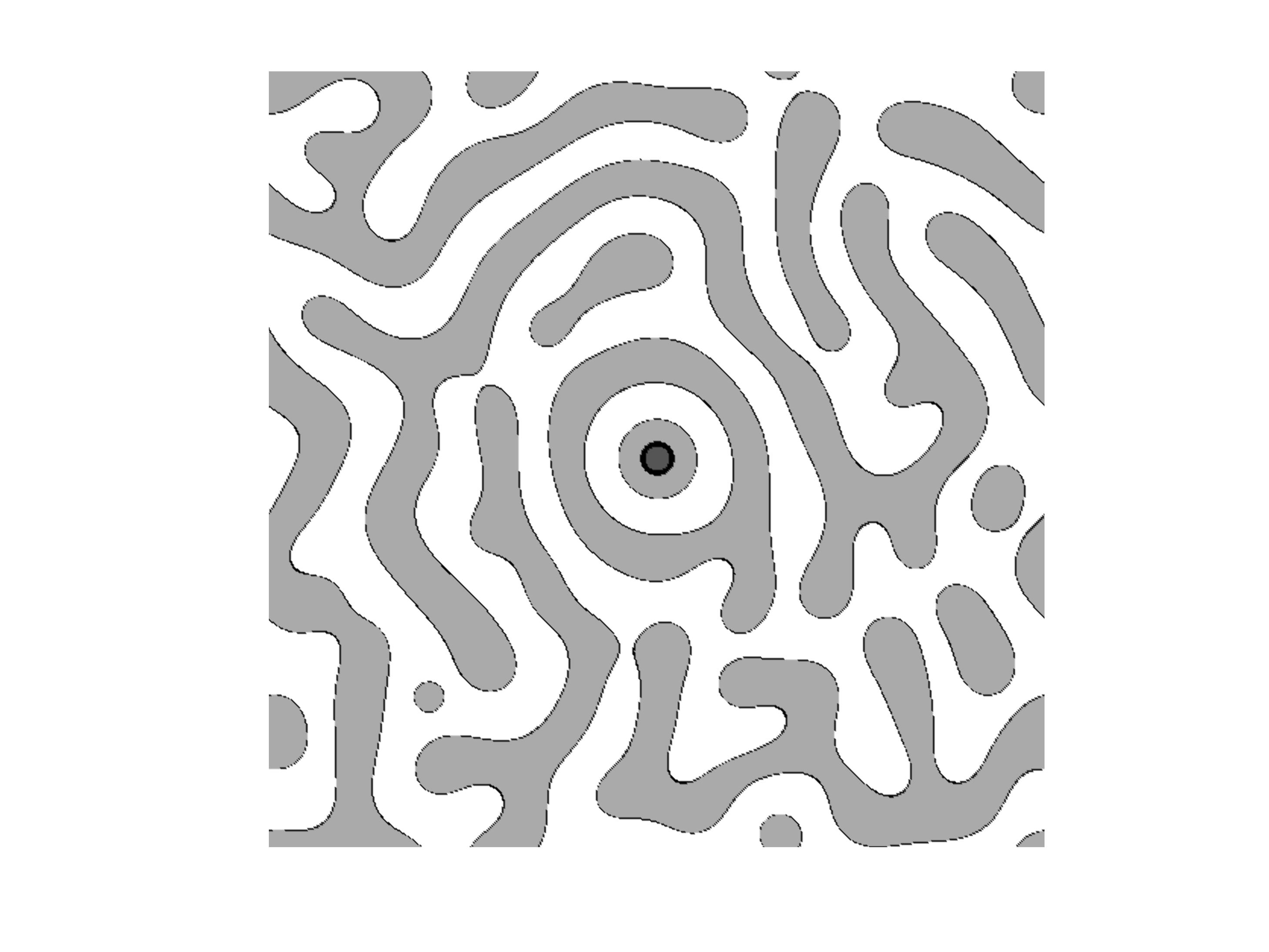}}\hspace{-8mm}
\subfloat[$t = 3000$]{\label{comp2}\includegraphics[scale=0.42]{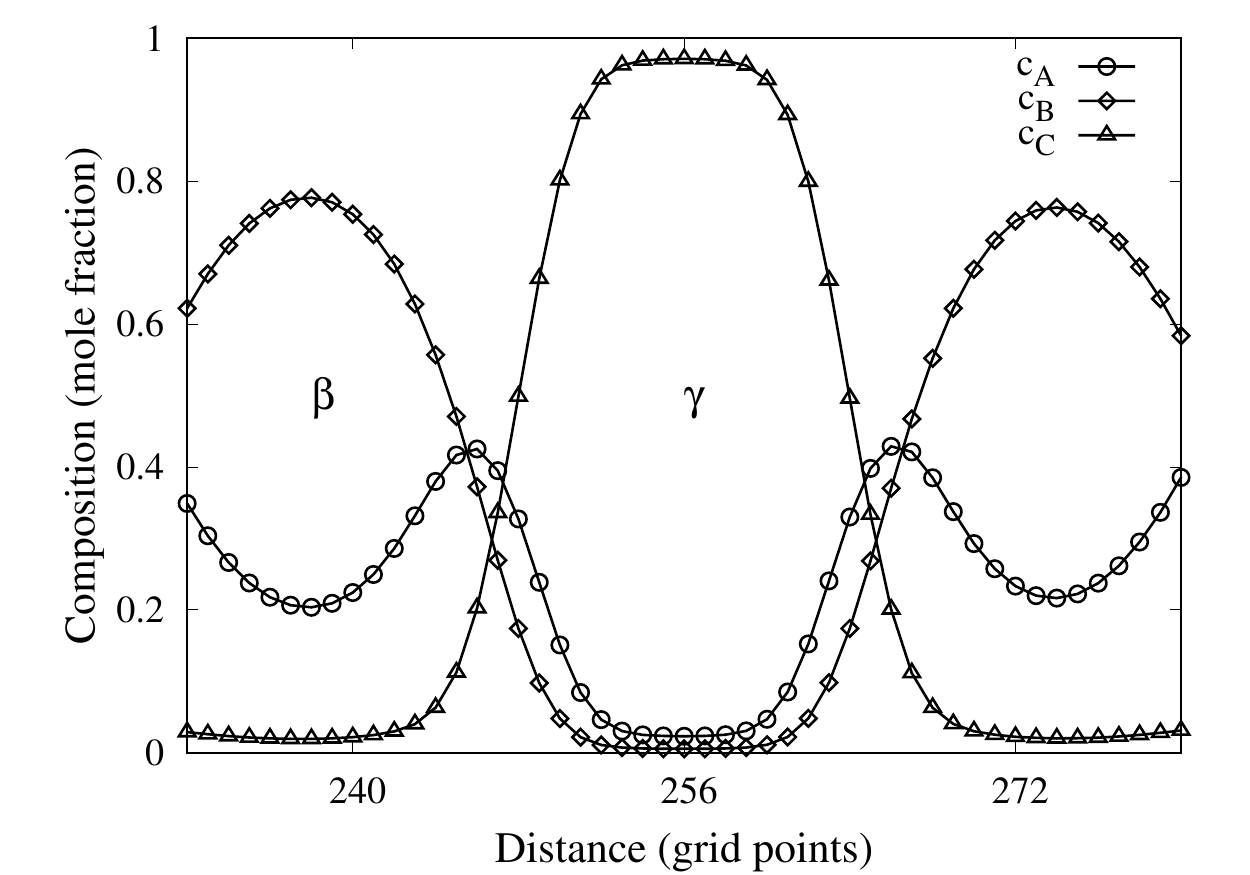}}
%\subfloat[$S_{A}$ vs. $k$]{\label{target2}\includegraphics[scale=0.5]{sp_sf}}
\caption{(a) At early times, the target pattern forms around the particle with $R = 8$ units. (b) The alternate concentration profile across the $\gamma$-particle (in Fig.~\ref{target1}) is shown. (c) At later times, the background spinodal pattern reigns far away from the particle. (d) The concentration profile across the $\gamma$-particle (in Fig.~\ref{target2}) is shown. In Figs.~\ref{target1} and \ref{target2}, $\alpha$, $\beta$, and $\gamma$ are illustrated by white, light gray, and dark gray, respectively. The initial matrix composition is ($c_A$, $c_B$, $c_C$) = (0.48, 0.48, 0.04) and the $\gamma$-particle composition is ($c_A$, $c_B$, $c_C$) = (0.035, 0.008, 0.957).}
\label{fig_target}
\end{figure}

\begin{figure}[h]
\centering
\subfloat[$R = 8$, $t = 1500$]{\label{target11}\includegraphics[trim={20cm 15 20 15},clip,scale=0.05]{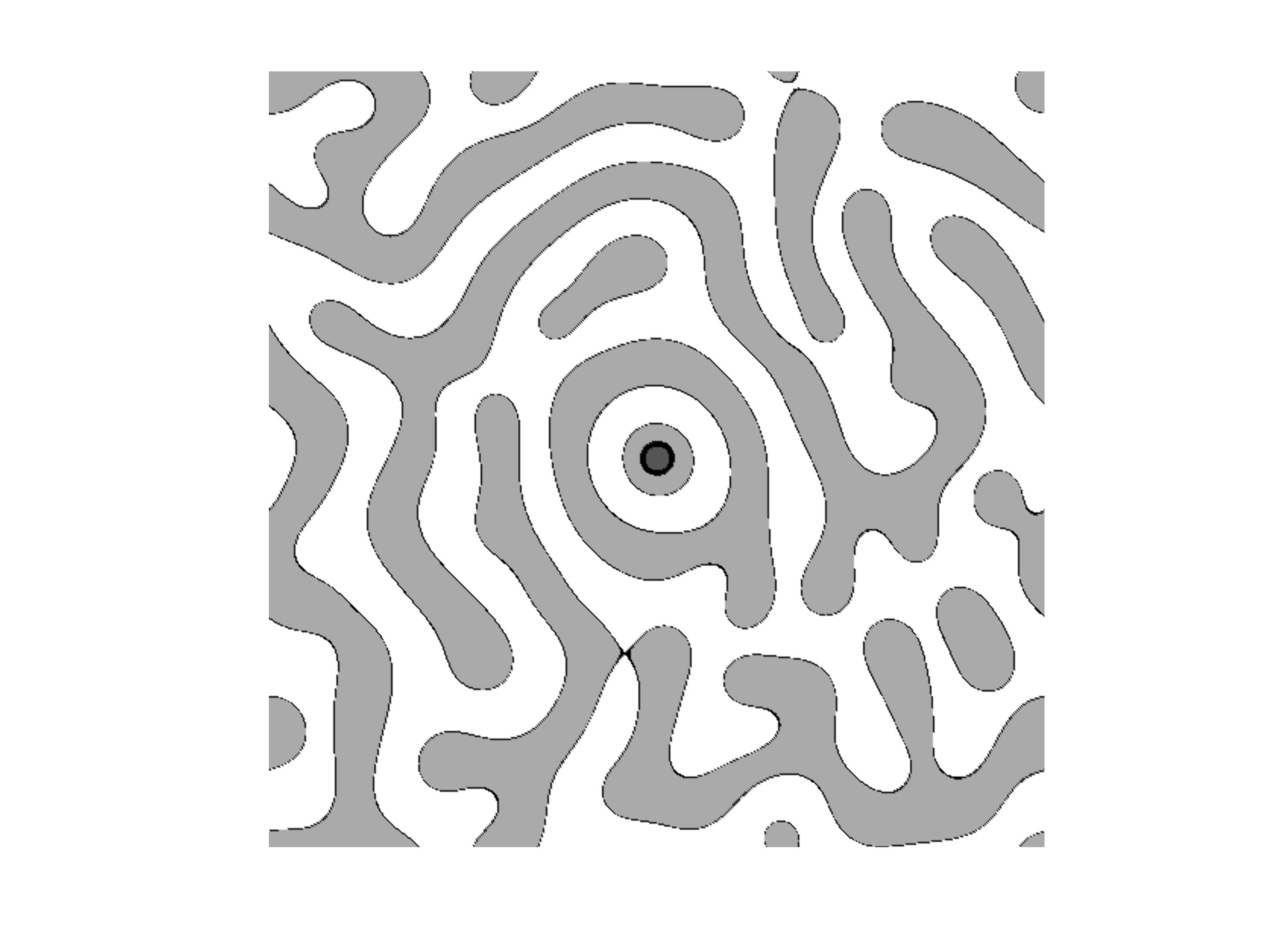}}\hspace{-2mm}
\subfloat[$R = 16$, $t = 1500$]{\label{target22}\includegraphics[trim={20cm 15 20 15},clip,scale=0.05]{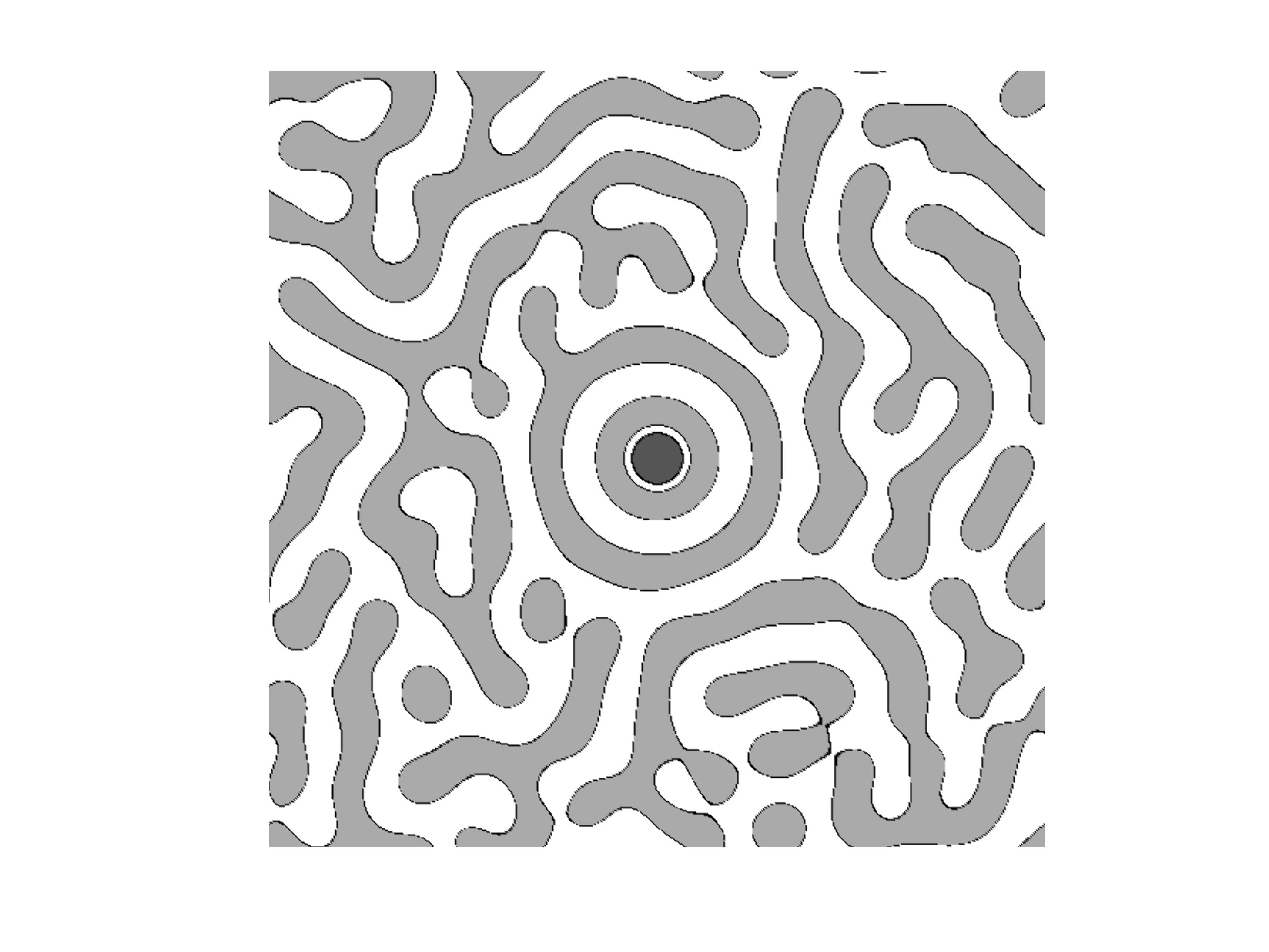}}\hspace{-2mm}
\subfloat[$R= 16$, $t = 3000$]{\label{target33}\includegraphics[trim={20cm 15 20 15},clip,scale=0.05]{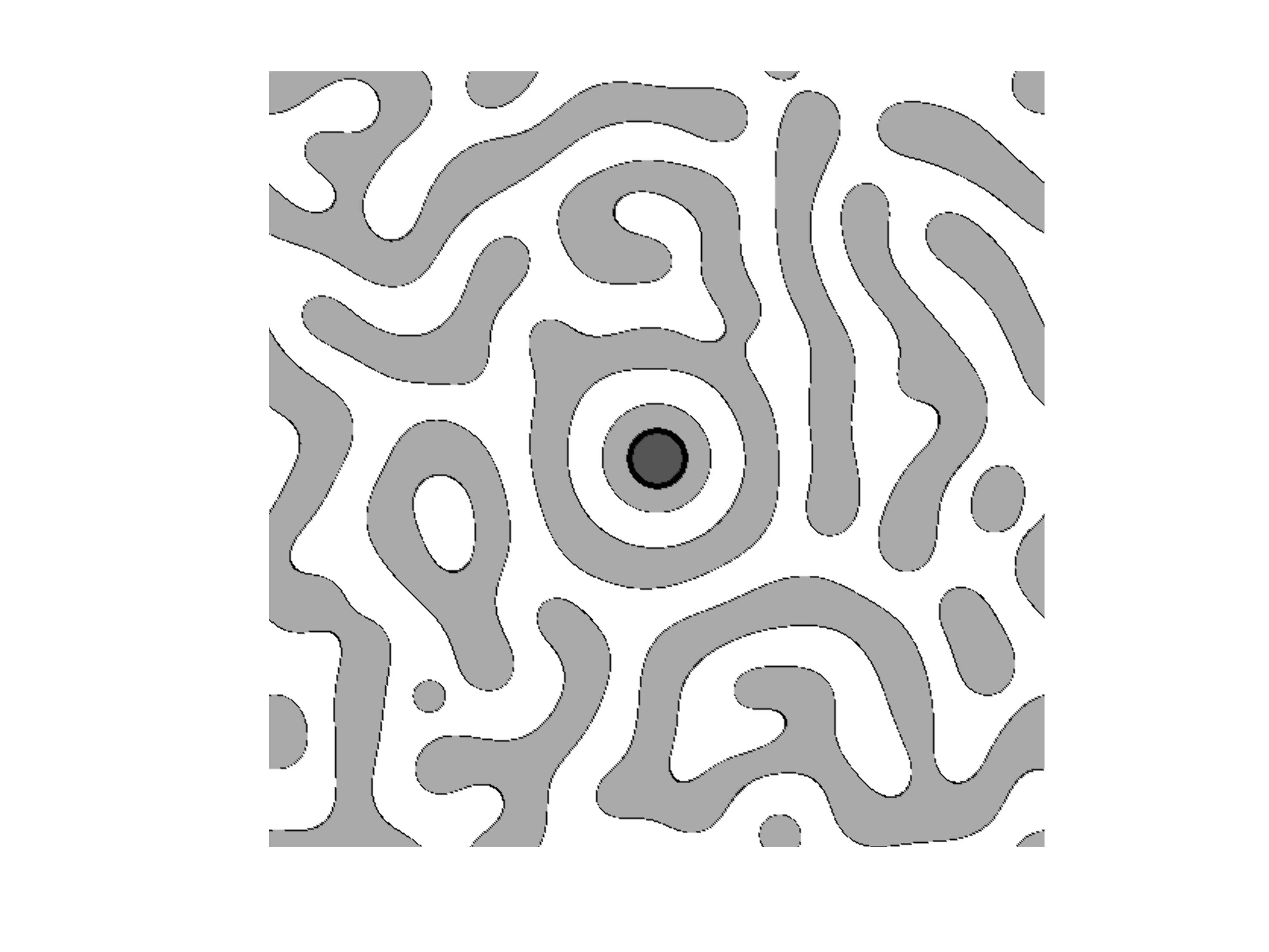}}\hspace{-8mm}
\subfloat[$S_{A}$ vs. $k$]{\label{target3}\includegraphics[scale=0.42]{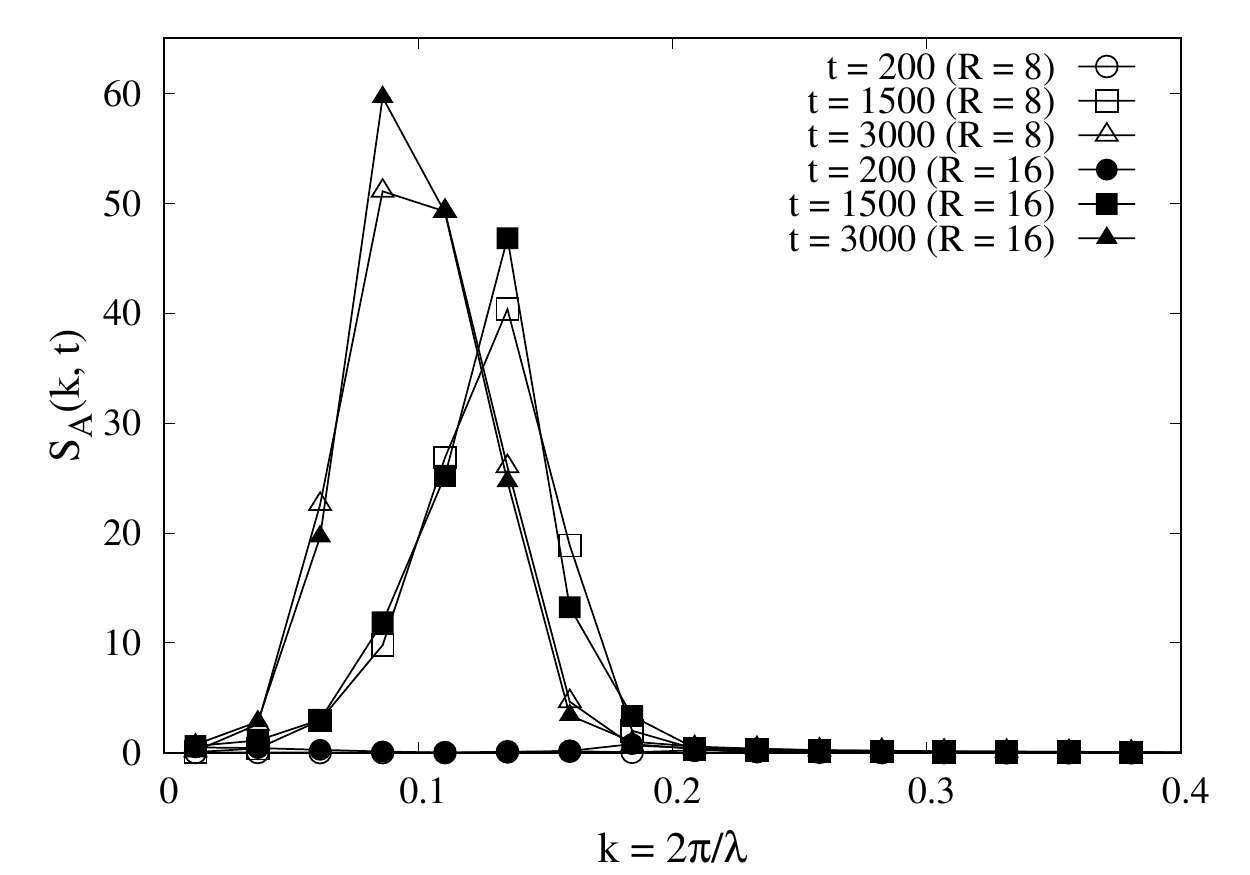}}
\caption{The target and bulk spinodal patterns develop around the particle with (a) $R = 8$ units and (b, c) $R = 16$ units.  The $\alpha$, $\beta$, and $\gamma$ are illustrated by white, light gray, and dark gray, respectively. (d) The length scale of the spinodal pattern at different times is shown for both particle radii.}
\label{fig_target2}
\end{figure}

Domain formation and growth in the target pattern are characterized using circularly-averaged structure function~\cite{chakrabarti1989late, zhu1999coarsening} $S_{i}(k,t)$ of component $i$ given by,
\begin{equation}\label{eq_sf}
S_{i}(k,t) = \frac{1}{N}\left\langle c_{i}^*(\textbf{k},t)c_{i}(\textbf{k},t)\right\rangle,
\end{equation}
where $N$ is the lattice size and $k$ the magnitude of the wave vector $\textbf{k}$. The profiles of $S_i(k,t)$ compare to the power spectrum analysis of the measurement data obtained from related scattering experiments~\cite{Karim}. Figure~\ref{target3} shows the evolution of $S_{A}$ over time. The length scale of the target pattern is given by the tiny peak of the curve at $t = 200$; this is a reference time when only the target pattern exists with no phase separation in bulk (Fig.~\ref{target1}). At later times, spinodal decomposition occurs in bulk, and the scale of the resultant pattern grows with time. When the length scale of bulk exceeds that of the target pattern, the innermost $\alpha$ ring in the target pattern begins to shrink and eventually disappear (Fig.~\ref{target11}) while the background continues to coarsen over time to realize Figs.~\ref{target2},~\ref{target33}. The net effect of particle size on domain dynamics increasingly becomes negligible as time lapses (Fig.~\ref{target3}). Henceforth, we limit our simulations with $R = 8$ units.

\subsection{Multi-Particle Systems}\label{sec_mp_systems}
 
\subsubsection{Particle Effects in a Critical Blend}
Technological applications often involve multi-particle systems with particles distributed in specific configurations. A periodic template of particles that are arranged symmetrically with a periodicity $\lambda$ in a homogeneous $A_{50}B_{50}$ matrix is simulated. The matrix phase separates to A-rich $\alpha$ and B-rich $\beta$ microdomains, the distribution and size of which depend on $\lambda$. The value of $\lambda$ is varied in our simulations, which correspond to low-$\lambda$ ($\lambda = 32, 48$), intermediate-$\lambda$ ($\lambda = 64$), and high-$\lambda$ ($\lambda = 96$) systems, when compared to the spinodal wavelength involved in these systems. The rationale behind using these values of $\lambda$ will be explained later in the text. We note that $\lambda$ can be construed as particle loading in our simulations. The values of $\lambda$ = 32, 48, 64, and 96 correspond to respective particle loadings of $\approx$ 20~\%, 9~\%, 5~\%, and 2~\%. The particle fraction in each simulation remains nearly constant because particles are rendered immobile following Sec.~\ref{sec_parameters}.
%We note that, similar to $\lambda$, our simulations can be classified between low and high particle-filled systems since the particle loading varied between dilute (2~\%) and large (20~\%) amounts.

\begin{figure}[h]
\centering
\subfloat[$\lambda = 32$]{\label{sa50b50_32}\includegraphics[trim={20cm 15 20 15},clip,scale=0.04]{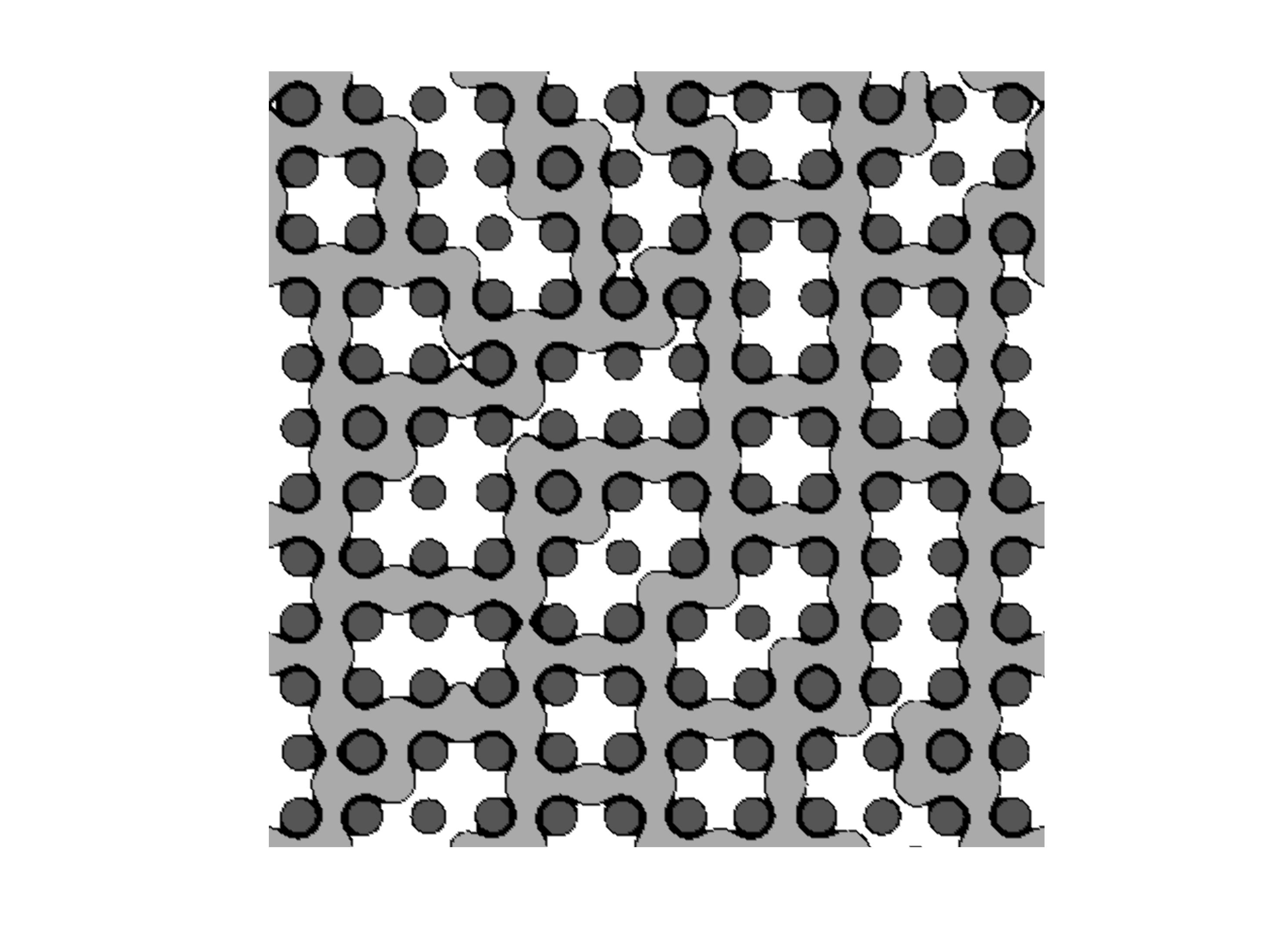}}
\subfloat[$\lambda = 48$]{\label{sa50b50_48}\includegraphics[trim={20cm 15 20 15},clip,scale=0.04]{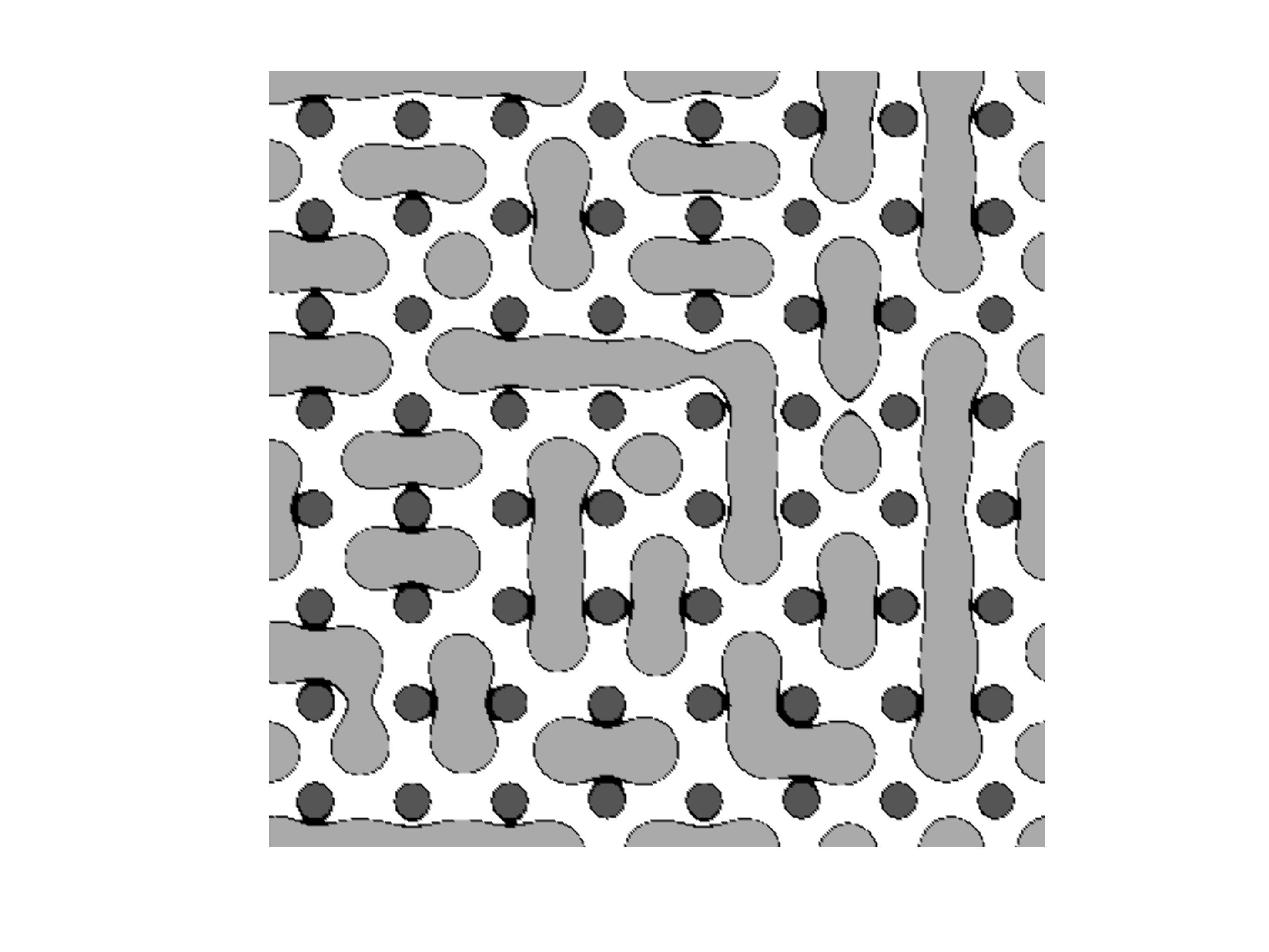}}
\subfloat[$\lambda = 64$]{\label{sa50b50_64}\includegraphics[trim={20cm 15 20 15},clip,scale=0.04]{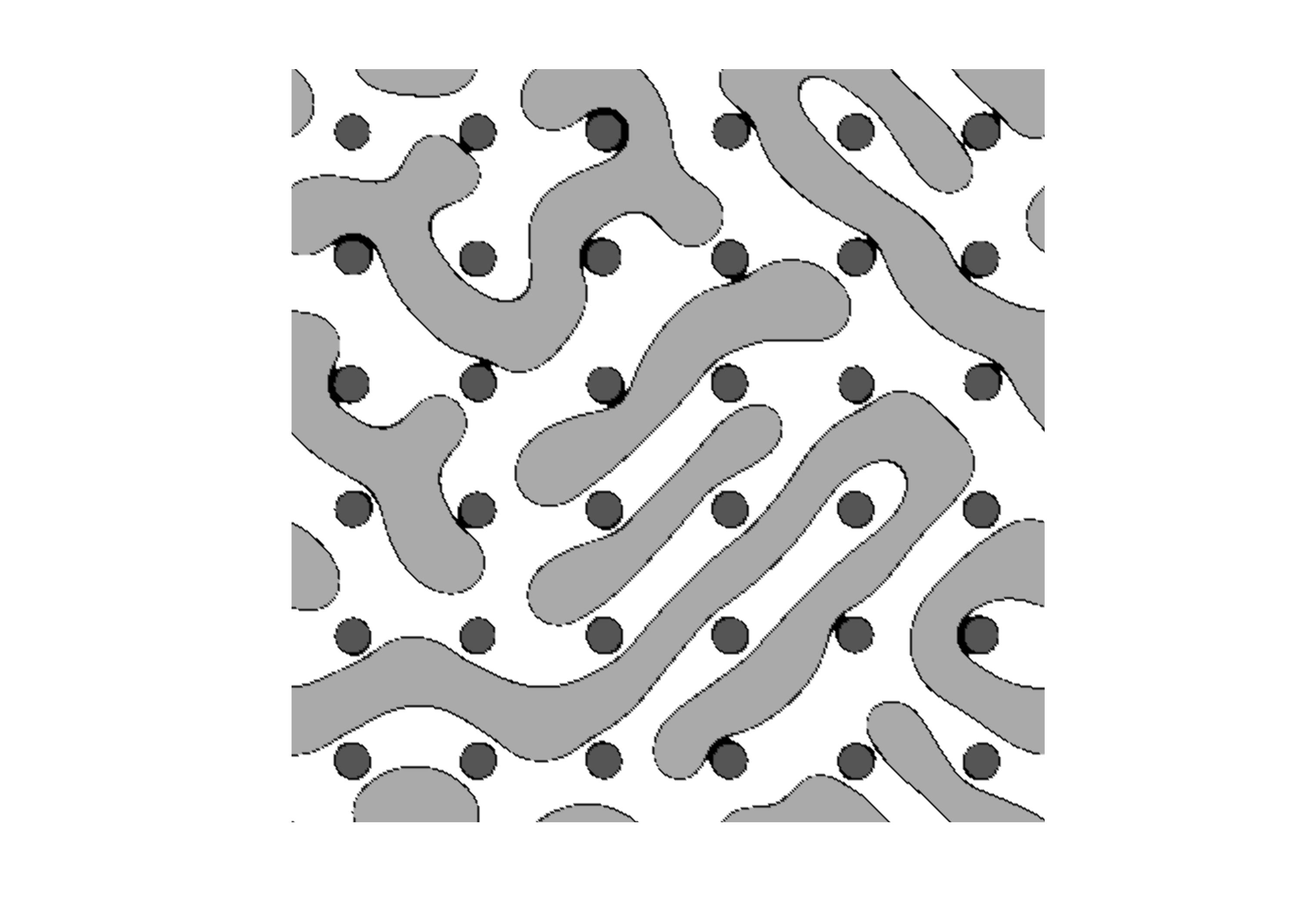}}
\subfloat[$\lambda = 96$]{\label{sa50b50_96}\includegraphics[trim={20cm 15 20 15},clip,scale=0.04]{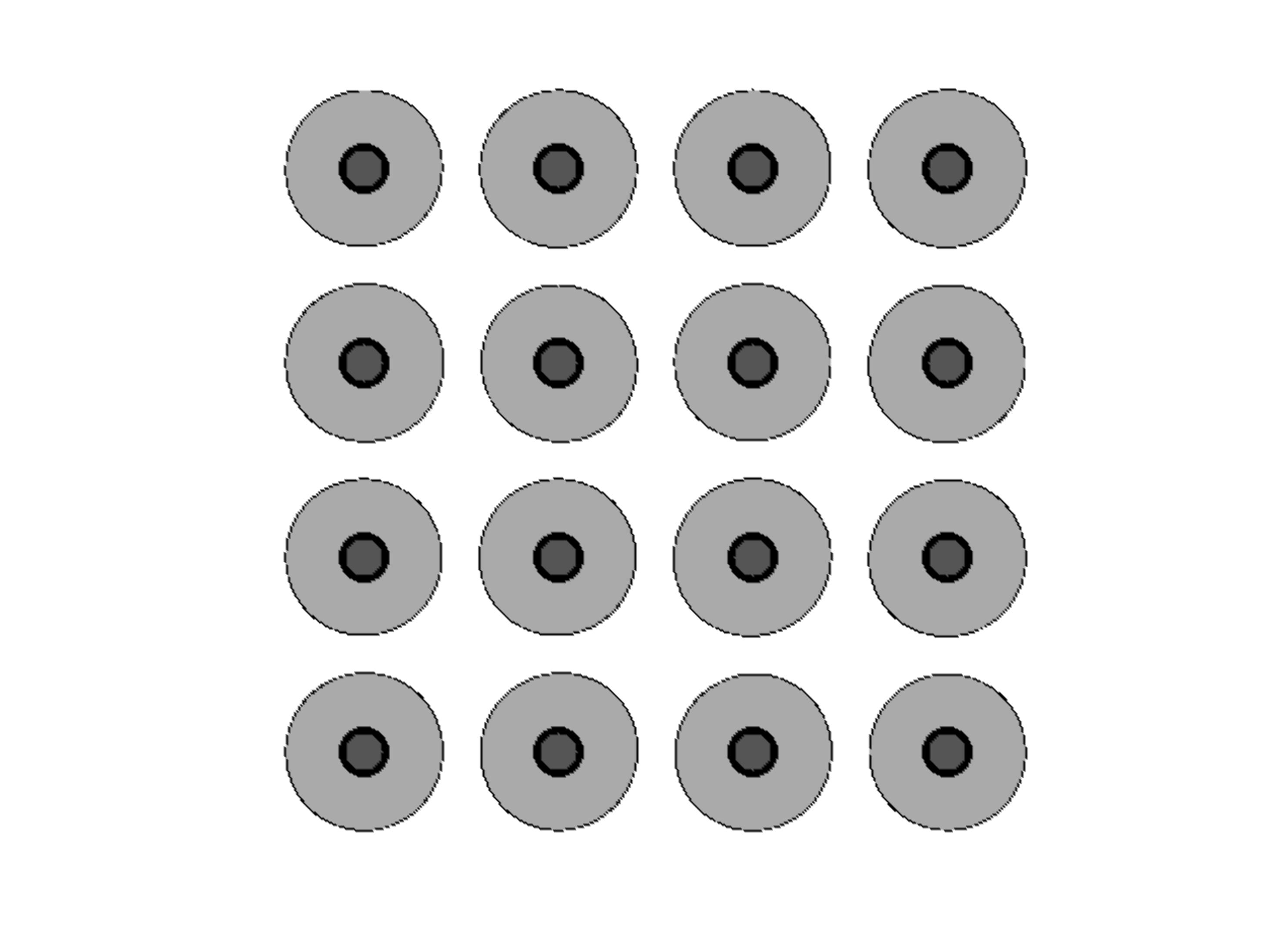}}
\subfloat[$\lambda = 128$]{\label{sa50b50_128}\includegraphics[trim={20cm 15 20 15},clip,scale=0.04]{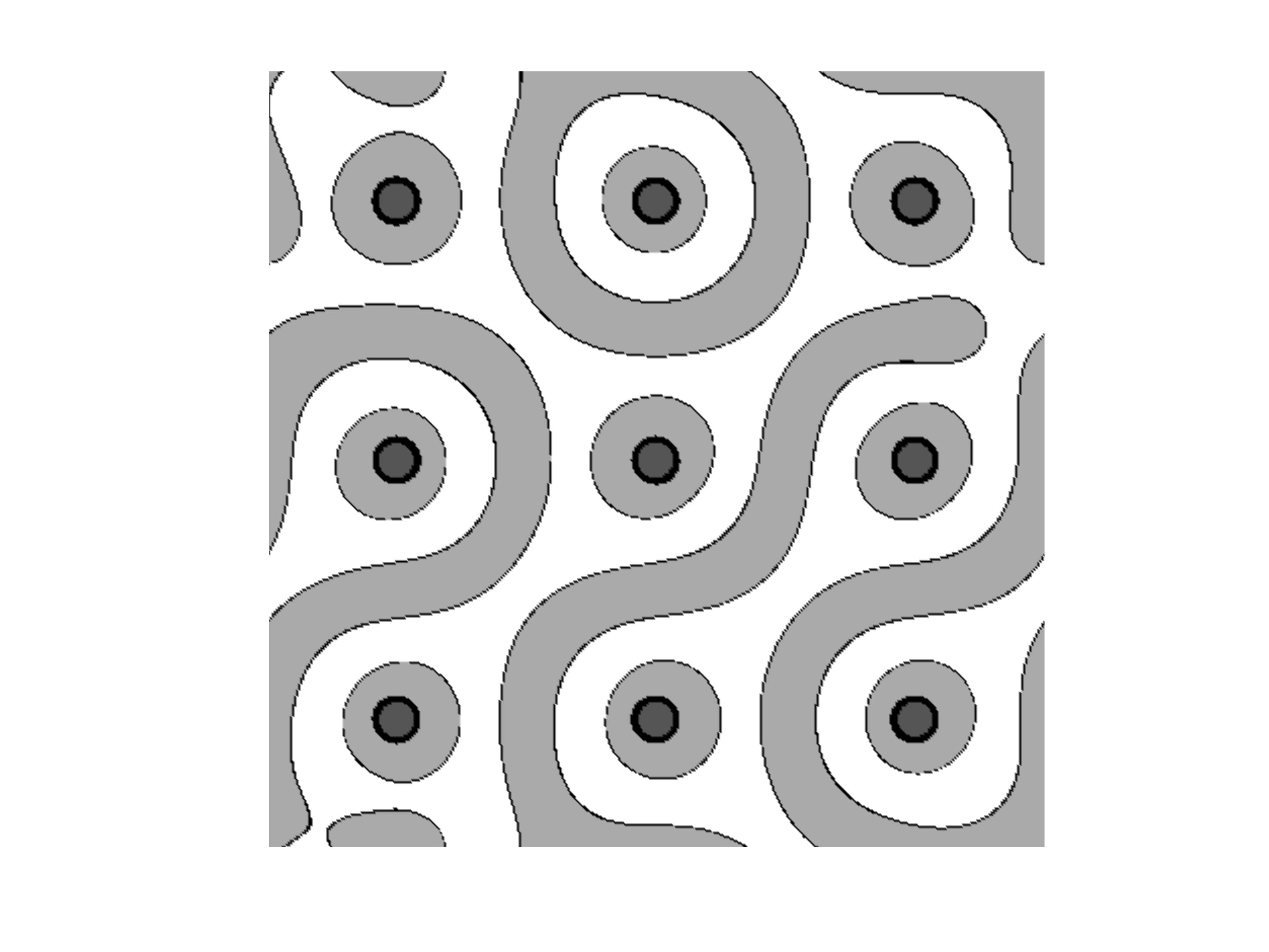}}
\caption{$A_{50}B_{50}$: SDSD microstructures that form in a critical blend in the presence of \textit{symmetric} distribution of particles with varying $\lambda$: (a) $\lambda = 32$ (b) $\lambda = 48$ (c) $\lambda = 64$ (d) $\lambda = 96$ (e) $\lambda = 128$ are presented. All lengths are in grid units. The snapshots correspond to the dimensionless times (a) $t = 10000$ and (b, c, d, e) $t = 4000$. The $\alpha$, $\beta$, and $\gamma$ phases are illustrated by white, light gray, and dark gray, respectively.}
\label{fig_symmetric}
\end{figure}

In low-$\lambda$ systems, the mixture $A_{50} B_{50}$ rapidly phase separates to irregular microdomains of $\alpha$ (with the particles inside them) in continuous $\beta$ (Fig.~\ref{fig_symmetric}a) or irregular domains of $\beta$ staggered between $\gamma$ particle arrays in continuous $\alpha$ (Fig.~\ref{fig_symmetric}b). The growth of $\beta$ in these systems is guided by the circular particles, blocking the local $\beta$-composition waves. The energy minimizing shapes of $\beta$ in low-$\lambda$ systems are due to the area constraint within and around the particle arrays. In intermediate-$\lambda$ systems, the target pattern around each particle disappears, forming a typical transition pattern (Fig.~\ref{fig_symmetric}c). With the increasing value of $\lambda$, the diffusion of species will be of long-range, leading to enhanced phase coarsening compared to that of in low-$\lambda$ systems. 

Pattern evolution is significantly different in high-$\lambda$ systems (Fig.~\ref{fig_symmetric}d). Due to the large value of $\lambda$, many alternate rings of $\alpha$ and $\beta$ form around each particle at early times. This is followed by curvature-driven coarsening at later times. Due to the Gibbs-Thomson effect~\cite{Porter}, the ring with the smallest radius of curvature, i.e., $\alpha$ ring, shrinks and eventually disappears, bringing the $\beta$ ring next to the particle surface. Although not shown here, the outermost $\beta$ ring in the target pattern with many rings always breaks and then reconnects to the background phase separation at later times. Qualitatively, the SDSD pattern, in particular the target morphology, remains similar with increasing $\lambda$ or decreasing particle fraction. This is illustrated by the simulation with $\lambda= 128$ and particle loading $\approx 1~\%$ (Fig.~\ref{fig_symmetric}e). Henceforth, we limit our simulations up to $\lambda = 96$. 

\begin{figure}[h]
\centering
\subfloat[$\lambda = 32$]{\label{asa50b50_32}\includegraphics[trim={20cm 15 20 15},clip,scale=0.05]{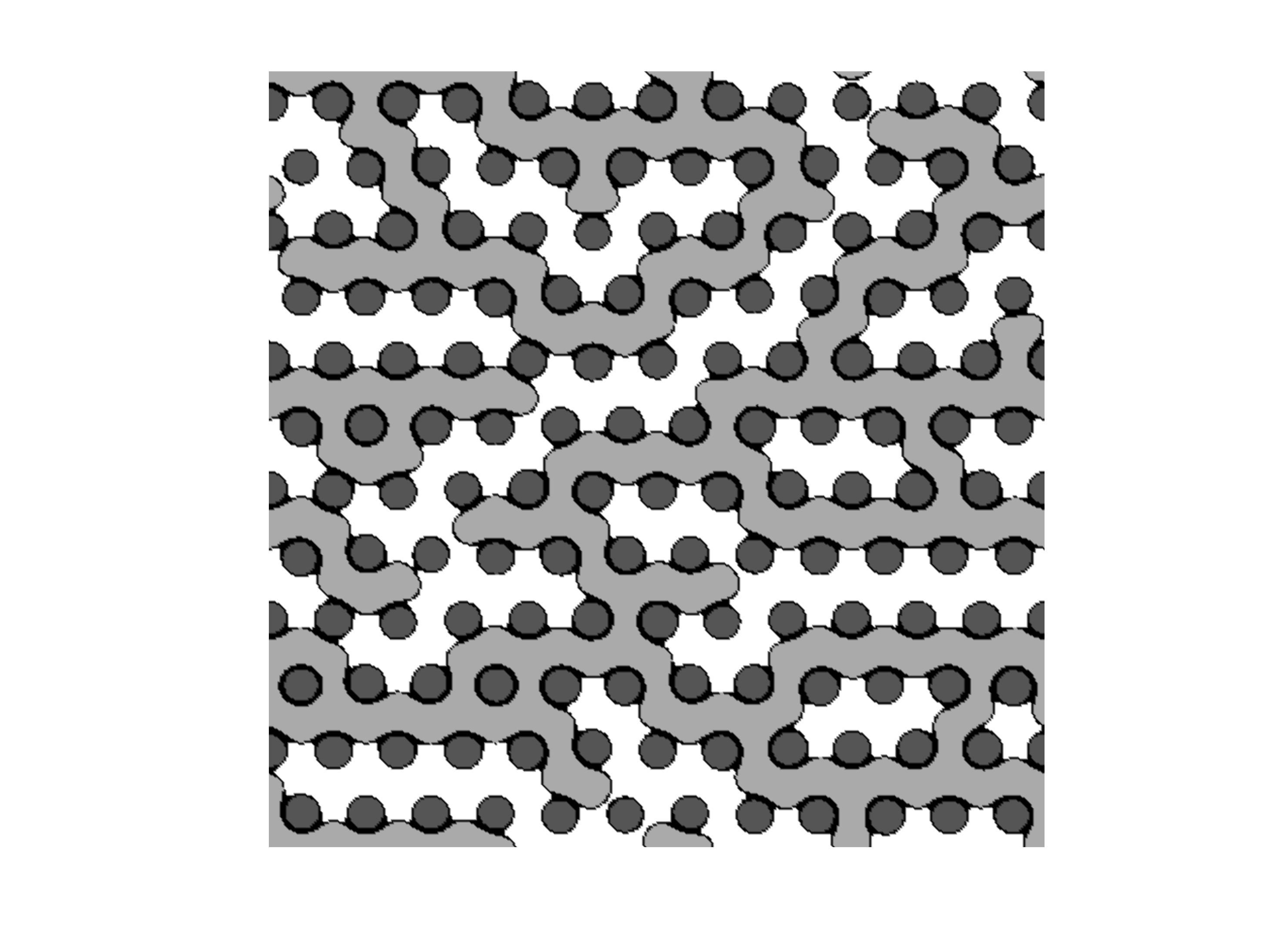}}
\subfloat[$\lambda = 48$]{\label{asa50b50_48}\includegraphics[trim={20cm 15 20 15},clip,scale=0.05]{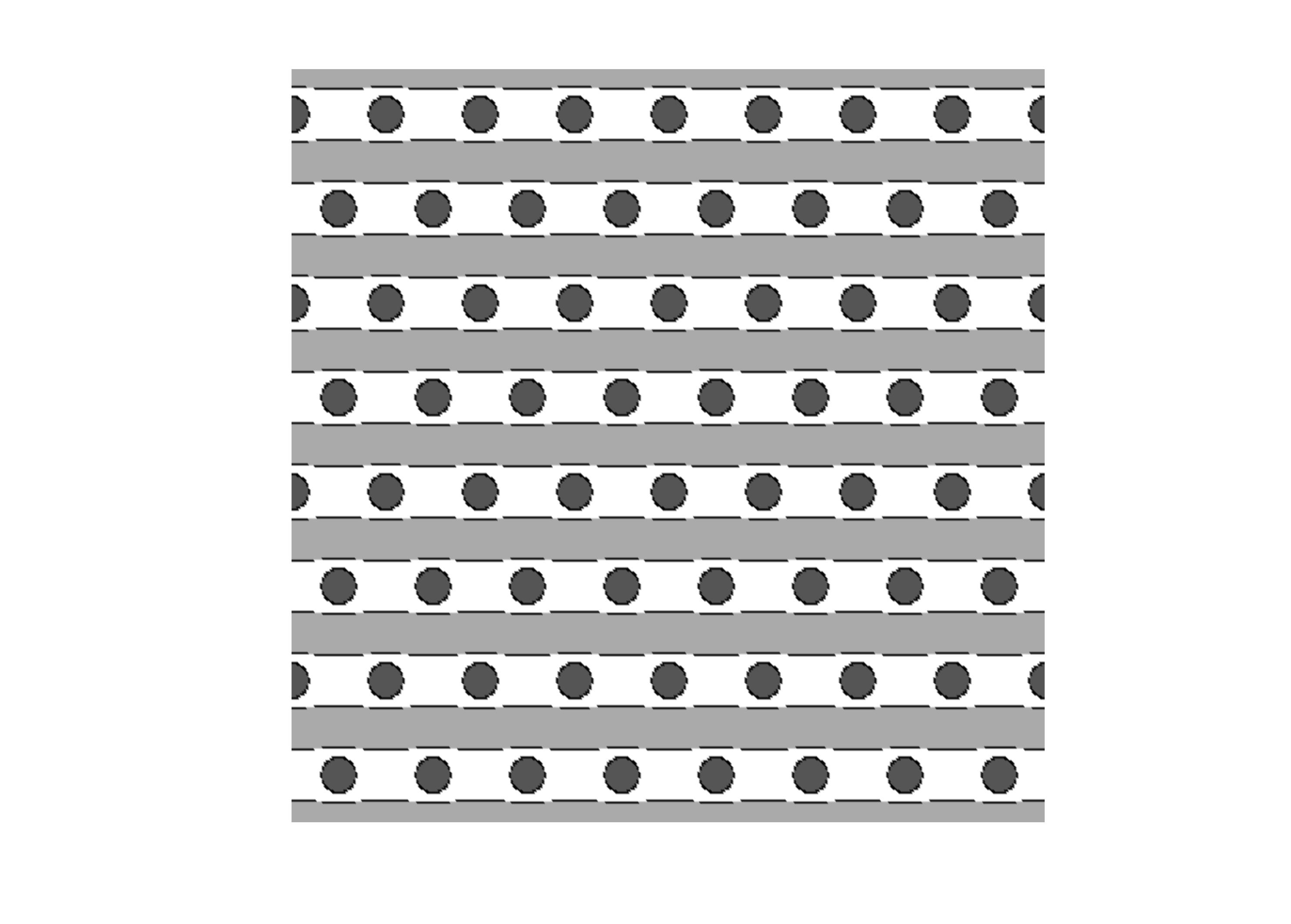}}
\subfloat[$\lambda = 64$]{\label{asa50b50_64}\includegraphics[trim={20cm 15 20 15},clip,scale=0.05]{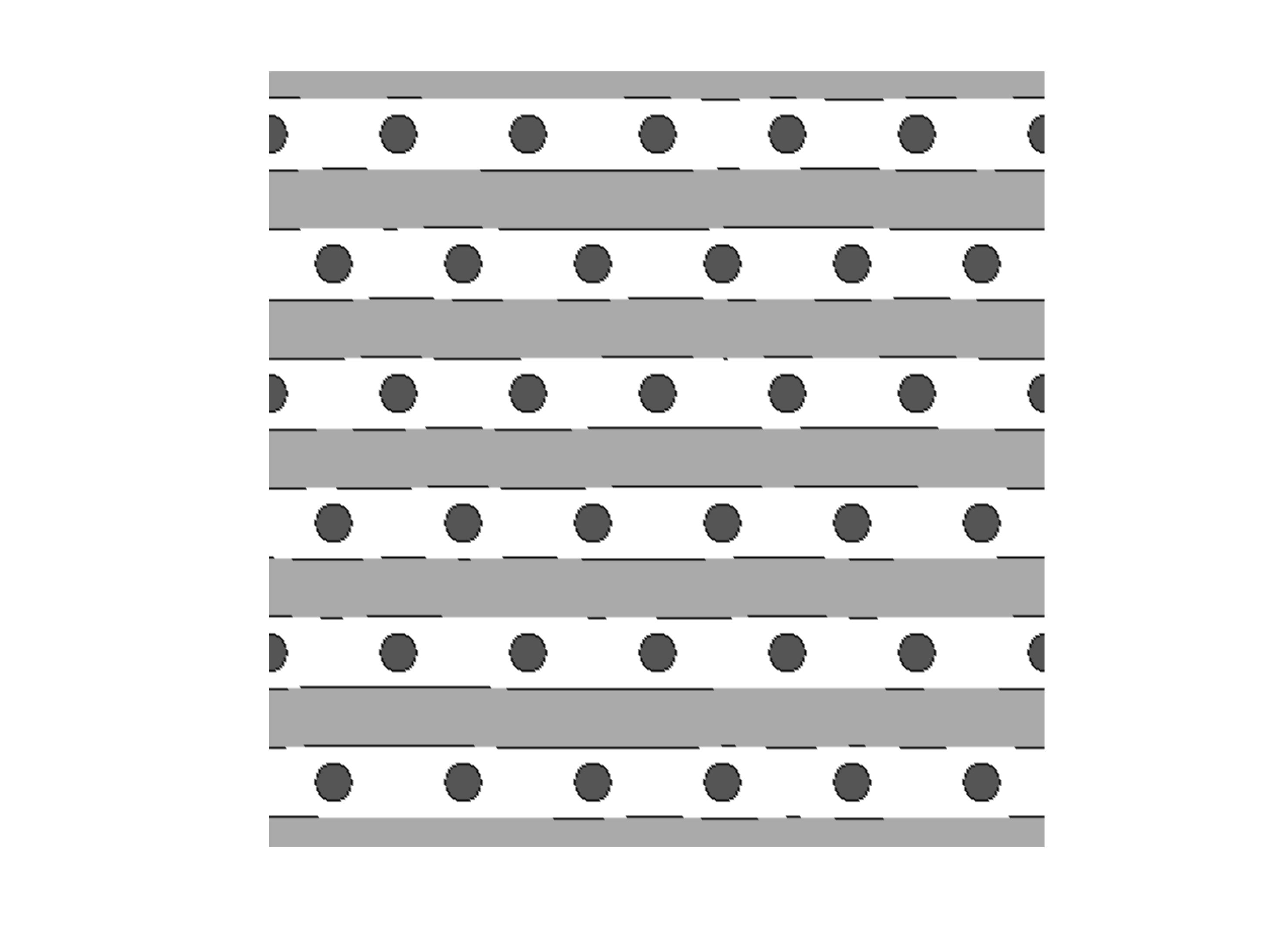}}
\subfloat[$\lambda = 96$]{\label{asa50b50_96}\includegraphics[trim={20cm 15 20 15},clip,scale=0.05]{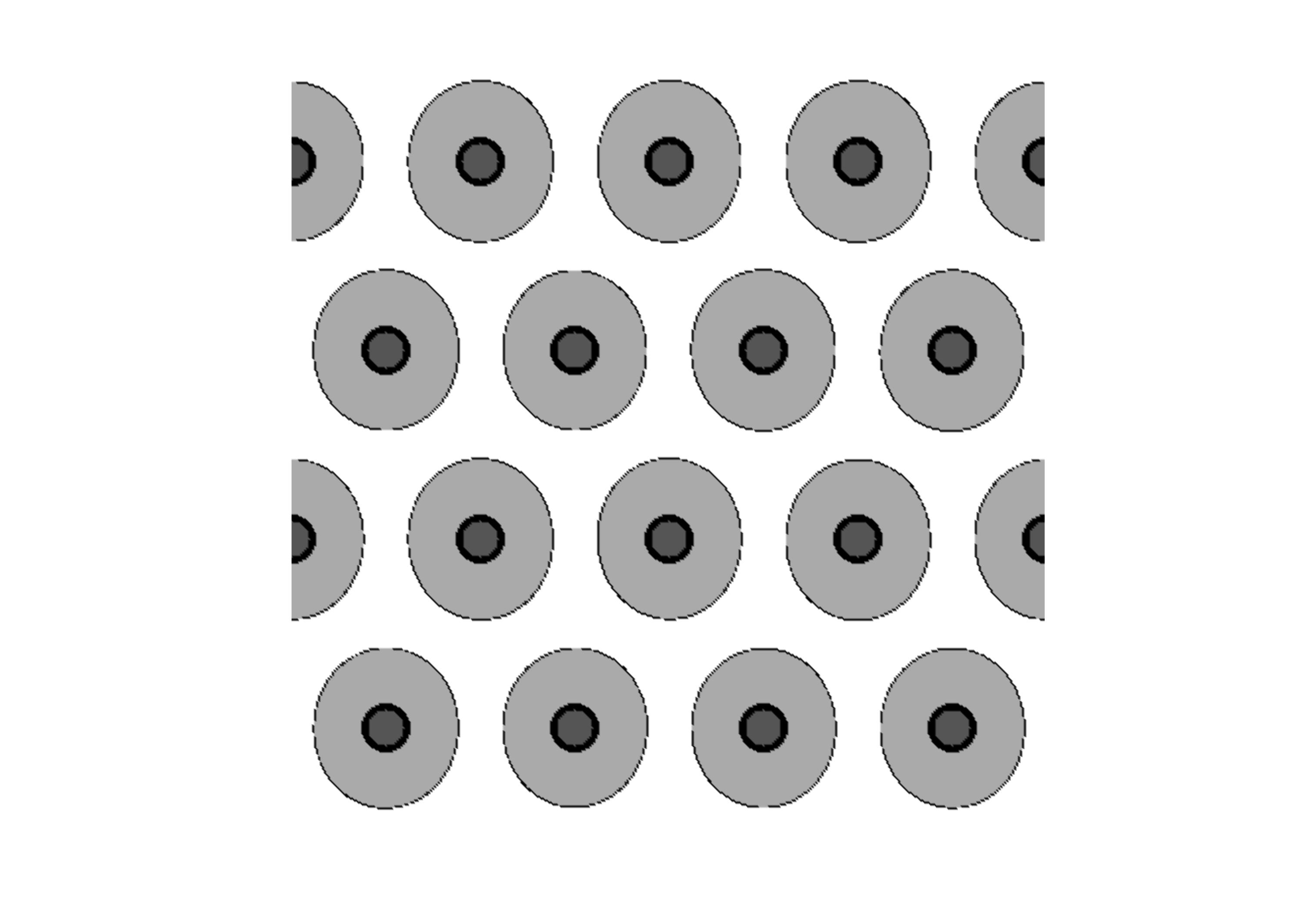}}
\caption{$A_{50}B_{50}$: SDSD microstructures that form in a critical blend in the presence of \textit{asymmetric} distribution of particles with varying $\lambda$: (a) $\lambda = 32$ (b) $\lambda = 48$ (c) $\lambda = 64$ (d) $\lambda = 96$ are presented. All lengths are in grid units. The snapshots correspond to the dimensionless times (a) $t = 10000$ and (b, c, d) $t = 4000$. The $\alpha$, $\beta$, and $\gamma$ phases are illustrated by white, light gray, and dark gray, respectively.}
\label{fig_asymmetric}
\end{figure}

Next, SDSD patterns are simulated with asymmetric arrangement of particles (Fig.~\ref{fig_asymmetric}). In low-$\lambda$ systems, the matrix phase separates to either irregular-shaped (Fig.~\ref{fig_asymmetric}a) or lamella-like (Figs.~\ref{fig_asymmetric}b,~\ref{fig_asymmetric}c) $\alpha$ and $\beta$ microdomains. The $\gamma$ particles are confined within the $\alpha$ domains due to the low $\sigma_{\alpha\gamma}$. On average, the width of the $\alpha$ domains is of the same order with the particle size, while $\beta$ domains are relatively thinner. Depending on $\lambda$, the morphology of these domains ranges between straight and wavy lamella.

In high-$\lambda$ asymmetric particle systems (Fig.~\ref{fig_asymmetric}d), SDSD patterns are similar to those of symmetric particle systems, starting with the formation of the target pattern followed by phase inversion, which dissolves the $\alpha$ rings to develop the continuous phase and $\beta$ rings to surround the particle surface.

\subsubsection{Particle Effects in Off-Critical Blends}
\begin{figure}[h]
\centering
\subfloat[$\lambda = 32$]{\label{sa40b60_32}\includegraphics[trim={20cm 15 20 15},clip,scale=0.05]{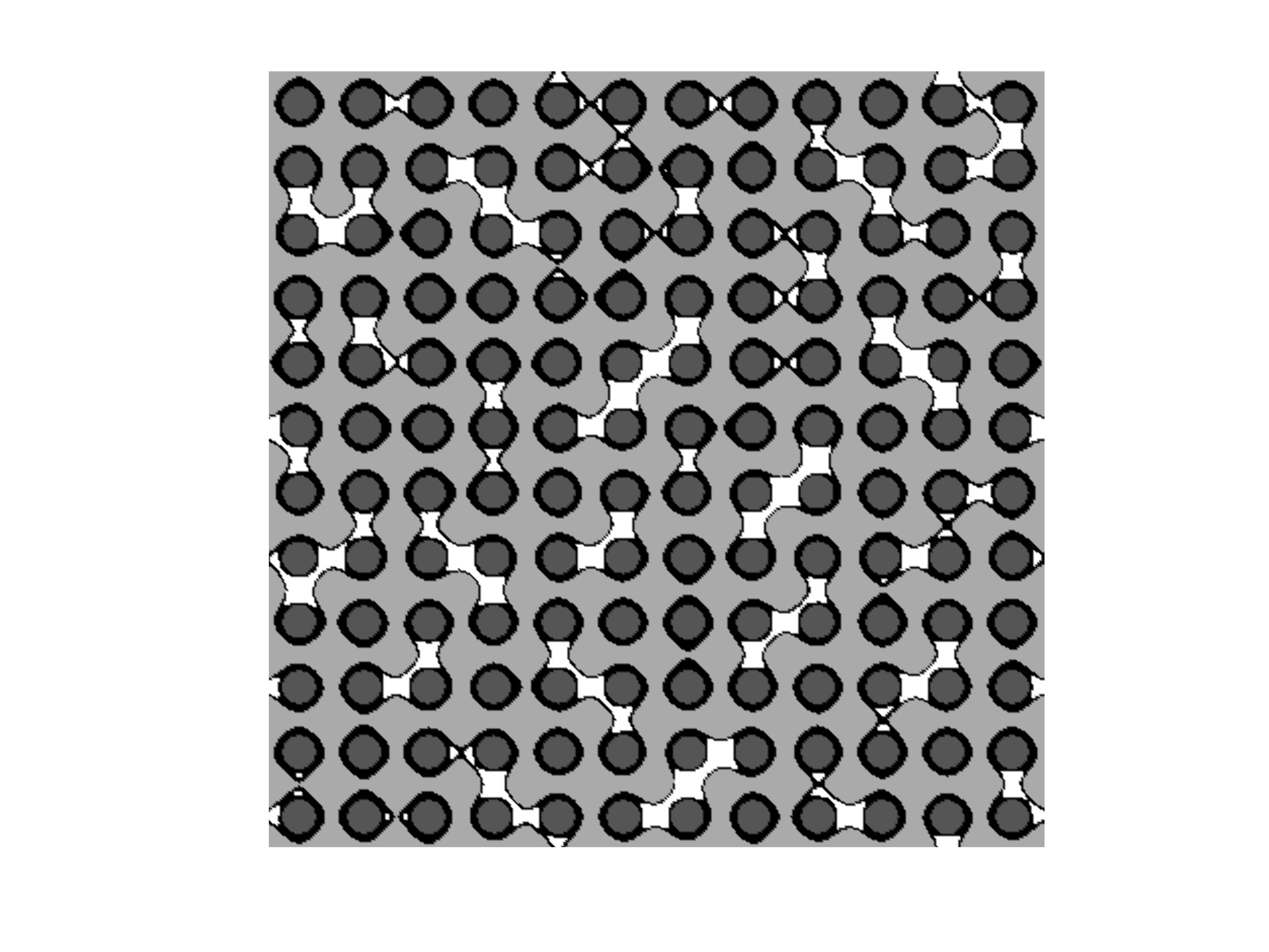}}
\subfloat[$\lambda = 48$]{\label{sa40b60_48}\includegraphics[trim={20cm 15 20 15},clip,scale=0.05]{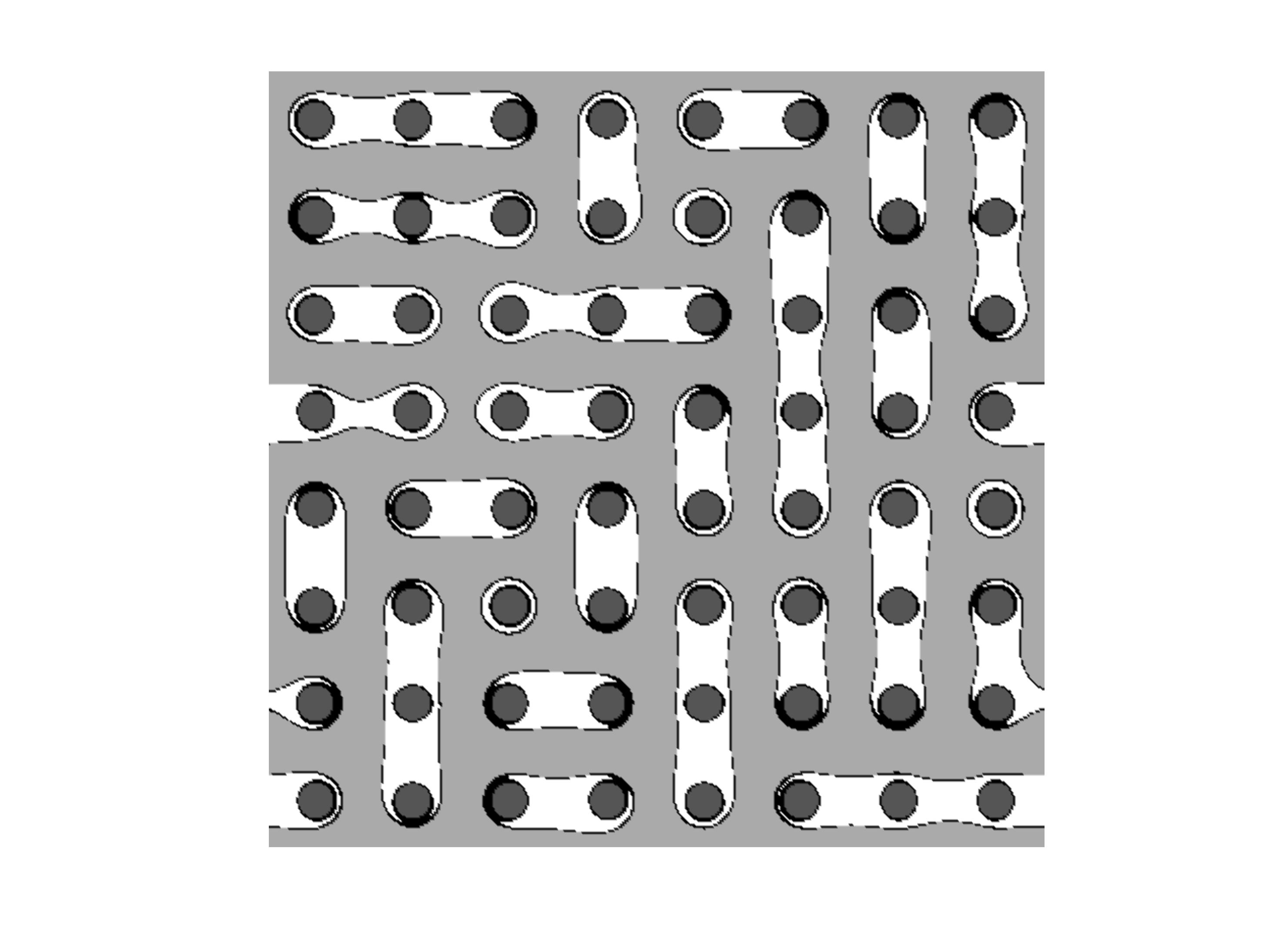}}
\subfloat[$\lambda = 64$]{\label{sa40b60_64}\includegraphics[trim={20cm 15 20 15},clip,scale=0.05]{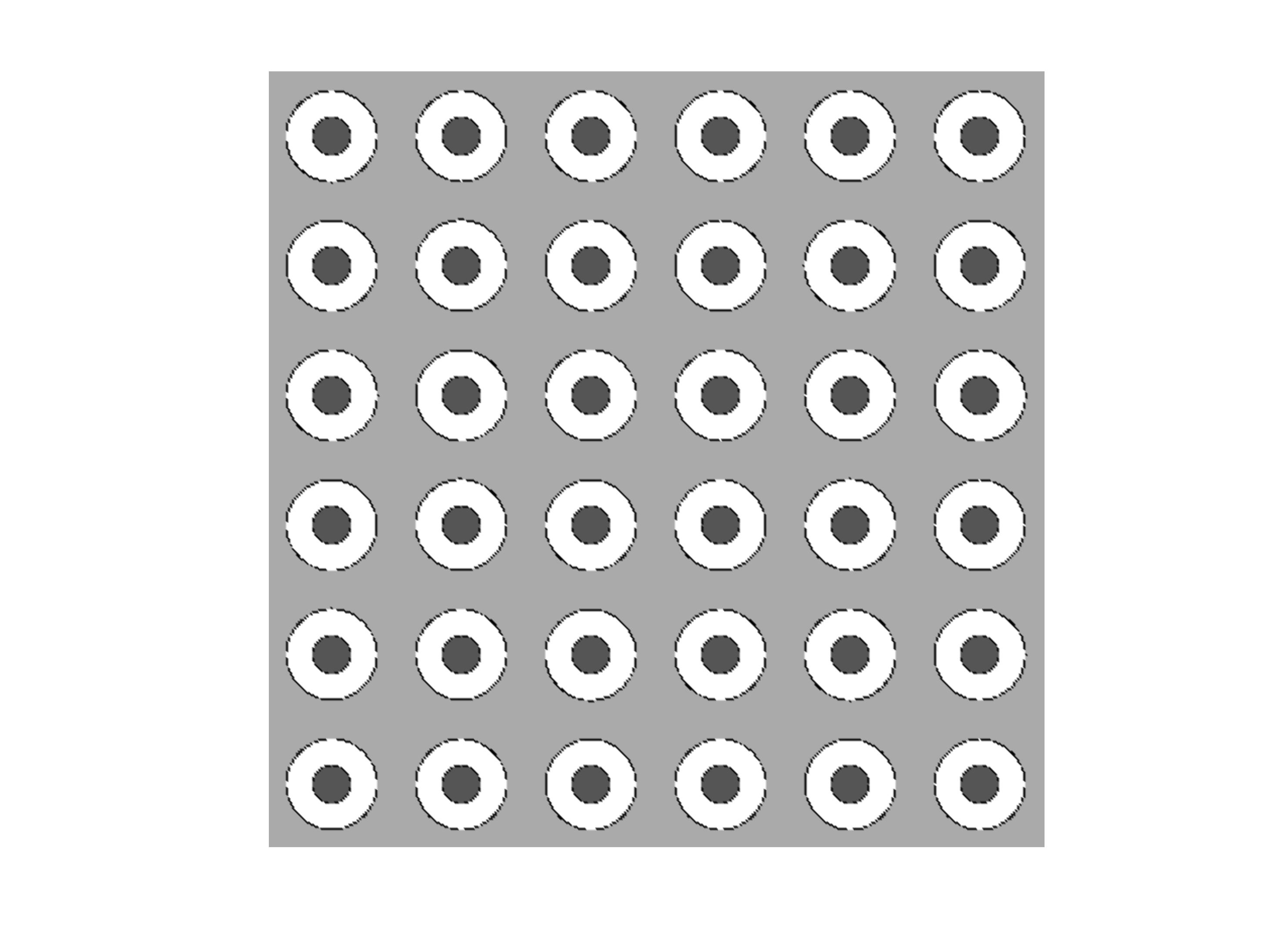}}
\subfloat[$\lambda = 96$]{\label{sa40b60_96}\includegraphics[trim={20cm 15 20 15},clip,scale=0.05]{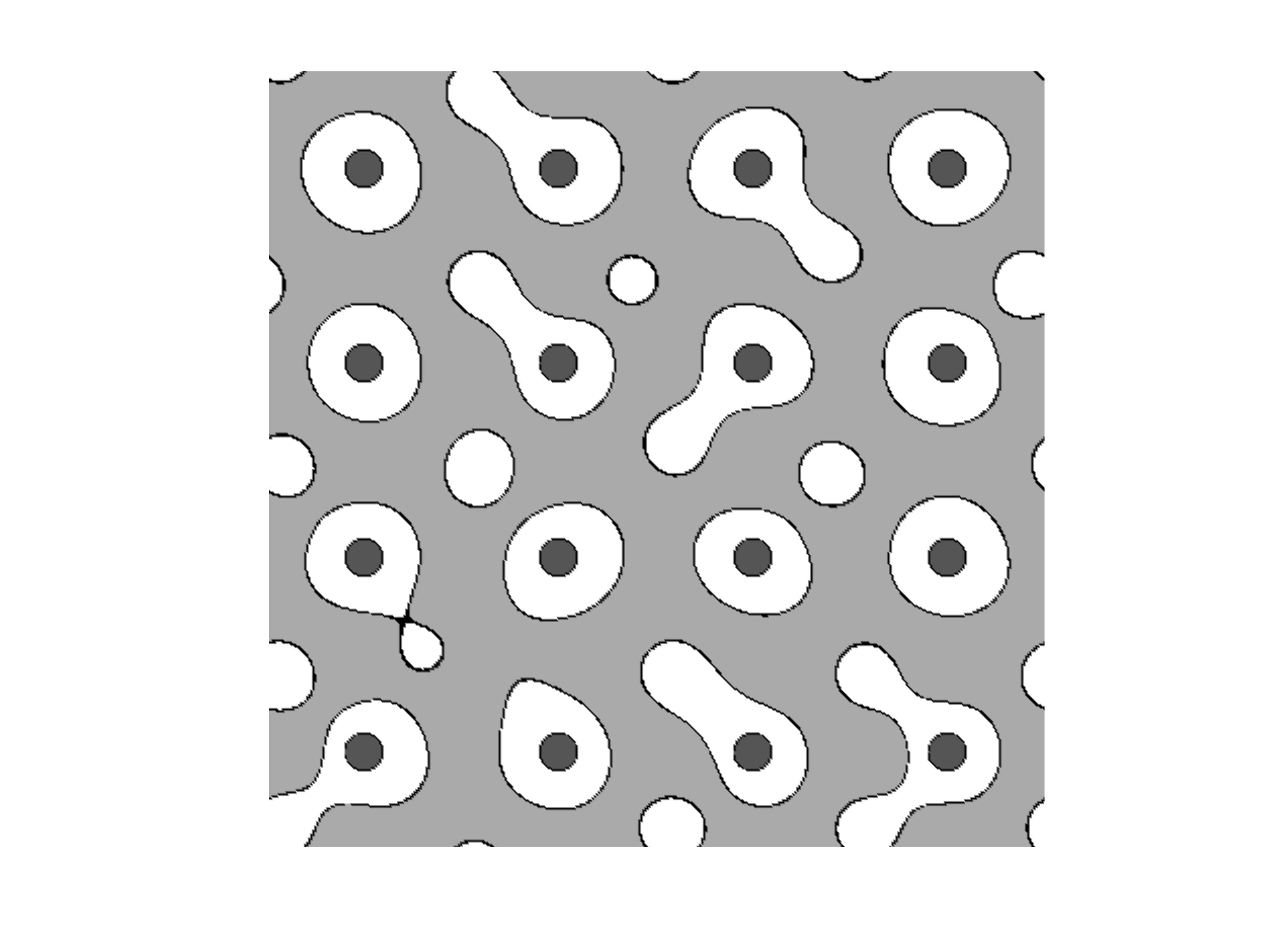}}
\caption{$A_{40}B_{60}$: SDSD microstructures that form in the presence of \textit{symmetric} distribution of particles with varying $\lambda$: (a) $\lambda = 32$ (b) $\lambda = 48$ (c) $\lambda = 64$ (d) $\lambda = 96$ are presented. All lengths are in grid units. The snapshots correspond to the dimensionless times (a) $t = 10000$ and (b, c, d) $t = 4000$. The $\alpha$, $\beta$, and $\gamma$ phases are illustrated by white, light gray, and dark gray, respectively.}
\label{fig_symmetric_a40b60}
\end{figure}

Two off-critical blends are studied: $A_{40}B_{60}$ and $A_{60}B_{40}$. In $A_{40}B_{60}$, the minority component A is attracted to the particle surface. In the low- and intermediate-$\lambda$ systems, the majority $\beta$ forms the continuous phase while $\alpha$ develops as a thin network (Fig.~\ref{fig_symmetric_a40b60}a), broken lamellae (Fig.~\ref{fig_symmetric_a40b60}b), and target rings (Fig.~\ref{fig_symmetric_a40b60}c). Such low-$\lambda$ morphologies are significantly different to that of the critical blend (Fig.~\ref{fig_symmetric}b) because $\sigma_{\alpha\gamma} < \sigma_{\beta\gamma}$. In high-$\lambda$ systems, the majority $\beta$ is the continuous phase with a layer of $\alpha$ surrounds each particle in the long-time limit (Fig.~\ref{fig_symmetric_a40b60}d), in contrast to a critical blend (Fig.~\ref{fig_symmetric}d).

\begin{figure}[h]
\centering
\subfloat[$\lambda = 32$]{\label{sa60b40_32}\includegraphics[trim={20cm 15 20 15},clip,scale=0.04]{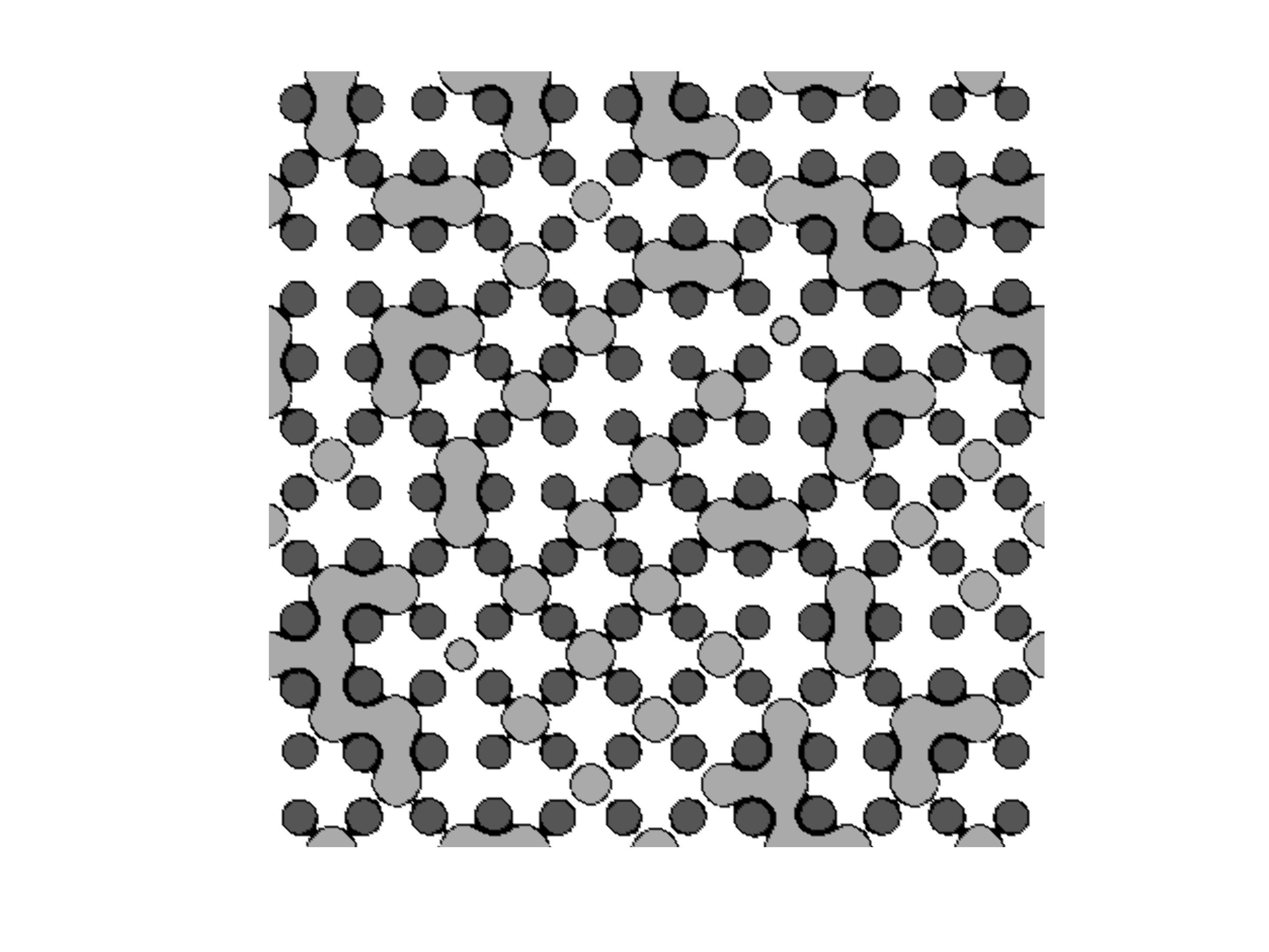}}
\subfloat[$\lambda = 48$]{\label{sa60b40_48}\includegraphics[trim={20cm 15 20 15},clip,scale=0.04]{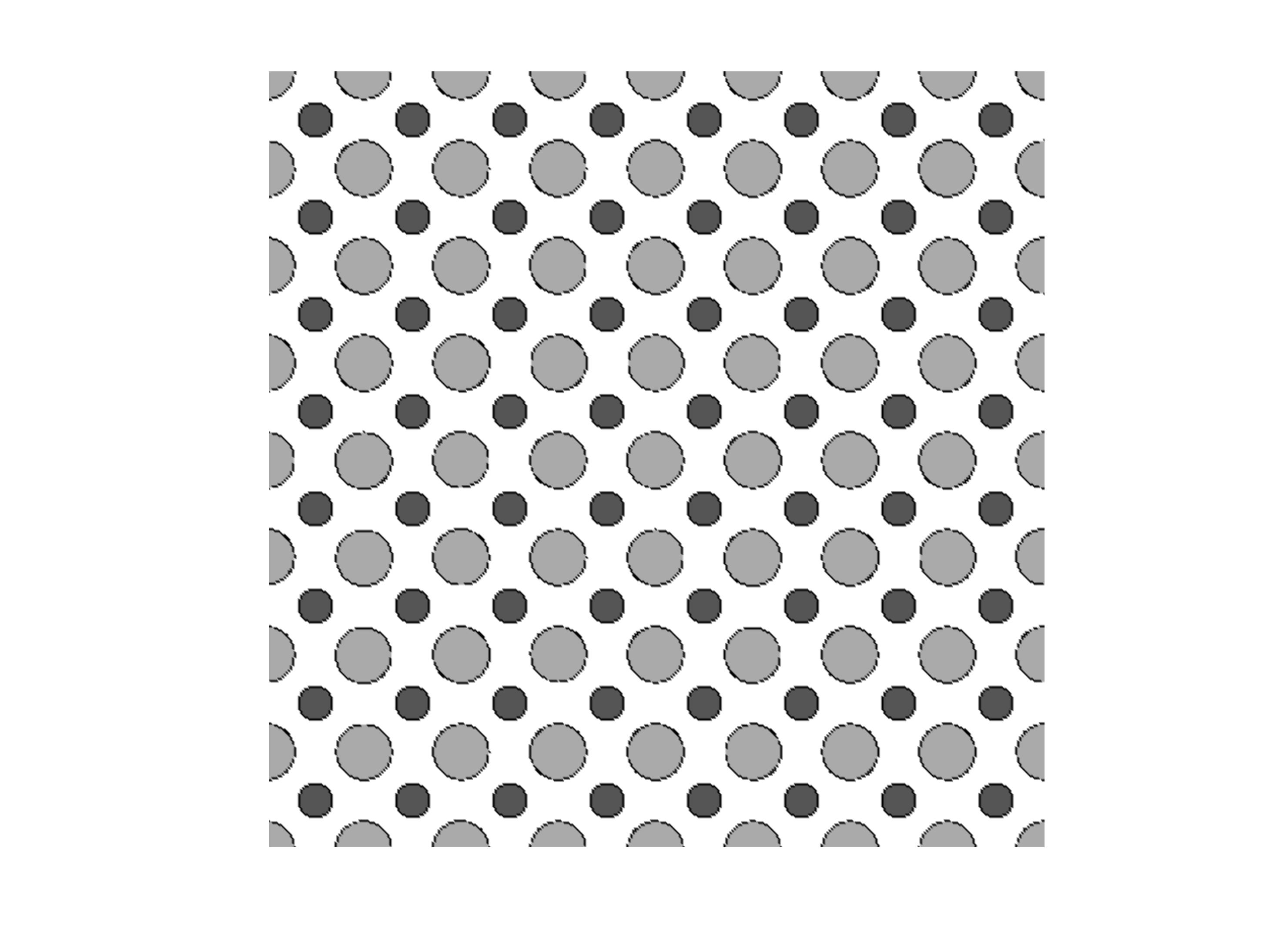}}
\subfloat[$\lambda = 64$]{\label{sa60b40_64}\includegraphics[trim={20cm 15 20 15},clip,scale=0.04]{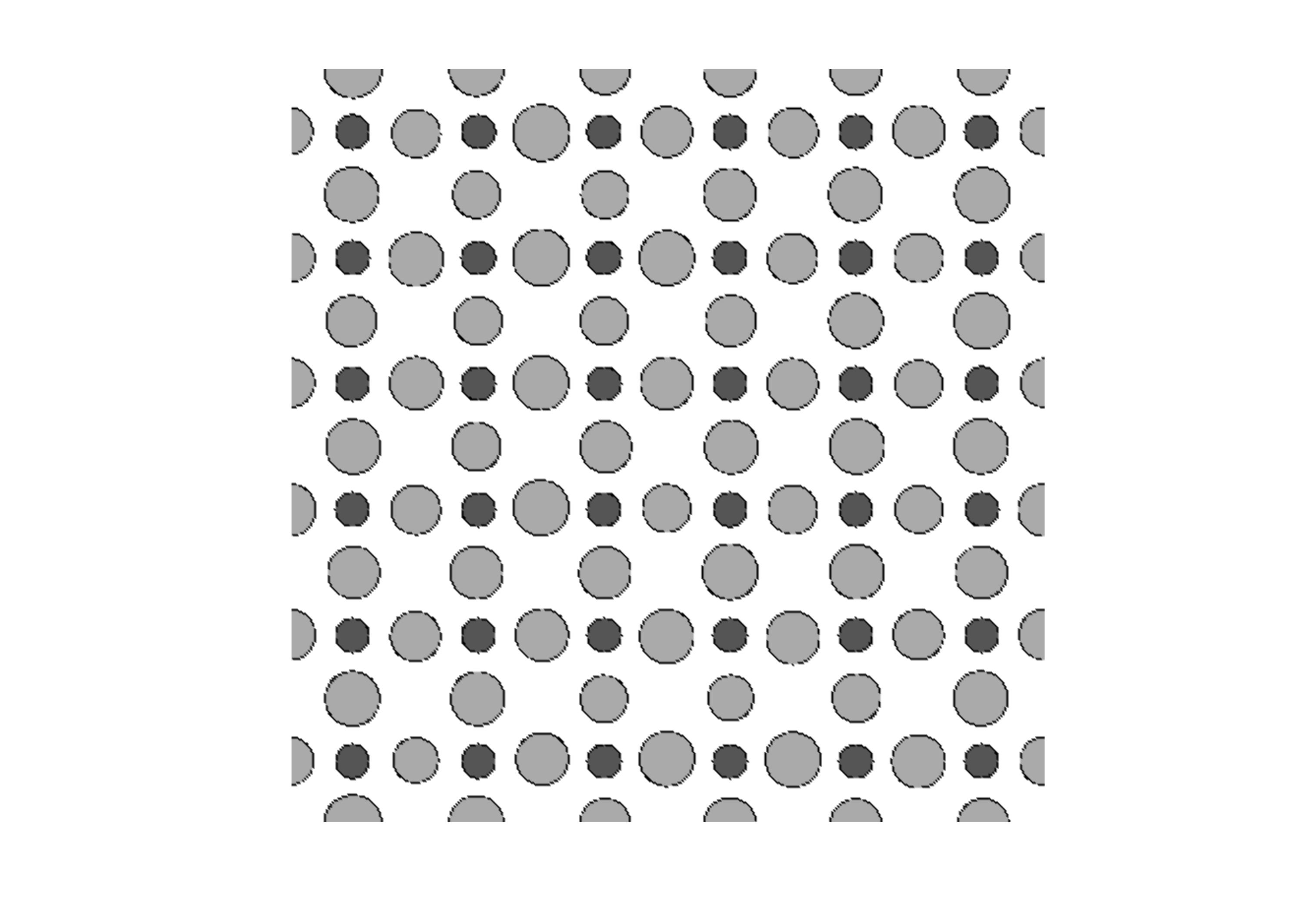}}
\subfloat[$\lambda = 96$]{\label{sa60b40_96}\includegraphics[trim={20cm 15 20 15},clip,scale=0.04]{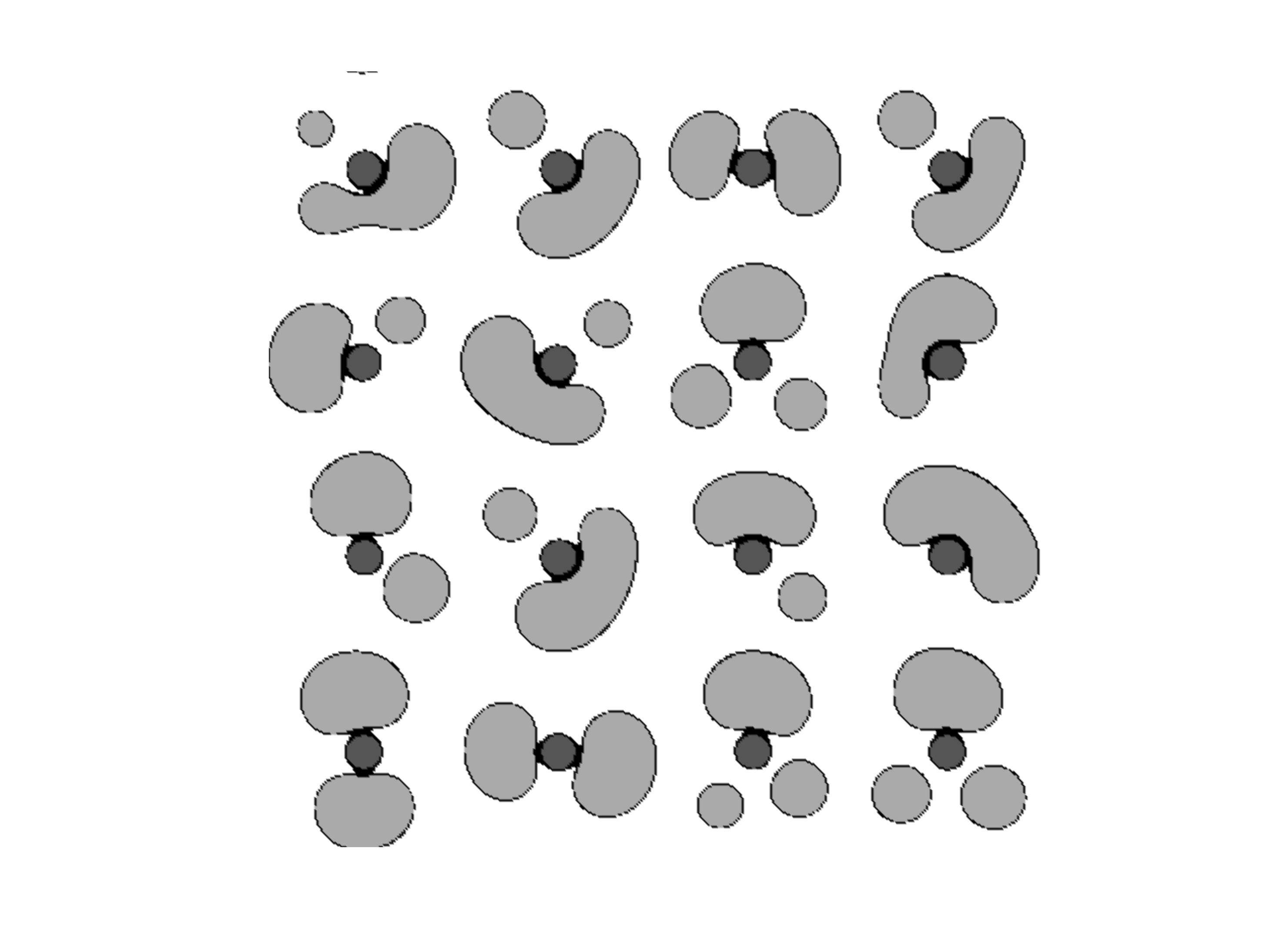}}
\subfloat[$\lambda = 96$]{\label{sa60b40_96_2}\includegraphics[trim={20cm 15 20 15},clip,scale=0.04]{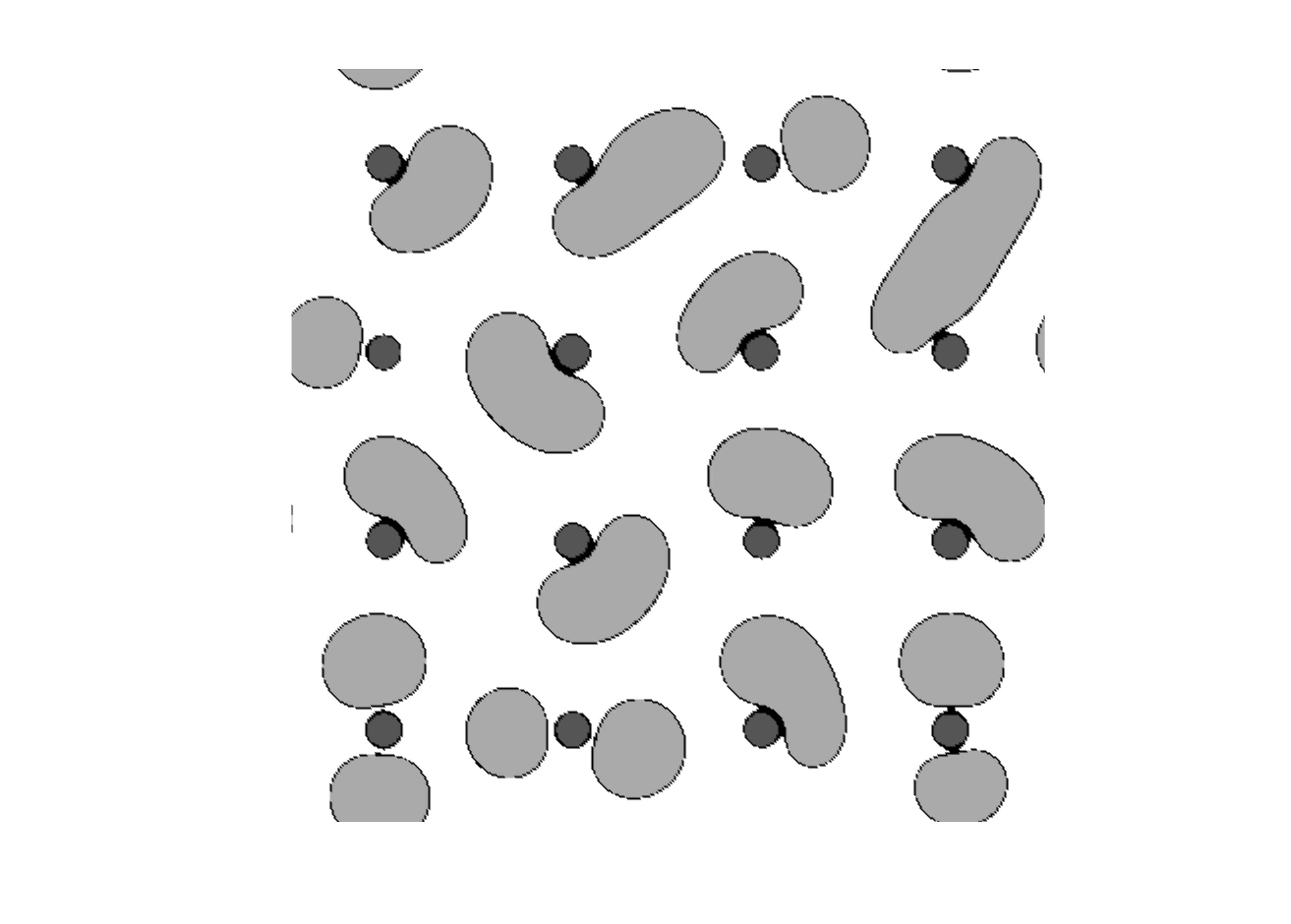}}
\caption{$A_{60}B_{40}$: SDSD microstructures that form in the presence of \textit{symmetric} distribution of particles with varying $\lambda$: (a) $\lambda = 32$ (b) $\lambda = 48$ (c) $\lambda = 64$ (d) $\lambda = 96$ are presented. All lengths are in grid units. The snapshots correspond to the dimensionless times (a) $t = 10000$, (b, c, d) $t = 4000$, and (e) $t = 10000$. The $\alpha$, $\beta$, and $\gamma$ phases are illustrated by white, light gray, and dark gray, respectively.}
\label{fig_symmetric_a60b40}
\end{figure}

In $A_{60}B_{40}$, the majority A is attracted to the particle surface. As a result, SDSD morphologies are completely different compared to $A_{50}B_{50}$ and $A_{40}B_{60}$ (Fig.~\ref{fig_symmetric_a60b40}). In low-$\lambda$ systems, the minority $\beta$ droplets form as staggered (Fig.~\ref{fig_symmetric_a60b40}b) and inline (Fig.~\ref{fig_symmetric_a60b40}c) arrays in between particles in a continuous $\alpha$. In high-$\lambda$ systems, unlike to that of other blends, no $\beta$ rings survive as they partially engulf the particles (Fig.~\ref{fig_symmetric_a60b40}d) at early to intermediate times. Such a partial wetting~\cite{Hore,sprenger2003hierarchic} scenario seemingly possible due to $\sigma_{\alpha\beta} + \sigma_{\beta\gamma} > \sigma_{\alpha\gamma}$ (Table~\ref{tab_param}); however, over time, $\beta$ tends to break up into spherical droplets and drifts away from the particle surface for interface energy minimization (Fig.~\ref{sa60b40_96_2}).

\begin{figure}[h]
\centering
\subfloat[$\lambda = 32$]{\label{asa40b60_32}\includegraphics[trim={20cm 15 20 15},clip,scale=0.05]{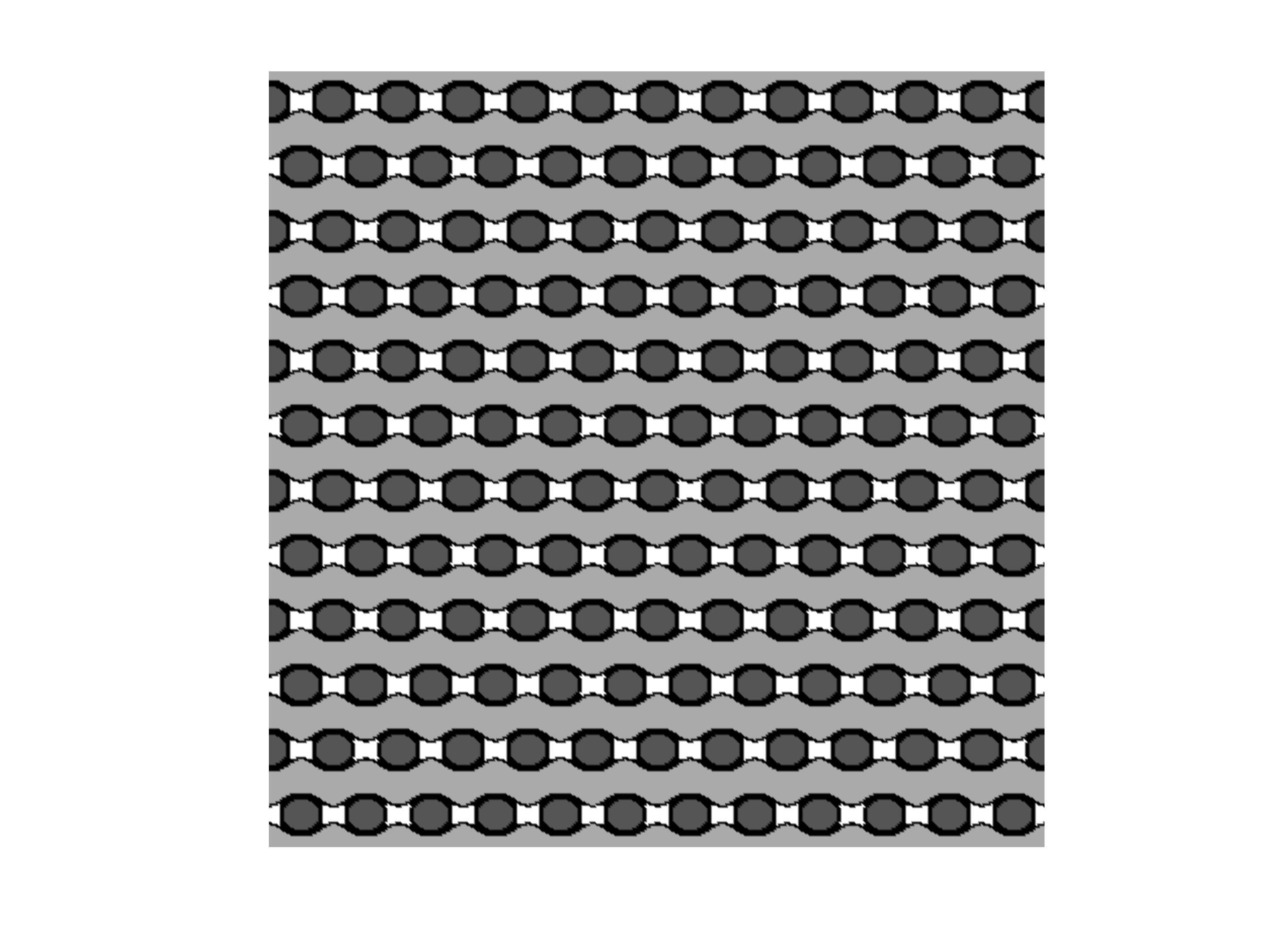}}
\subfloat[$\lambda = 48$]{\label{asa40b60_48}\includegraphics[trim={20cm 15 20 15},clip,scale=0.05]{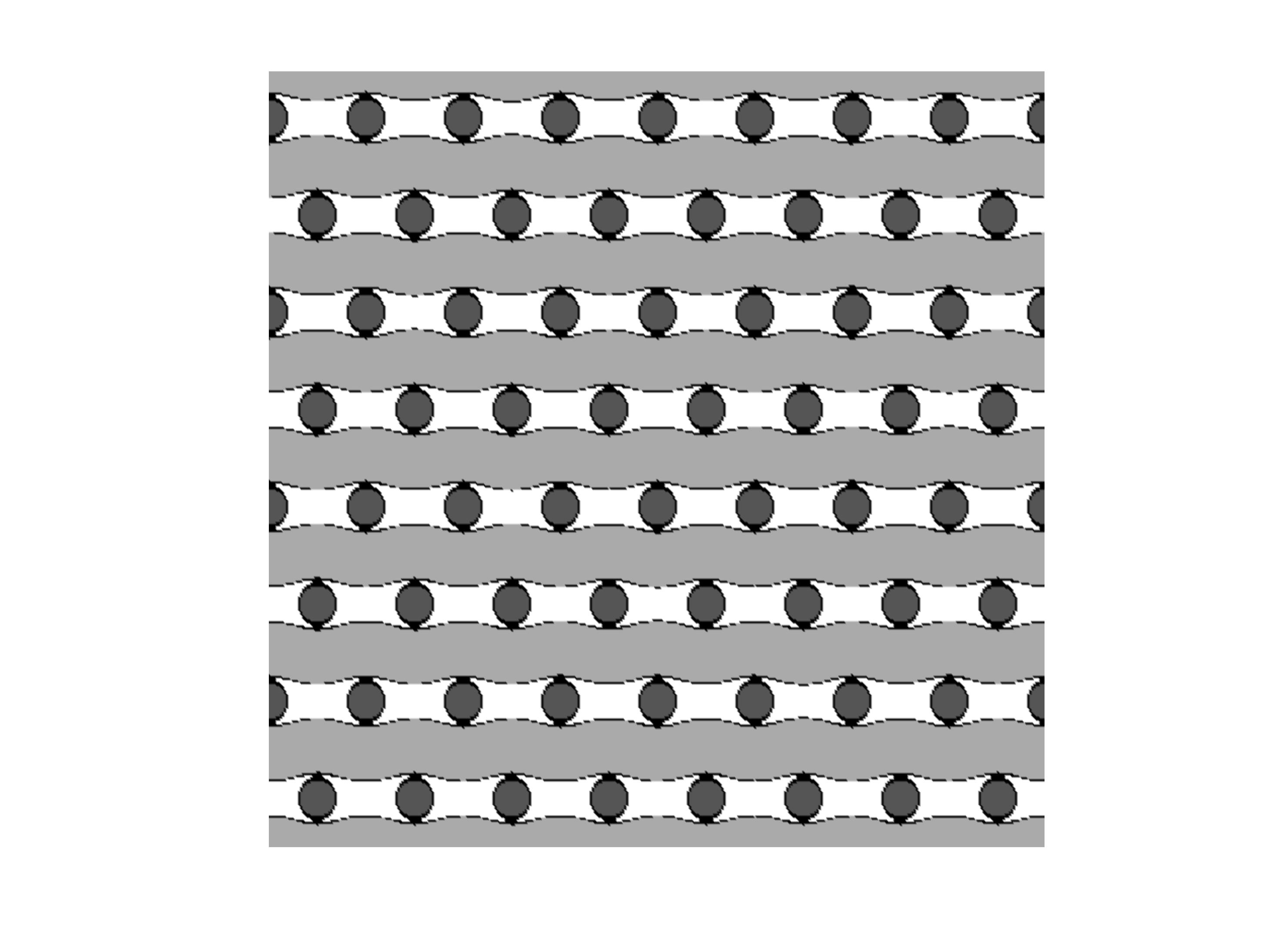}}
\subfloat[$\lambda = 64$]{\label{asa40b60_64}\includegraphics[trim={20cm 15 20 15},clip,scale=0.05]{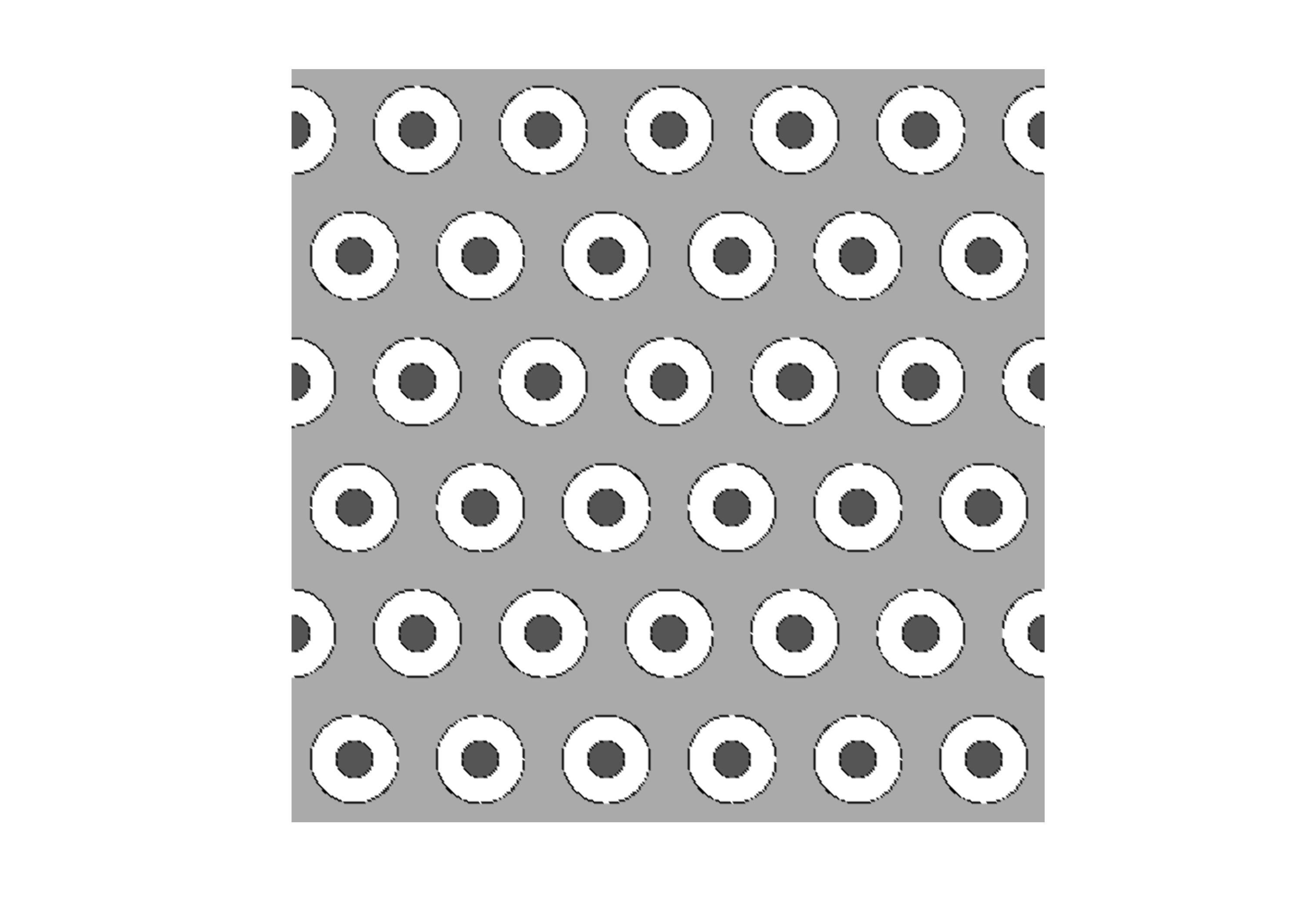}}
\subfloat[$\lambda = 96$]{\label{asa40b60_96}\includegraphics[trim={20cm 15 20 15},clip,scale=0.05]{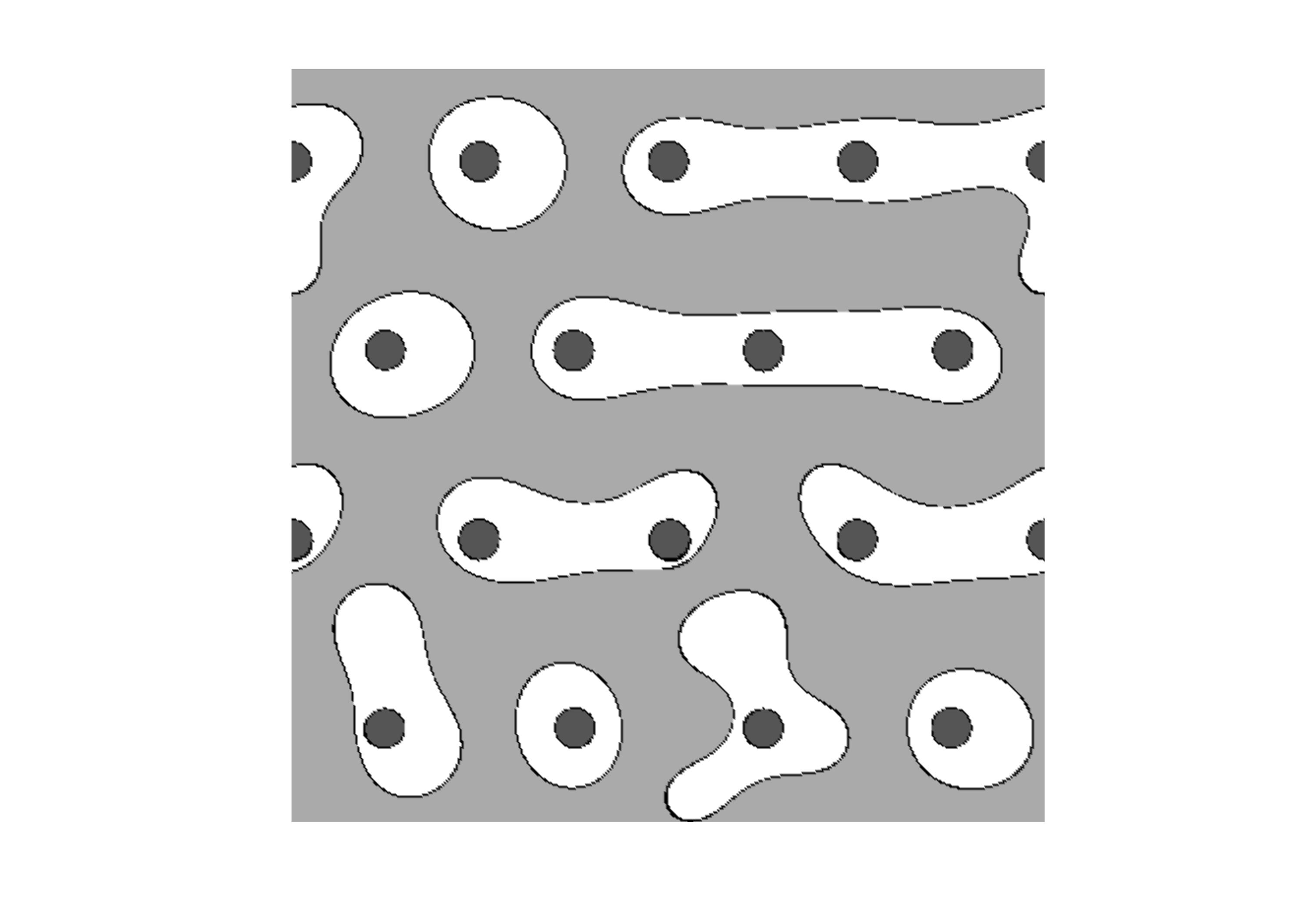}}
\caption{$A_{40}B_{60}$: SDSD microstructures that form in the presence of \textit{asymmetric} distribution of particles with varying $\lambda$: (a) $\lambda = 32$ (b) $\lambda = 48$ (c) $\lambda = 64$ (d) $\lambda = 96$ are presented. All lengths are in grid units. The snapshots correspond to the dimensionless times (a) $t = 10000$, (b, c) $t = 4000$, and (d) $t = 7500$. The $\alpha$, $\beta$, and $\gamma$ phases are illustrated by white, light gray, and dark gray, respectively.}
\label{fig_asymmetric_a40b60}
\end{figure}

\begin{figure}[h]
\centering
\subfloat[$\lambda = 32$]{\label{asa60b40_32}\includegraphics[trim={20cm 15 20 15},clip,scale=0.05]{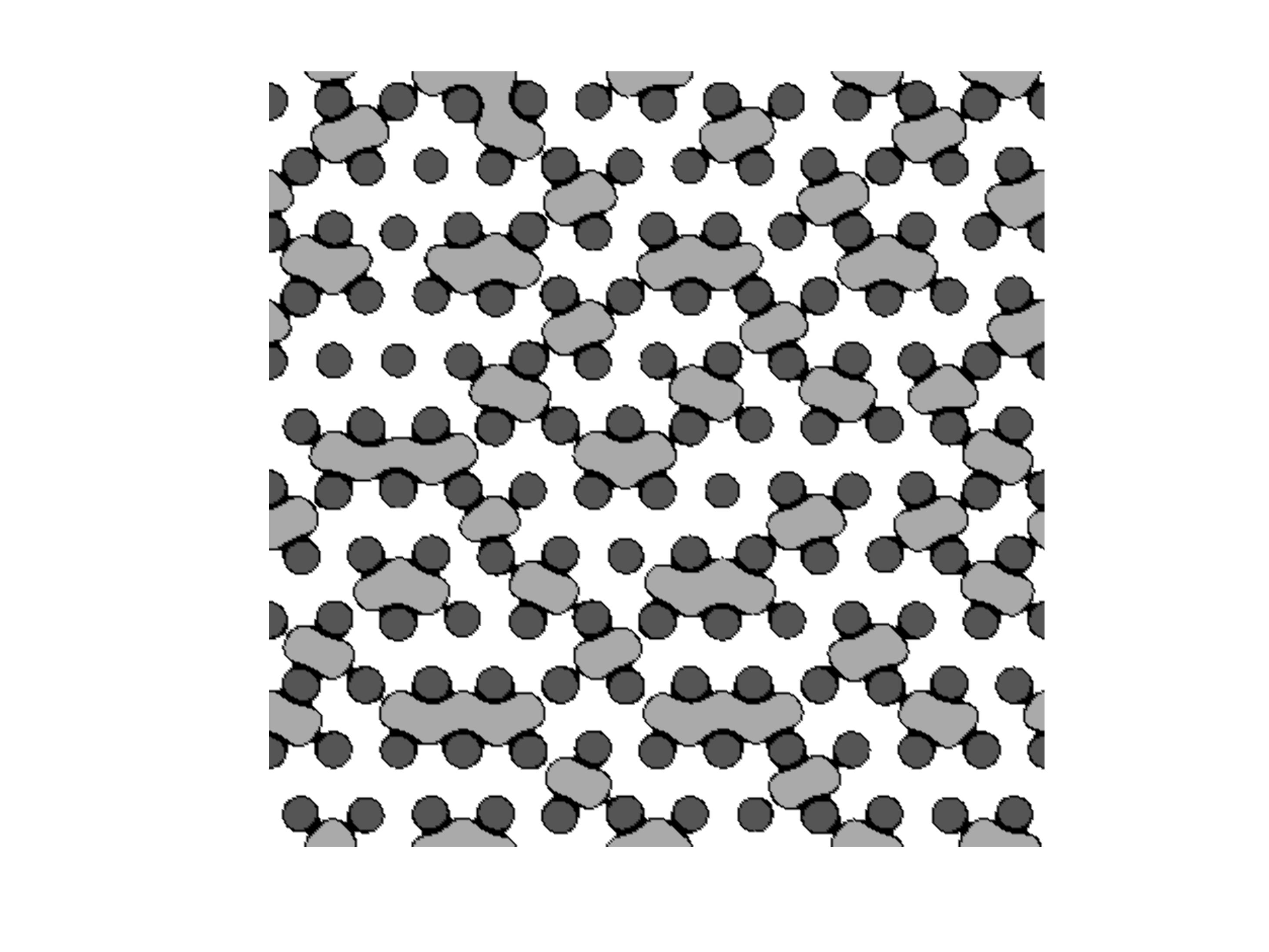}}
\subfloat[$\lambda = 48$]{\label{asa60b40_48}\includegraphics[trim={20cm 15 20 15},clip,scale=0.05]{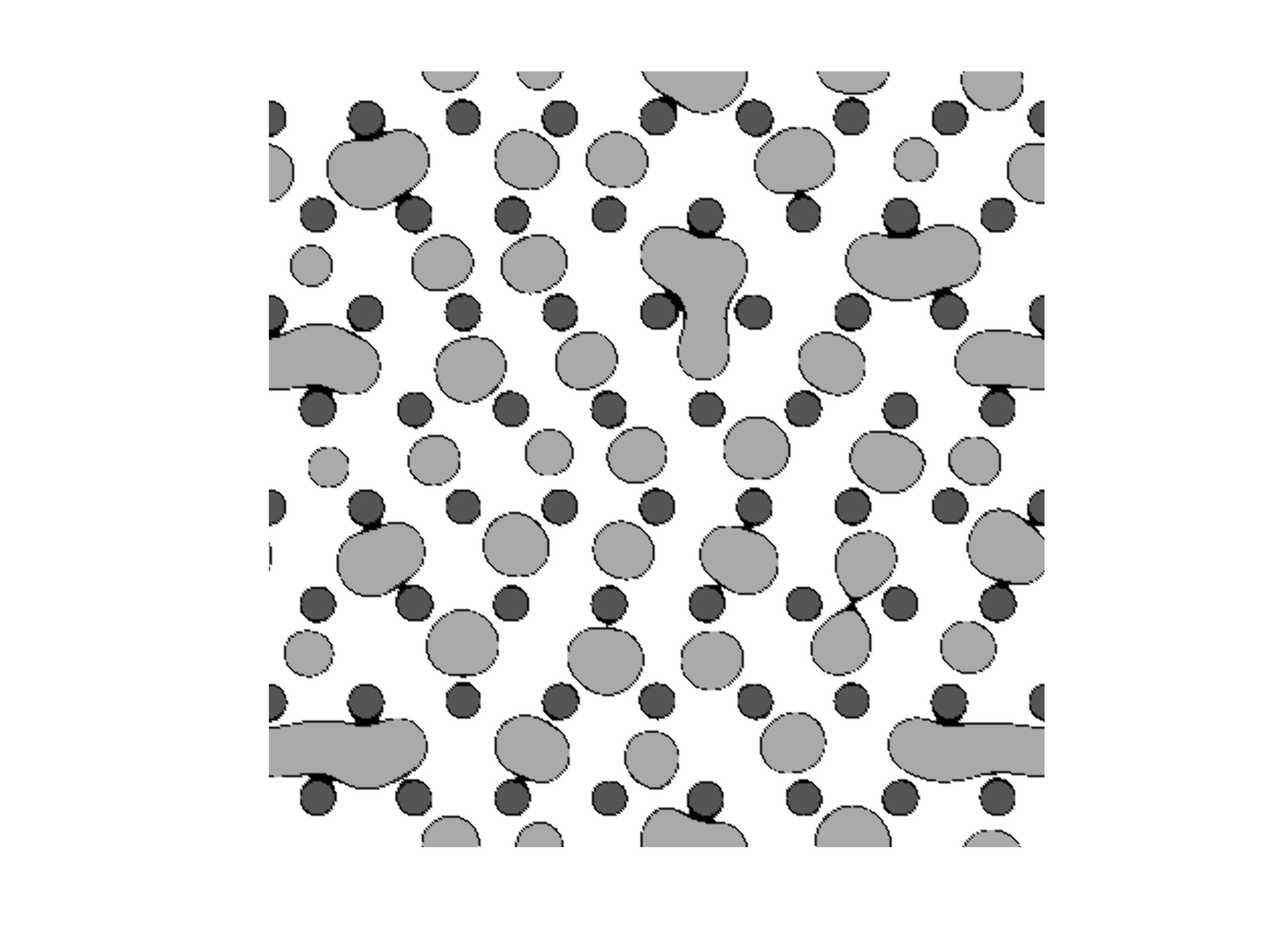}}
\subfloat[$\lambda = 64$]{\label{asa60b40_64}\includegraphics[trim={20cm 15 20 15},clip,scale=0.05]{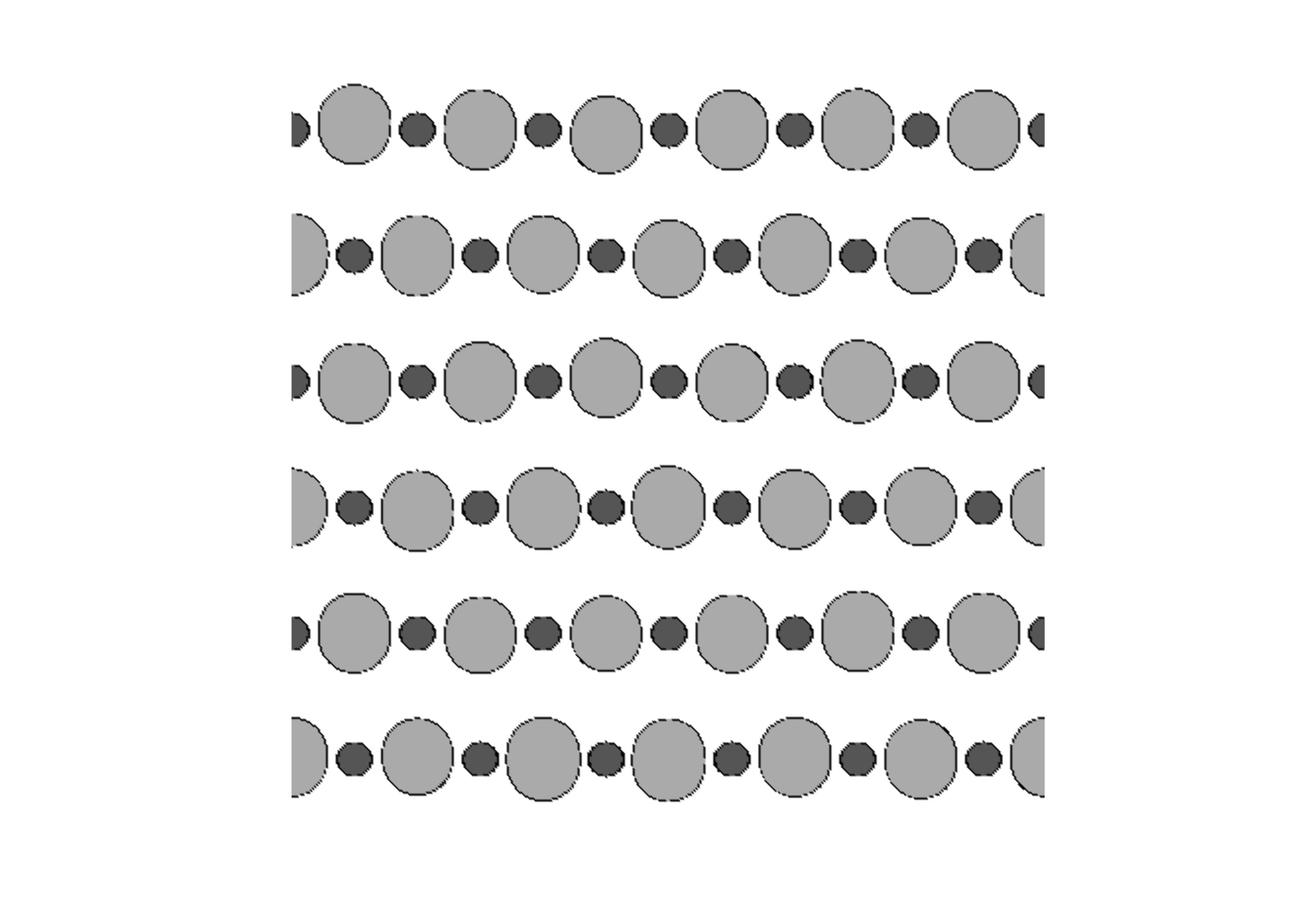}}
\subfloat[$\lambda = 96$]{\label{asa60b40_96}\includegraphics[trim={20cm 15 20 15},clip,scale=0.05]{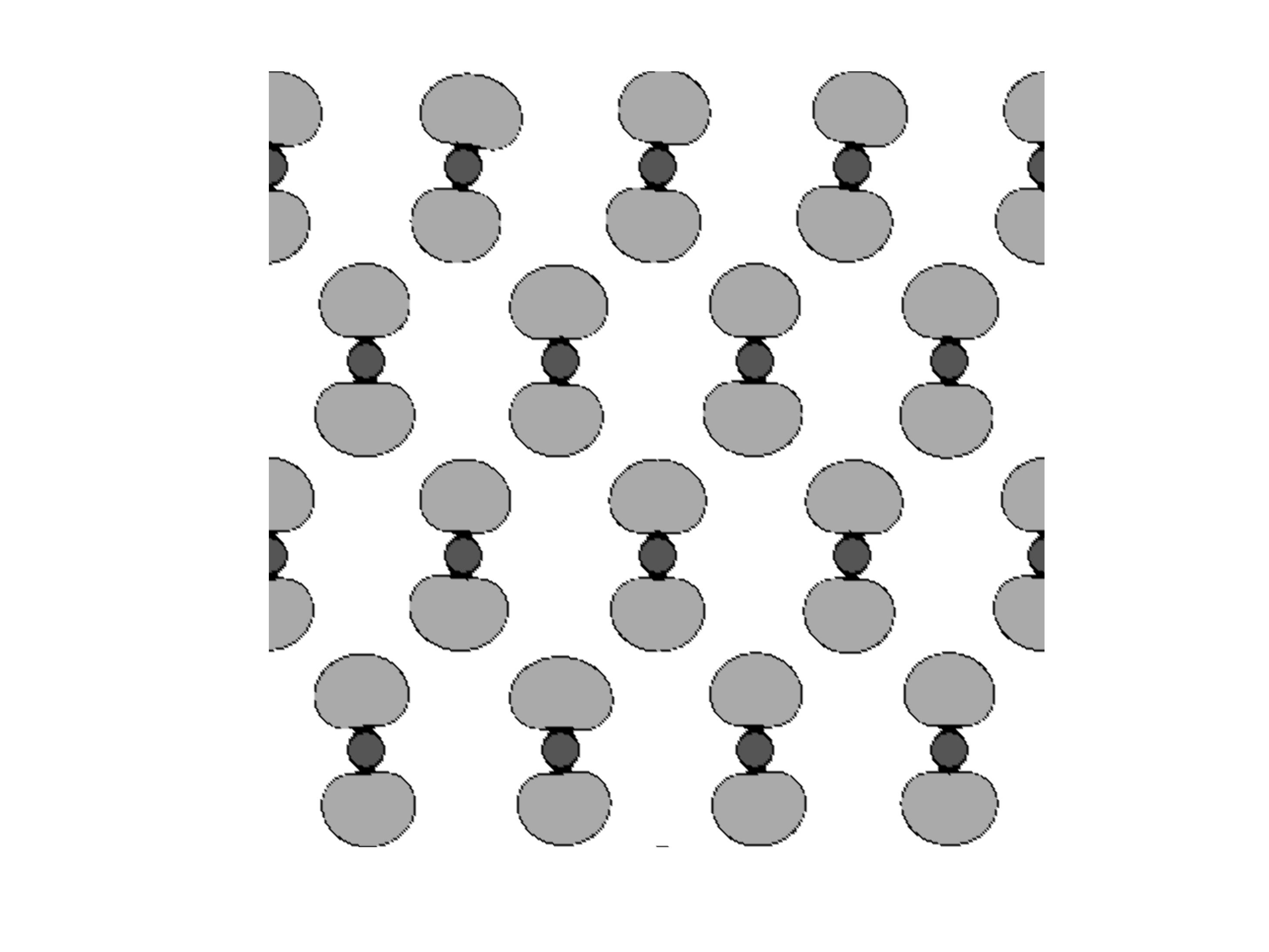}}
\caption{$A_{60}B_{40}$: SDSD microstructures that form in the presence of \textit{asymmetric} distribution of particles with varying $\lambda$: (a) $\lambda = 32$ (b) $\lambda = 48$ (c) $\lambda = 64$ (d) $\lambda = 96$ are presented. All lengths are in grid units. The snapshots correspond to the dimensionless times (a) $t = 10000$ and (b, c, d) $t = 4000$. The $\alpha$, $\beta$, and $\gamma$ phases are illustrated by white, light gray, and dark gray, respectively.}
\label{fig_asymmetric_a60b40}
\end{figure}

The SDSD in off-critical blends but with an asymmetric distribution of particles remains similar. In $A_{40}B_{60}$, the lamella-like microdomains in low-$\lambda$ systems (Fig.~\ref{fig_asymmetric_a40b60}b) transform to the target pattern in intermediate- and high-$\lambda$ systems (Figs.~\ref{fig_asymmetric_a40b60}c,~\ref{fig_asymmetric_a40b60}d). In these target morphologies, the $\alpha$ ring survives around each particle in the long-time limit, and $\beta$ forms the continuous matrix. Also, the wetting layer of $\alpha$ tends to form an interconnected structure so that particles are essentially bridged by it (Fig.~\ref{fig_asymmetric_a40b60}d). In $A_{60}B_{40}$, no target pattern survives as the $\beta$ droplets are prevalent in continuous $\alpha$ (Fig.~\ref{fig_asymmetric_a60b40}). Although not shown here, similar to Fig.~\ref{sa60b40_96_2}, $\beta$ also forms droplets in Fig.~\ref{asa60b40_96} at very late times. The length scale and shape of $\beta$ droplets are dictated by the size of the particle and the value of $\lambda$. 

\subsubsection{Stability of Target Patterns}\label{sec_stability_target}
Referring to SDSD morphologies in high-$\lambda$ systems, it is evident that the physical mechanism of the formation and kinetics of the target pattern remains equivalent, irrespective of the particle arrangement in the matrix. Since the SDSD morphologies in our simulations are diffusion-controlled, we aim to explore the characteristic length, time, and composition measures to determine the stability of such patterns in critical blends.

Spinodal decomposition occurs \textit{via} amplification of the composition waves originating from involved phases with the critical wavelength~\cite{cahn1961spinodal}:
\begin{equation}
\lambda_{sp} \approx \sqrt{\frac{8\pi^2(\kappa_{i}+\kappa_{j})}{-\partial^2 f/\partial c^2}}.
\end{equation} 
Since we work on the spinodal decomposition of mixture A:B, the above expression results in a dimensionless scale $\lambda_{sp} \approx 31$. In our simulations, we varied $\lambda$ on multiples of $\lambda_{sp}$ to explore its effect on SDSD. Therefore, our results can be represented as a function of the dimensionless ratio of $\lambda/\lambda_{sp}$. The critical value of $\lambda$ that stabilizes the target pattern, including either the $\alpha$ or $\beta$ ring or the both, happens to be at least $\approx 3 \, \lambda_{sp}$. Note that the critical value of $\lambda/\lambda_{sp}$ will vary for arbitrary material parameters (e.g., $\sigma_{ij}$) and process conditions (e.g., $\chi_{ij}$).

\begin{figure}[htbp]
\centering
\subfloat[]{\label{cp50b50_1}\includegraphics[scale=0.42]{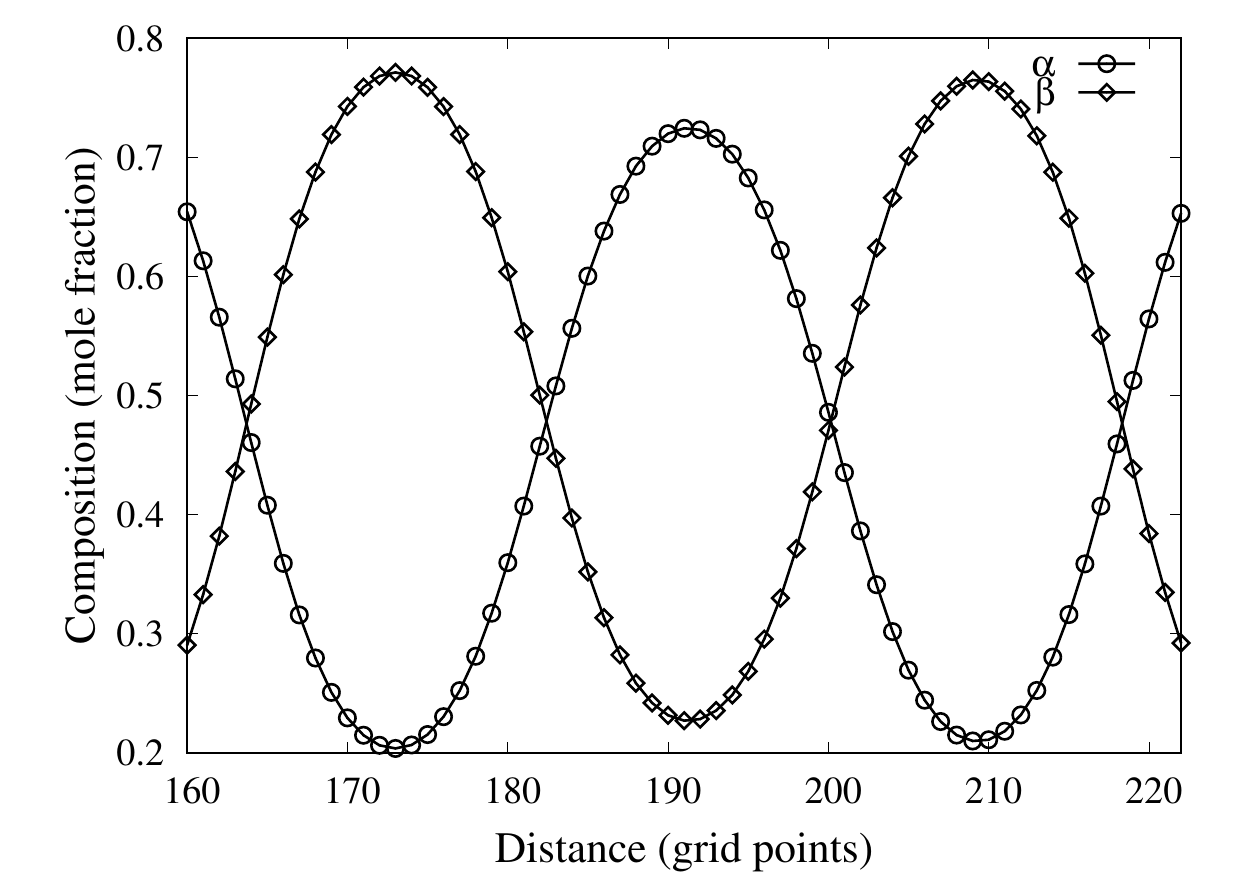}}
\subfloat[]{\label{cp40b60_1}\includegraphics[scale=0.42]{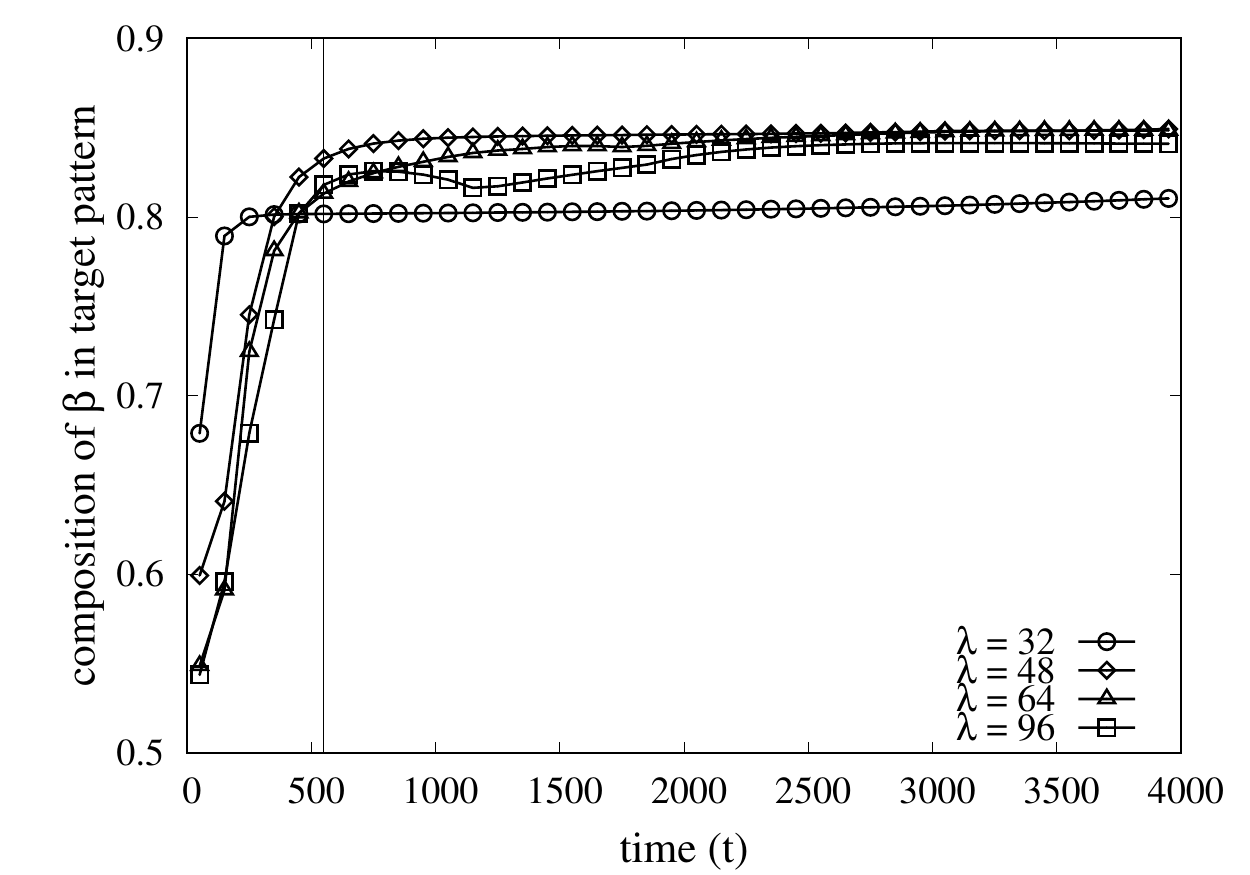}}
\subfloat[]{\label{cp40b60_2}\includegraphics[scale=0.42]{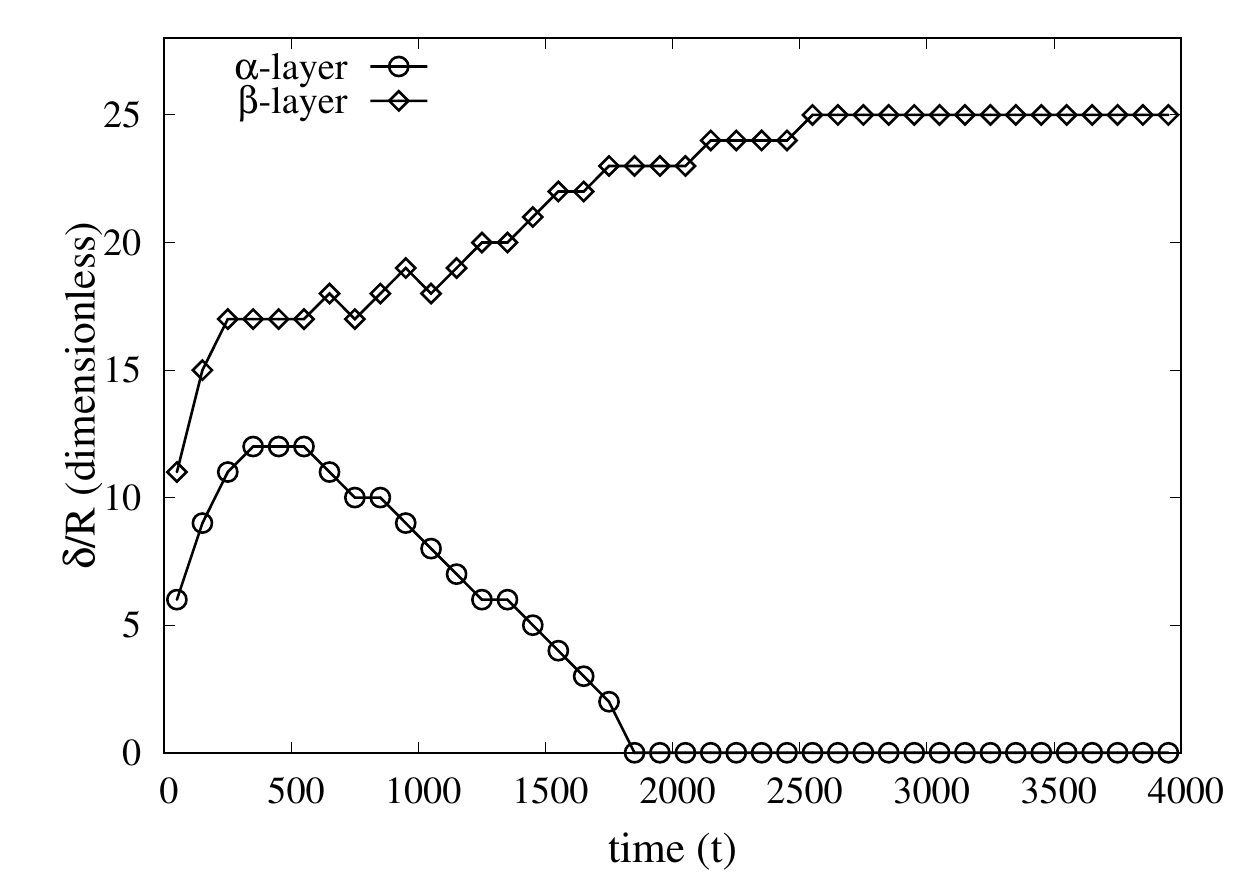}}
\caption{(a) The composition profile of B ($c_B$) across the alternate $\alpha$ and $\beta$ rings in the target pattern is illustrated from Fig.~\ref{fig_symmetric}d. (b) At early times, the metric of maximum composition of B in $\beta$ within the target pattern increases as the phase separation progresses. A steady state is reached when the target pattern formation is complete, signifying the time scale for the onset of the subsequent coarsening process as the $\alpha$ ring at the particle surface begins to shrink (indicated by the vertical line). (c) The thickness $\delta$ (scaled by the particle size $R$) of $\alpha$ and $\beta$ layers in the target pattern varies with time.}
\label{fig_critical}
\end{figure} 

The characteristic time and composition scales are determined following the alternate composition profiles developed due to concentric $\alpha$ and $\beta$ layers around each particle (Fig.~\ref{fig_critical}a). Due to the coarsening process at later times, the species of $\alpha$ at the particle surface continues to decrease until it disappears from the target pattern. Such a phase inversion process that brings the non-preferred $\beta$ to the particle surface dictates the characteristic composition for which the target pattern remains stable. This critical composition $c_{sp}$ can be determined following the rule of mixture: $c_{sp} = V_\alpha c_\alpha + V_\beta c_\beta$, where $V$ is the volume fraction of the mixture components, and $c$ is the local equilibrium composition of the phases estimated at a particular quench temperature in the two-phase region of the spinodal phase diagram. In Fig.~\ref{fig_critical}a, which corresponds to the SDSD morphology in Fig.~\ref{fig_symmetric}d, the representative values are given by $V_\alpha = V_\beta = 0.5$, $c_\alpha = 0.76$, and $c_\beta = 0.2$, yielding $c_{sp} = 0.48$.

A characteristic time scale for the growth of spinodal wavelength $\lambda_{sp}$ can be approximated by the relation~\cite{cahn1961spinodal}: 
\begin{equation}
\tau_{sp} \approx \frac{4(\kappa_{i}+\kappa_{j})}{|\partial^2 f/\partial c^2|^2 \, M_{ij} \, c_{sp}}.
\end{equation}
The physical interpretation of $\tau_{sp}$ is the time needed for the diffusion of species across the distance $\lambda_{sp}$ divided by the thermodynamic driving force, which is on the order of $T/T_c$ (refer to Sec.~\ref{sec_parameters}). Substituting the parameter values for a critical blend approximates dimensionless $\tau_{sp} \approx 80$. In Fig.~\ref{fig_critical}b, the time when the target pattern begins to form is given by $t/\tau_{sp} \approx 1$. Following this, the species of $\beta$ continues to build up around particles before reaching the local equilibrium (denoted as a vertical line in Fig.~\ref{fig_critical}b). The coarsening of the bulk phases begins around $t/\tau_{sp} \approx 6$, beyond of which phase inversion occurs within the target pattern, altering the sequence of phases around particles. Phase inversion is further illustrated by plotting the thickness of the wetting layer ($\delta$) in the target pattern (Fig.~\ref{fig_critical}c). The preferred $\alpha$-layer rapidly grows at early times before reaching a plateau and finally disappearing (i.e., $\delta_\alpha = 0$) from the particle surface due to coarsening. As a result, $\beta$ accumulates around the particles and becomes thicker with time before reaching a plateau with a fixed $\delta_{\beta}$. This $\beta$-layer of thickness $\delta_\beta$ persists in the target pattern beyond $t/\tau_{sp} > 50$ and $\lambda/\lambda_{sp} \geq 3$ limits, while $\alpha$ forms the continuous phase (Figs.~\ref{fig_symmetric}d, \ref{fig_asymmetric}d). The plot of $\delta$ vs. $t$ compares excellently with related experimental measurements reported in Refs.~\cite{puri2013,aichmayer2003}.

\subsection{Finite Particle Clusters in Phase-Separating Mixtures}
\begin{figure}[h]
\centering
\subfloat[$A_{50}B_{50}$]{\label{cp50b50}\includegraphics[trim={20cm 15 20 15},clip,scale=0.05]{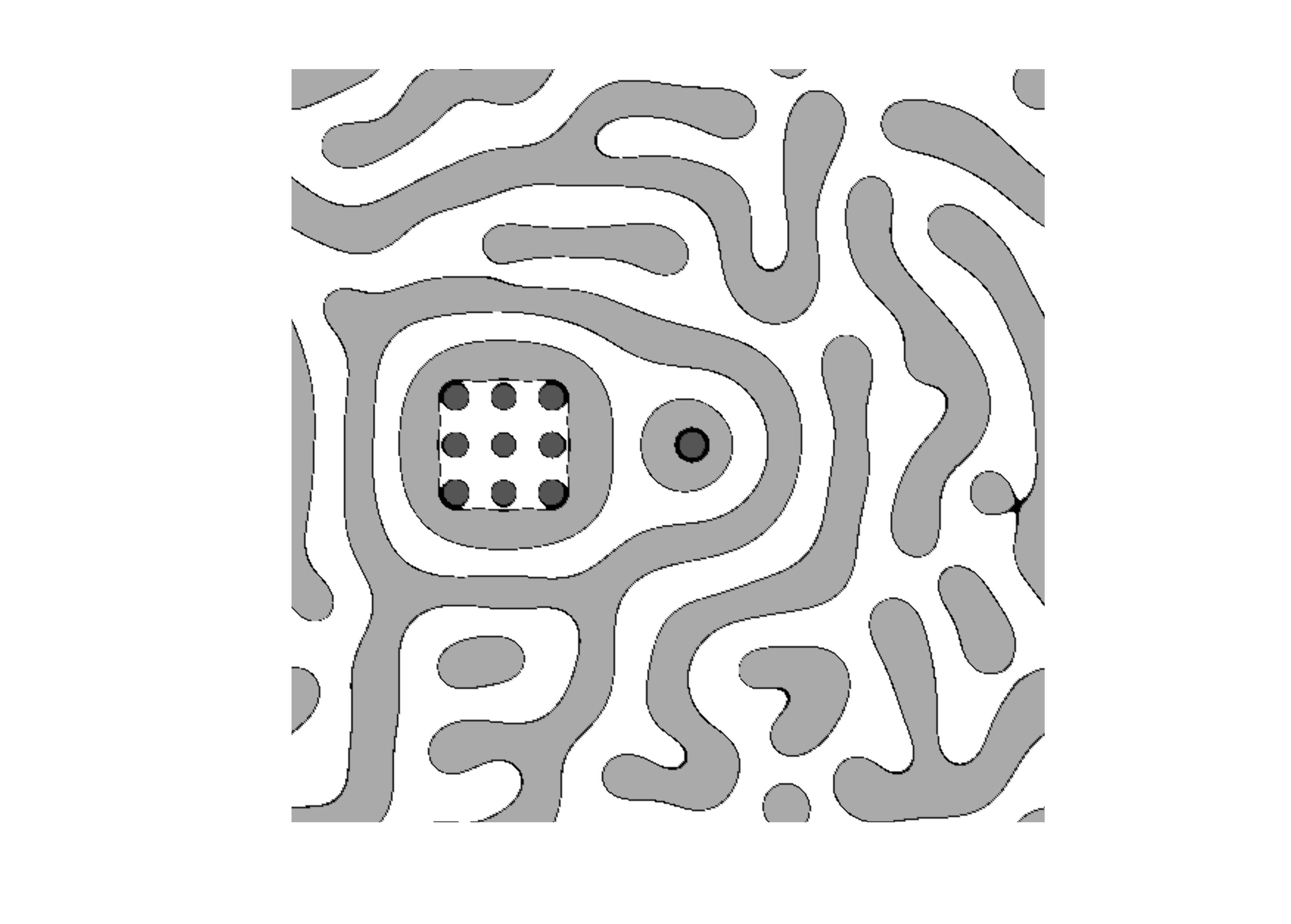}}
\subfloat[$A_{40}B_{60}$]{\label{cp40b60}\includegraphics[trim={20cm 15 20 15},clip,scale=0.05]{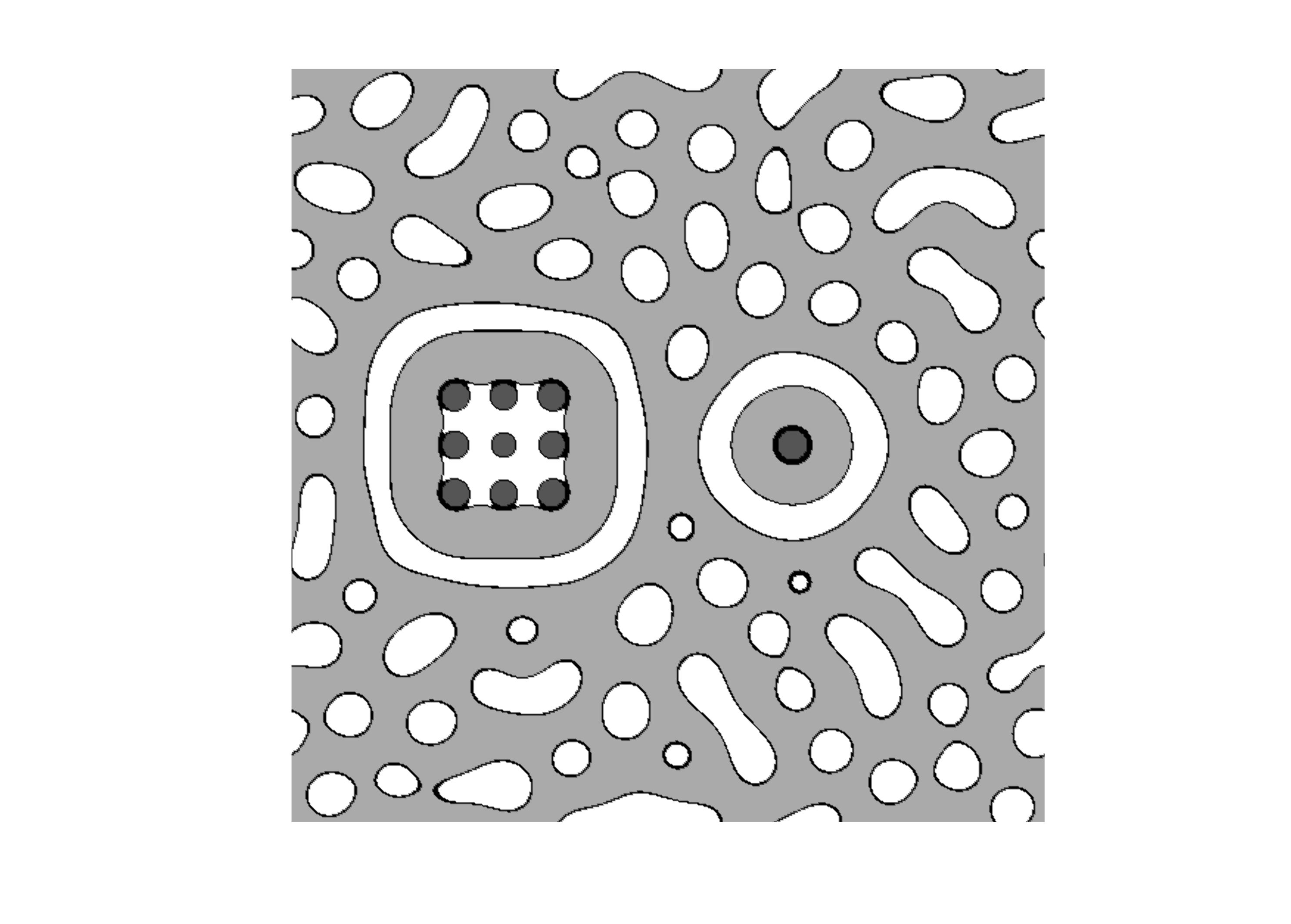}}
\subfloat[$A_{60}B_{40}$]{\label{cp60b40}\includegraphics[trim={20cm 15 20 15},clip,scale=0.05]{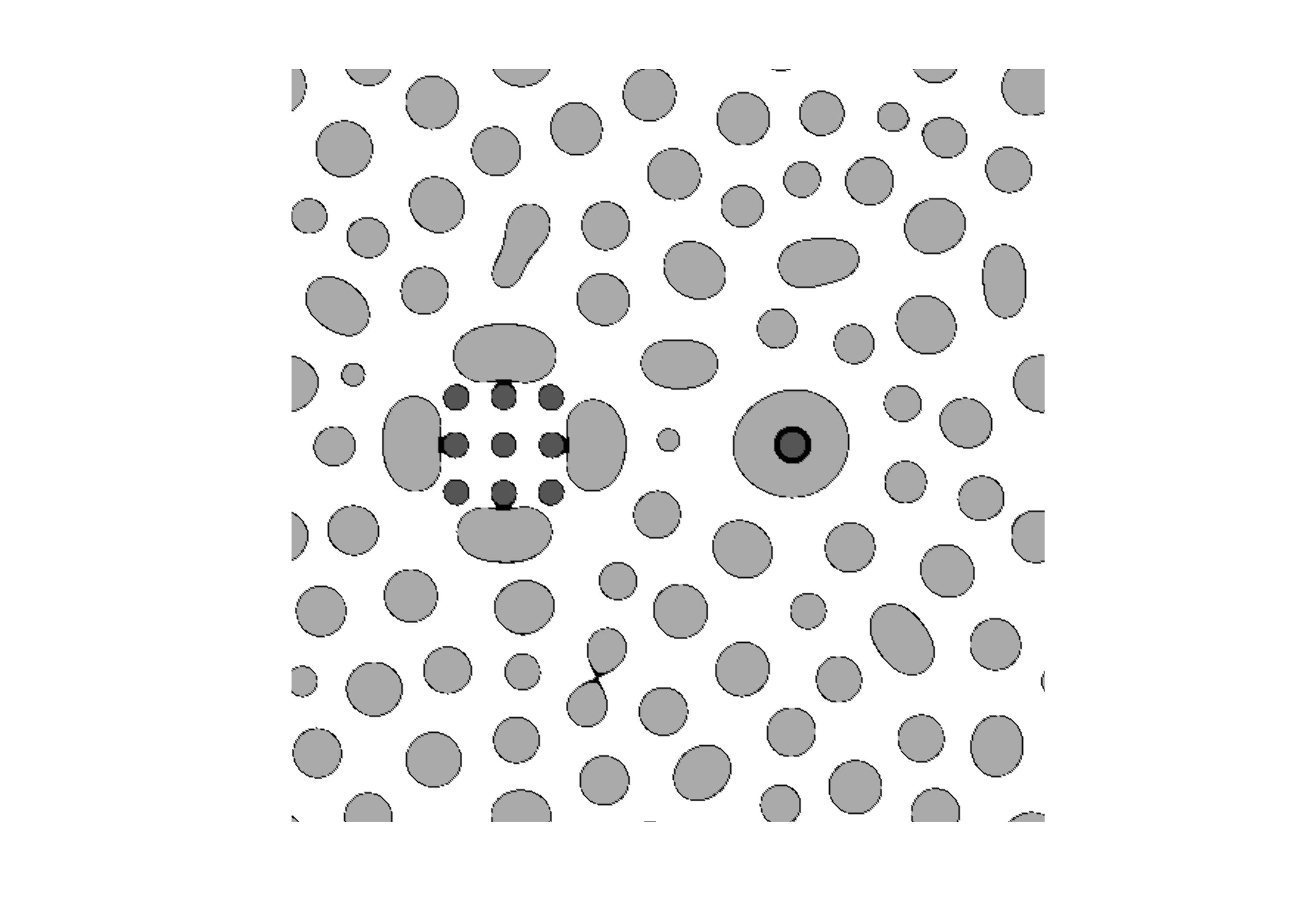}}
\caption{SDSD microstructures are simulated in (a) critical blend and (b, c) off-critical blends using cluster-particle substrates with various values of cluster-particle distance $\lambda_1/\lambda_2$. The value of $\lambda_1/\lambda_2 = 3$ in (a) and $\lambda_1/\lambda_2 = 5$ in (b, c). The snapshots correspond to the dimensionless time $t = 3000$. The $\alpha$, $\beta$, and $\gamma$ phases are illustrated by white, light gray, and dark gray, respectively.}
\label{fig_cluster_particle}
\end{figure}

\begin{figure}[htbp]
\centering
\includegraphics[scale=0.75]{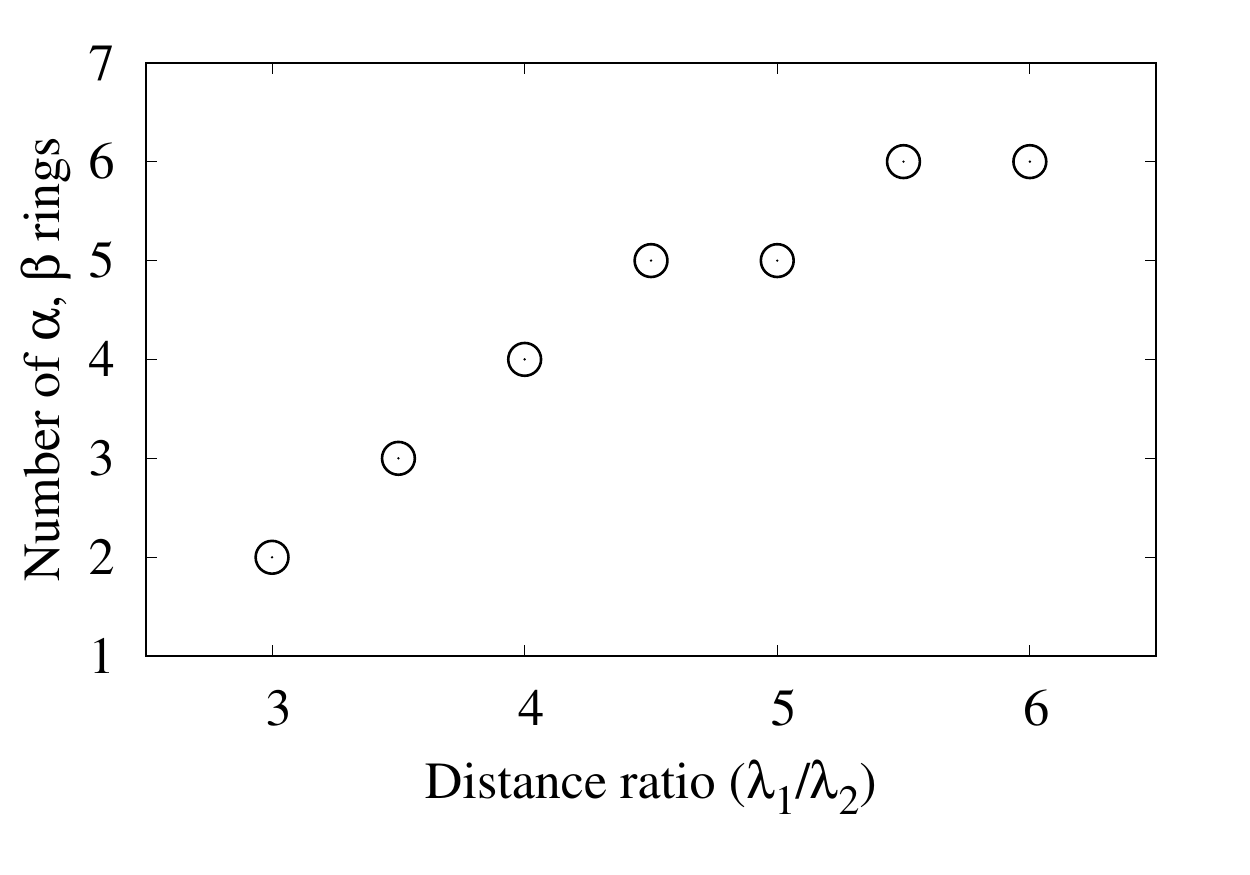}
\caption{The average number of rings in the target pattern around cluster increases with the increasing value of interparticle distance ratios ($\lambda_1/\lambda_2$) in cluster-particle systems. Besides $\lambda_1/\lambda_2$, the number of rings in these systems depends on quench depth, interaction strength between particle and component phases, and time. For a definition of $\lambda_1$ and $\lambda_2$, please refer to the text.}\label{fig_ring}
\end{figure}

The SDSD around a cluster of particles ($\gamma$) should be relevant to many industrial applications, including thin films and polymer blends~\cite{Jiang,segalman2005block}. The presence of filler particles in the form of a connected network could potentially interfere with the target pattern induced by isolated filler particles, depending on the separation between them. This interference may control the phase separation morphology as it results in the confinement of regions of preferred and non-preferred phases in and around the particle network. In such cluster-particle systems, the key parameters that control phase separation and the subsequent phase coarsening are various average interparticle spacings within the cluster network and between the cluster and isolated particles. Let us consider that the particle network contains a $3 \times 3$ array of particles, where $\lambda_1$ defines the minimum distance between the network and isolated particles, and $\lambda_2$ denotes the average distance between the particles within the network. The long-time morphological evolution of such particle systems with $\lambda_1/\lambda_2 = 3$ is simulated in a critical blend (Fig.~\ref{fig_cluster_particle}a). The picture that emerges from these simulations is qualitatively similar to that of a blend without particle clusters. At early to intermediate times, several concentric rings of $\alpha$ and $\beta$ not only develop around the $\gamma$ particle and cluster, but they may also enclose the entire $\gamma$ region (cluster + particle). The number of rings that form around $\gamma$ particles in the presence of the $\gamma$ cluster increases with increasing $\lambda_1/\lambda_2$, as shown in Fig.~\ref{fig_ring}. At later times, these rings undergo coarsening as phase inversion leads the innermost $\alpha$ ring in the target pattern to disappear, bringing the surrounding $\beta$ ring to the $\gamma$ surface. These rings of $\alpha$ and $\beta$ around the $\gamma$ region survive for extended times. In bulk, the spinodal pattern resembles that of single-particle simulations (Fig.~\ref{target2}), noting that, no bulk $\beta$ or $\alpha$ exists in multi-particle high-$\lambda$ systems (Figs.~\ref{fig_symmetric}d, \ref{fig_symmetric_a40b60}d). 

Although there are many similarities, we speculate that the stability of the target pattern is more pronounced in cluster-particle systems compared to multi-particle systems. In cluster-particle systems, the target pattern around $\gamma$ remains stable for extended timescales as the coarsening process is significantly delayed. In contrast to multi-particle systems, where only the innermost $\beta$ ring survives around $\gamma$, the target pattern having both the $\alpha$ and $\beta$ rings survive for extended periods (before breaking off the outermost $\beta$ ring) in these systems. Such enhanced stability of the target rings in cluster-particle systems depends on several factors. The composition waves emanating from both the cluster and isolated particles meet at a certain distance from their respective sources. The average cluster-particle distance dictates whether these waves will represent a constructive or destructive interference, leading to improved or reduced stability of the resultant target pattern. Such a critical distance depends on the material properties and process conditions, in particular, the spinodal wavelength and polarity between involved phases.

Since we already determined (in Sec.~\ref{sec_mp_systems}) the critical value of $\lambda$ ($\approx 3\lambda_{sp}$) for which the target pattern around an isolated particle in a critical blend becomes stable in the long-time limit, we set the baseline values for $\lambda_1 = 3\lambda_{sp}$ and $\lambda_2 = \lambda_{sp}$, yielding $\lambda_1/\lambda_2 = 3$. We note that as long as $\lambda_2 \leq \lambda_{sp}$, the finite multi-particle cluster in Fig.~\ref{fig_cluster_particle} behaves like a single-particle (Fig.~\ref{target1}) about which the target pattern develops, adopting the symmetry of the cluster. Although not shown here, when $\lambda_2 > \lambda_{sp}$, the target pattern not only develops around the cluster but also surrounds each particle within the cluster at early times. We refer to these target patterns outside of the cluster and particle as ``cluster target'' and ``particle target,'' respectively. Depending on $\lambda_2$, blend composition, and time, the confined region inside the cluster target undergoes a phase life cycle process similar to that of multi-particle configurations simulated in Figs.~\ref{fig_symmetric}--\ref{fig_asymmetric_a60b40} (Sec.~\ref{sec_mp_systems}). Specifically, in these high-$\lambda_2$ systems, the particles within the cluster behave like an isolated particle around which the target pattern develops in which the number of rings primarily depends on $\lambda_2$ ($f(\lambda_{sp})$), blend composition, interaction energy ($\chi_{ij}$), and time. Here we focus on cluster systems with $\lambda_2 < 3\lambda_{sp}$ for which the formation of particle target inside the cluster tends to be suppressed, and thus the interference of the composition rings emanating from the cluster and an isolated particle can be realized into preferred and non-preferred regions of the bulk phases.

The target pattern around off-critical particle clusters is significantly different. At early to intermediate times in $A_{40}B_{60}$ blend, concentric $\alpha$ and $\beta$ rings form around both the cluster and isolated particles (Fig.~\ref{fig_cluster_particle}b). However, unlike the target pattern in a critical blend, both the rings of $\alpha$ and $\beta$ survive in this blend morphology with the preferred phase $\alpha$ forming the outermost ring. In the background, as expected, the non-preferred majority $\beta$ forms the continuous phase in which the minority $\alpha$ remains as thin isolated islands.

In $A_{60}B_{40}$ cluster-particle systems, the majority as well as the preferred phase $\alpha$ forms the continuous background while the minority $\beta$ forms as several rings of droplets around $\gamma$. The size of these droplets is comparable to the size of the particle, and gets bigger with time due to coarsening and with increasing distance from $\gamma$.

\subsection{Coarsening Kinetics}
Domain growth in SDSD microstructures is characterized using a structure function ($S_{i}$ in Eq.~(\ref{eq_sf})), the first moment ($k_1$) of which represents the average size ($R_1$) of $\alpha$ or $\beta$ domains:
\begin{equation}\label{eq_first_moment}
R_1(t) = \frac{1}{k_{1}(t)} = \frac{\sum S_{i}(k,t)}{\sum k S_{i}(k,t)}.
\end{equation}
%It is often observed in measurements and simulations that when the coarsening process during spinodal decomposition of a binary mixture is diffusion-controlled, as in the present case, the growth of domains follows the Lifshitz-Slyozov-Wagner (LSW) power-law: $R_1^3 \propto t$~\cite{lifshitz1961kinetics, Wagner1961}.

\begin{figure}[h]
\centering
\subfloat[]{\includegraphics[scale=0.45]{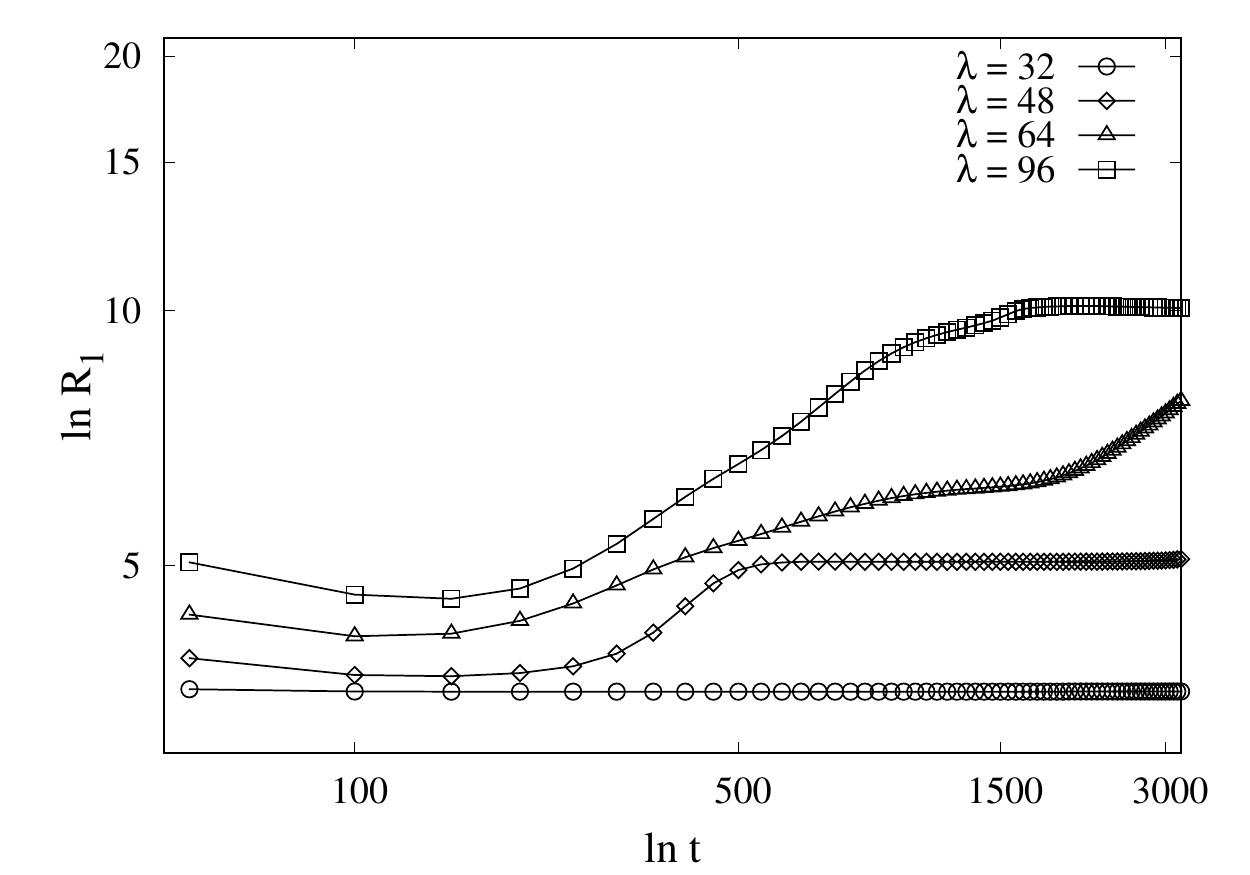}}
\subfloat[]{\includegraphics[clip,scale=0.45]{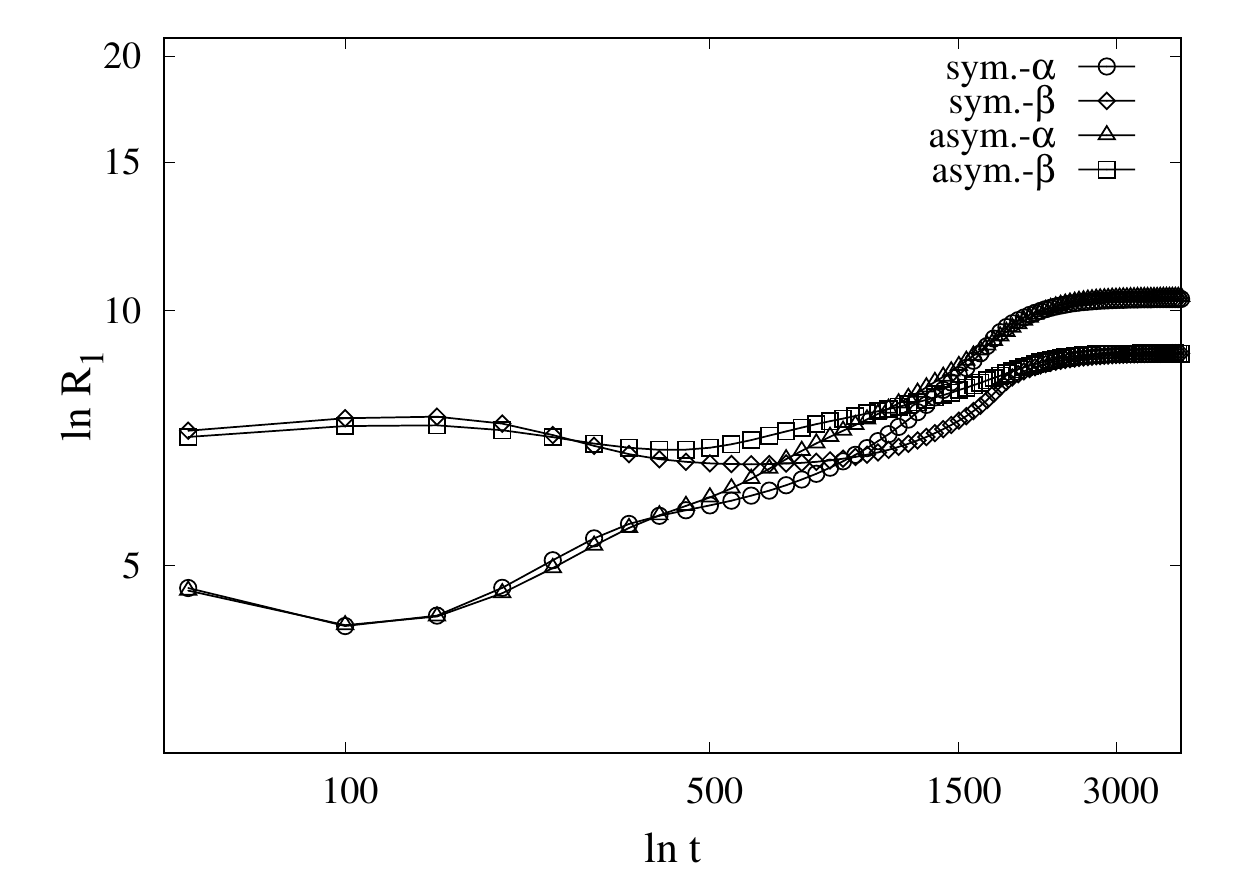}}
\subfloat[]{\includegraphics[clip,scale=0.45]{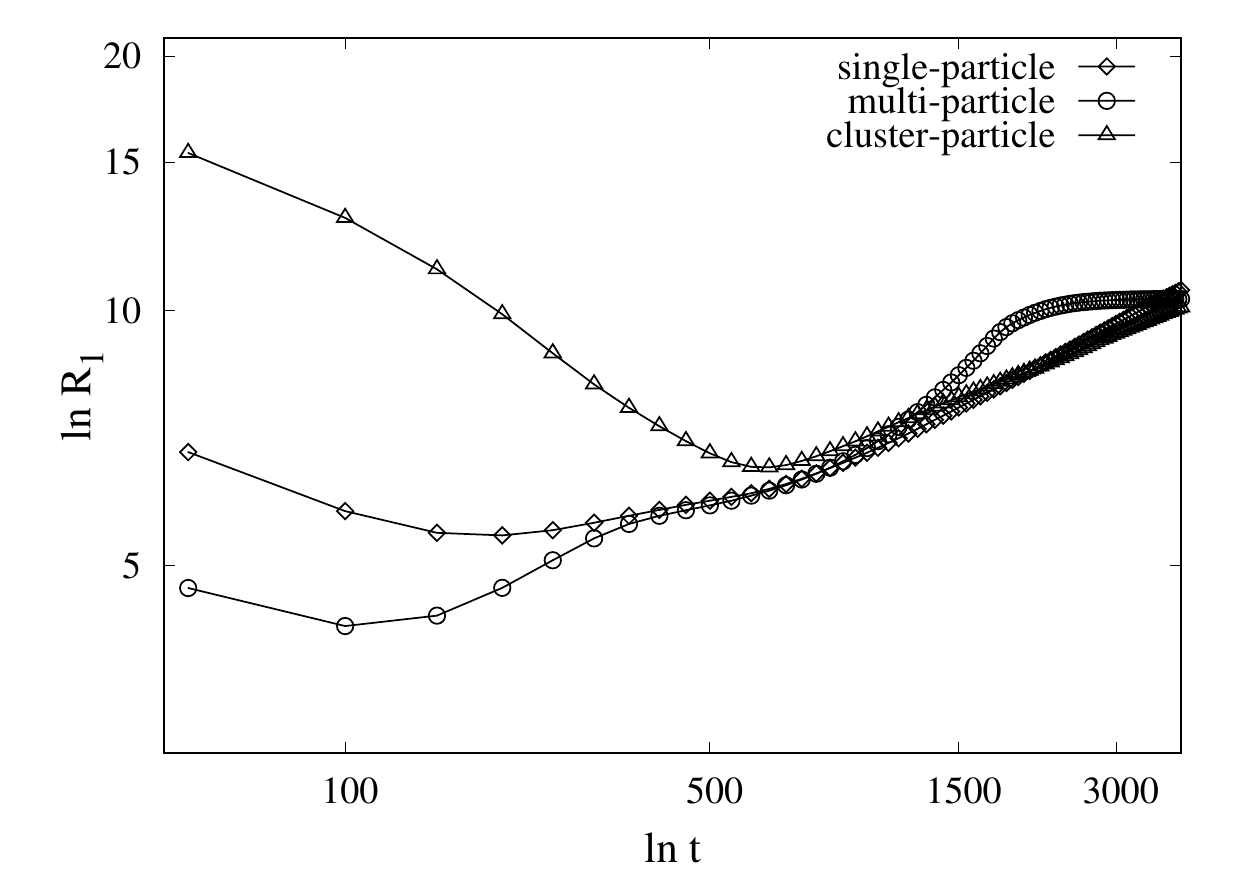}}
\caption{Average domain size $R_1(t)$ is plotted as a function of time $t$ in a double-logarithmic plot. (a) The time evolution of $\alpha$ domains is estimated from the SDSD microstructures (in Fig.~\ref{fig_symmetric}) in a symmetric multi-particle blend with varying values of $\lambda$. (b) The size of preferred $\alpha$ and non-preferred $\beta$ domains is compared using SDSD microstructures with symmetric and asymmetric multi-particle critical blends (comparison of domains in Fig.~\ref{fig_symmetric}d and Fig.~\ref{fig_asymmetric}d). (c) Domain size and coarsening rate of $\alpha$ are compared among single-particle (Fig.~\ref{fig_target2}), multi-particle (Fig.~\ref{fig_symmetric}d), and cluster-particle (Fig.~\ref{fig_cluster_particle}a) critical blends.}
\label{fig_domain}
\end{figure}

Figure~\ref{fig_domain} summarizes the coarsening kinetics in simulated SDSD patterns. Critical blends with symmetric or asymmetric particle distributions begin coarsening at a very early time (Fig.~\ref{fig_domain}a). And, as expected, the size of the bulk domains increases with increasing $\lambda$. At later times, these patterns become \textit{steady} as the domain growth slows down and pins to a finite size due to the high particle fraction. On average, the domain size remains invariant of particle distribution with a constant size ratio of the bulk domains. This is illustrated in Fig.~\ref{fig_domain}b in which the average domain size follows $R_1^\alpha : R_1^\beta = 2:1$, which corresponds to the SDSD morphology in Fig.~\ref{fig_symmetric}d. Although not shown here, a similar trend in domain size ratios can be established for all other systems described in this work.

The coarsening kinetics and the average domain size are significantly different in cluster-particle systems (Fig.~\ref{fig_domain}c). Coarsening of bulk domains is delayed considerably in such systems. Unlike the multi-particle systems, where steady patterns form, dynamical coarsening of the bulk domains dominate in cluster-particle systems, exhibiting a power-law growth behavior~\cite{lifshitz1961kinetics, Wagner1961}. Finally, the average size of the domains becomes comparable in all systems after extended times. We do not present the coarsening kinetics of the off-critical systems since the established trends are similar to that of the critical blend presented here.   

\section{Discussion}\label{sec_discussion}
We discuss our results highlighting the following observations during SDSD:

\begin{itemize}

\item \textit{Effect of particle substrate:} The presence of a particle template affects the target and bulk patterns greatly during phase separation. When the particles at high loadings were distributed randomly in Ref.~\cite{ghosh_pccp}, the resultant target pattern was transient, breaking into bicontinuous or isolated domains at later times. In contrast, when the particle arrangements are periodic, as in the present case, either the target of (non-preferred) $\beta$ remains around each particle in a continuous $\alpha$ (Figs.~\ref{fig_symmetric}d,~\ref{fig_asymmetric}d) or the target of (preferred) $\alpha$ remains around each particle in a continuous $\beta$ (Figs.~\ref{fig_symmetric_a40b60}c,~\ref{fig_symmetric_a40b60}d). Also, the coarsening kinetics and average size of the domains are different in above particle systems.

\item \textit{Effect of preferential wetting:} In our work, particles have a strong preference for A. This is why at early times, A-rich $\alpha$ rings form first around particles surrounded by the non-preferred $\beta$ layers, thus forming the typical target pattern. However, over time, the coarsening and phase inversion processes break the $\alpha$ ring in the target pattern, bringing $\beta$ to the particle surface. In off-critical blends with minority A (i.e., $A_{40}B_{60}$), the preferred phase $\alpha$ survives in the target pattern (Figs.~\ref{fig_symmetric_a40b60}c,~\ref{fig_symmetric_a40b60}d). The reinforcement of such a \textit{selective} target phase around particles can be tailored to yield targeted properties in applications similar to those in block copolymer films~\cite{epps2016block,mai2012block,hu2014block,bang2009block,hamley2009block}.

\item \textit{Stability of target pattern:} The target pattern stabilizes for characteristic values of length, time, composition, and layer thickness measures. A linear stability analysis yields the critical length $\lambda_{sp}$ and time $t_{sp}$ scales for which A:B phase separates~\cite{cahn1961spinodal}. We find that the target pattern begins to form around $t/\tau_{sp} \approx 1$ and develops until the onset of coarsening, leading to inversion of the phase sequence in the target pattern. In the long-time limit, within the target pattern, the non-preferred $\beta$ survives in a critical blend and the preferred $\alpha$ survives in an off-critical blend often beyond $\lambda/\lambda_{sp} \geq 3$ and $t/\tau_{sp} \geq 50$ limits with critical composition $c_{sp}$ and critical layer thickness $\delta$ at the particle surface.

\item \textit{Effect of particle clusters:} The target pattern forms not only around isolated $\gamma$ particles, but also around clusters of these particles. Two length scales, the interparticle distance in the cluster and cluster-particle distance, interact with the spinodal length scale in bulk, guiding SDSD in such systems. Qualitatively at least, we find that by controlling the cluster-particle distance, the number of rings around particles can be controlled and the stability of these rings can be improved. The coarsening rate and average domain size in these systems are smaller compared to those of multi-particle systems. 

\item \textit{ Effect of mixture composition:} Multi-particle blends often phase separate to either continuous $\alpha$ with a stable $\beta$ target around each particle (Fig.~\ref{fig_symmetric}d) or continuous $\beta$ with a stable $\alpha$ target around each particle (Fig.~\ref{fig_symmetric_a40b60}d). In cluster-particle $A_{40}B_{60}$ blends, continuous $\beta$ with two rings of $\alpha$ and $\beta$ survive around $\gamma$, with $\alpha$ and $\beta$ forming the outer and inner rings, respectively (Fig.~\ref{fig_cluster_particle}b). These $\alpha$ rings are thinner compared to that of $\beta$ due to the smaller volume fraction of $\alpha$. In $A_{60}B_{40}$ blends, the target pattern does not form. Instead, concentric rings of non-preferred $\beta$ droplets form around $\gamma$. In cluster-particle systems, the distribution of these droplets ($\lambda_{\beta}$) is guided by the interparticle spacing ($\lambda_2$) in the cluster, and its size increases with increasing distance ($\lambda_1$) from the cluster (Fig.~\ref{fig_cluster_particle}c). 

\item \textit{Effect of interparticle spacing:} With the varying values of $\lambda$, SDSD leads to droplet, continuous, transition, and lamellar microdomains. The formation of such domains is controlled by particle configuration and selective preference between the particle and bulk phases. While these domains saturate to a finite size in multi-particle systems, particle clusters exhibit dynamical coarsening at late times (Fig.~\ref{fig_domain}).

\item \textit{Effect of particle loading:} In our simulations, particle loading reaches as high as 20~\% as we vary $\lambda$. Particles at a high loading (i.e., low-$\lambda$) can be used to reinforce one phase while the other phase remains dispersed. In dilute blends (i.e., high-$\lambda$ systems), particles are surrounded by a selective phase in the target pattern, depending on the interfacial energy between the particle and bulk phases, while the other phase remains continuous (Figs.~\ref{fig_symmetric}d,~\ref{fig_symmetric_a40b60}d). In this context, the effect of constant particle loading on pattern stability by varying particle size and thus interfacial extent is currently under investigation. While the continuous phase controls mechanical robustness and overall transport properties of the material for applications, the target phase around particles can be tuned to impart desired properties. For instance, the thermal or electrical conductivity of the target phase can be modified using metal particles around which the target morphology forms~\cite{amoabeng2017}.

\item \textit{Effect of phase inversion:} Assuming volume diffusion and obeying the Gibbs-Thomson effect~\cite{Porter,voorhees1985}, the rate of coarsening of the particle is proportional to $\sigma_{ij}/R^2$ \cite{Porter,voorhees1985}. Target pattern, having the symmetry of the particle, a change in particle radius will have a similar effect on the size and, hence, the radius of curvature of the associated target rings. This may delay the onset of the phase inversion process in the target pattern. However, on average, the net effect of altering the particle size on phase inversion among the target rings remains similar.

Similar to $R$, the spinodal structures can be designed by altering $\sigma_{ij}$ (Table~\ref{tab_param}), which controls the magnitude of effective interaction between particle and matrix phases. In particular, the onset (or, time scale) of wetting-induced phase separation around particles as well as later in the bulk and the rate and duration of phase coarsening are likely to be affected. Work in these directions is currently in progress.

\item \textit{Effect of ``quenched disorder'':} Quenched disorder~\cite{hashimoto1984time,yue2009suppression,paul2004domain,paul2005domain} can greatly alter phase transition behavior in ordering systems such as materials undergoing crystallization and block copolymer ordering, and a sensitivity to disorder can also be expected in phase-separating materials even though we are dealing with the case of a conserved order parameter defined in terms of material composition. Disorder sites often trap the coarsening of domains during ordering at late times in these systems, and domain growth proceeds \textit{via} thermal activation over disorder barriers primarily as a function of quench temperature and disorder strength~\cite{paul2004domain,paul2005domain}. The domain growth shows a crossover from a power-law to either logarithmic (if thermal fluctuations are present) or freezing (if there are no thermal fluctuations). Our study of phase separation with heterogeneity is somewhat reminiscent of latter situation since our simulations involve thermal quenches far away from critical points or spinodal lines, where the driving force for phase separation is much stronger than the effect of thermal fluctuations~\cite{huang1995phase}. Also, similarities may arise, at least in early times, when domain sizes are small and remain unaffected by any disorder in the system, before regular multi-particle arrays arrest domain growth in intermediate to late times in our phase-field simulations.

Previous work has focused on the influence of quenched disorder on block copolymer ordering, providing an interesting contrast to our own study of phase-separating fluid mixtures in the presence of particles. The influence of quenched disorder in block copolymer materials in the presence of particle non-uniformities has been reviewed and investigated experimentally~\cite{hashimoto1984time}. In particular, some measurements indicate that adding a relatively small concentration of fullerene nanoparticles into polystyrene-polyisoprene block copolymer materials that order as a fluctuation-induced first-order phase transition without the nanoparticles, causes the phase transition to be destroyed, yielding a kind of ``glass'' where the dynamics of ordering is completely structurally arrested. The general tendency of fullerenes to cluster in the material, creating quenched disorder, is apparently crucial in achieving such disruption of order-disorder transition, as other authors have shown that nanoparticles that do not associate, and which segregate to regions rich in the composition of one or the other polymers in block copolymers, lead to a shift of the order-disorder transition temperature and somewhat modifies the geometry of the ordered domains (e.g., domain spacing). Phase-field simulations of block copolymer ordering inherently do not treat the fluctuations in this type of system correctly which, in this case, drive the phase transition from being second-order to weakly first-order. Studies of how fluctuations affect ordering in this type of system will apparently require full molecular dynamics simulation of block copolymers in the presence of particle, particle array, and particle cluster impurities. This topic is extremely interesting to us, but far beyond the scope of the present study.

\end{itemize}

Our results suggest that adjusting the matrix composition, selective wetting, and interparticle spacing in patterned multi-particle and cluster-particle substrates, SDSD morphologies can be effectively designed and controlled. Recalling that we have neglected hydrodynamic interactions that are known to play a significant role in the very late stages of the phase separation, but we expect the effect of these interactions on target pattern formation to be very limited in the early to intermediate stages of phase separation~\cite{ACBalazs,clarke2002target}. However, hydrodynamic interactions can be incorporated in our scheme by coupling our phase-field simulations with lattice Boltzmann simulations to address long-time stability of the phase separation structures induced by particles. This method also allows us to incorporate processing related effects encountered in real applications such as fluid flow~\cite{wu2019flow,zoumpouli2016flow,DCBalazs,tanaka2000flow}. Three-dimensional simulations of the above particle systems will address intricate percolated patterns guided by preferential wetting and geometrical confinement~\cite{shimizu2017}. New SDSD morphologies should also arise by varying the size and, in particular, the mobility of the particles. The spinodal length scale ($\lambda_{sp}$) can be tuned in experiments by varying the quench depth into the two-phase region of the phase diagram. The greater the quench depth (or $\chi_{ij}$), smaller the $\lambda_{sp}$. Elastic interactions can also affect the compositional history and coarsening kinetics of SDSD microstructure phases that can be simulated, for example, using microelastic classes of phase-field models~\cite{saswata2020,khachaturyan2013,rajdip2009,sukriti2018}. The mechanical behavior of the resulting patterns can be estimated using a finite element~\cite{oof} or finite volume~\cite{carolan2015} based analysis of the representative microdomains. In future work we will elaborate further the effect of changing the above parameters. Our model and approach are first approximations toward treating spinodal decomposition guided by topographic templates for obtaining controlled morphologies during the fabrication of complex ``solid-like'' mesoscale structures~\cite{choo2018,Herzig}.

%Although there a number of parameters that can be varied for any given system, our phase-field simulations address
%since more time required to overwhelm the scale of the target pattern around larger particles.
\section{Summary}\label{sec_conclusions}

We have used a ternary phase-field model in two dimensions to simulate the spinodal decomposition of a binary mixture templated by multi-particle and particle-cluster systems. We modeled the average phase separation behavior of critical and off-critical blends for varying interparticle spacing ($\lambda$) with particles having a selective preference for one of the components. With a symmetrically periodic distribution of particles in a critical blend in the low-$\lambda$ limit, phase separation resulted in droplets of the non-preferred phase in the continuous preferred phase. In contrast, asymmetrically periodic distribution of particles resulted in lamellar microdomains; in these systems, the preferred phase tends to form an interconnected structure so that the particles are essentially bridged by it. The morphological evolution in the high-$\lambda$ limit in the above multi-particle systems was equivalent: at early to intermediate times, the target pattern having alternate rings of preferred and non-preferred phases develops around particles. These phases undergo coarsening and phase inversion at later times, bringing a selective target of either the non-preferred phase (in the continuous preferred phase in a critical blend) or the preferred phase (in the continuous non-preferred phase in an off-critical blend) around particles. The stability limits of the target pattern were determined using characteristic spacing, time, and composition scale measures that can potentially be used to design a spinodal structure. All multi-particle simulations reached steady-state with domain sizes saturated to a finite size, the value of which increased with increasing $\lambda$. 

When spinodal decomposition was simulated with cluster-particle configurations, the stability of the resultant target pattern was enhanced, meaning that more rings of both the preferred and non-preferred phases survived for extended timescales. The mixture composition played a significant role in these morphologies. When the preferred phase was minor in the blend, more rings survived around the particles. In contrast, when the preferred phase was major, no target pattern developed as the concentric rings of minority droplets reigned around particles. Overall, our phase-field simulations of target morphologies for varying mixture composition and particle fraction address the average dynamical interplay among spinodal decomposition, interparticle spacing, preferential wetting, and coarsening in particle-filled blends often encountered in applications. One important particle configuration that we have not treated in our study is the situation where the particles have aggregated into fractal aggregates where the particles are directly touching and exhibit a hierarchical structure. Preliminary calculations~\cite{Jiang} indicate that the fractal particle cluster ``wets'' itself with the preferred phase to create very complex phase separation morphologies exhibiting constructive and destructive interference between the composition waves about the particles in these structures. We plan to study this problem, which often arises from a general tendency for particles to form non-equilibrium aggregates, in a separate publication.   
%\section{Summary}
%%------------------------------------------------------------------------------------
%%%\section*{Acknowledgments}

%\section*{DATA AVAILABILITY}
%The data that support the findings of this study are available upon reasonable request.

\appendix

\section{Ternary Phase-Field Formulation}\label{sec_appendix}
The kinetics of spinodal decomposition in mixtures are described by the continuity equation,
\begin{equation}\label{app_continuity2}
\frac{\partial c_i}{\partial t}= - \nabla\cdot \bar{\mathbf{J}}_i,  
\end{equation} 
where $c$ is composition, $t$ is time, $\nabla$ is the gradient operator, $\nabla \cdot$ is the divergence operator, and $\bar{\mathbf{J}_i}$ is the net mass flux of component $i$ = A, B, C. The diffusion flux of each component, $\mathbf{J}_i$, relates to chemical potential, $\mu_i$, by~\cite{cahn1961spinodal}
\begin{equation}\label{app_mobility1}
\mathbf{J}_i= -M_i\nabla \mu_i, 
\end{equation}
where $ M_i $ is the Onsager mobility coefficient of $i$ and is always positive. 
In formulating the nonlinear diffusion equation for polymer mixtures, we used the approach by Kramer \emph{et al.}~\cite{kramer}, following Refs.~\cite{huang1995phase,Huang,Nauman}, which proposed there is a net vacancy flux operating during the lattice diffusion processes with the constraint of local thermal equilibrium of vacancies. Thus $\mathbf{\bar J}_i $ becomes the sum of the diffusion flux of $i$ plus $i$ transported by the vacancy flux $\mathbf{J}_V$,
\begin{equation}\label{app_flux}
\mathbf{\bar{J}}_i = \mathbf{J}_i + c_i \mathbf{J}_V,
\end{equation}
where the conservation of available lattice sites is denoted by
\begin{equation}\label{app_vacancy_flux}
   \mathbf{J}_V = -(\mathbf{J}_A + \mathbf{J}_B + \mathbf{J}_C). 
\end{equation}
Substituting Eq.~\eqref{app_vacancy_flux} into Eq.~\eqref{app_flux} yields 
\begin{equation}\label{app_ji2}  
\mathbf{\bar J}_i = \mathbf{J}_i - c_i \sum_{i = A, B, C} \mathbf{J}_i.
\end{equation}
Using $c_A +c_B +c_C = 1$ (Eq.~\eqref{eq_sum1}), Eq.~\eqref{app_ji2} becomes
\begin{equation}\label{app_j0}
\mathbf{\bar{J}}_A + \mathbf{\bar{J}}_B + \mathbf{\bar{J}}_C = 0.
\end{equation}  
Combining Eqs.~\eqref{app_mobility1} and \eqref{app_ji2}, $\mathbf{\bar J}_i$ can be written as,
\begin{eqnarray}\label{app_jabc}
\mathbf{\bar J}_A &=& -\left(1-c_A\right)M_A\nabla\mu_A + c_A M_B\nabla\mu_B + c_A M_C\nabla\mu_C,\nonumber\\
\mathbf{\bar J}_B &=& -\left(1-c_B\right)M_B\nabla\mu_B + c_B M_A\nabla\mu_A + c_B M_C\nabla\mu_C,\nonumber \; \text{and} \\
\mathbf{\bar J}_C &=& -\left(1-c_C\right)M_C\nabla\mu_C + c_C M_A\nabla\mu_A + c_C M_B\nabla\mu_B.
\end{eqnarray}
Due to the constraints in Eqs.~\eqref{eq_sum1} and \eqref{app_j0}, we only need two solutions, say, $c_A$ and $c_B$. Applying the Gibbs-Duhem equation locally~\cite{huang1995phase,Huang},
\begin{equation}\label{app_Gibbs_Duhem}
c_A \nabla\mu_A + c_B \nabla\mu_B + c_C \nabla\mu_C = 0,
\end{equation}
and rearranging Eq.~\eqref{app_Gibbs_Duhem} with Eq.~\eqref{eq_sum1} yield
\begin{eqnarray}
\nabla\mu_C &=& -c_A\nabla\mu_A^{eff}-c_B\nabla\mu_B^{eff}, \nonumber\\
\nabla\mu_A &=& (1-c_A)\nabla\mu_A^{eff}-c_B\nabla\mu_B^{eff}, \nonumber \; \text{and} \\
\nabla\mu_B &=& (1-c_B)\nabla\mu_B^{eff}-c_A\nabla\mu_A^{eff},
\end{eqnarray}
where $\nabla\mu_A^{eff} = \nabla\mu_A - \nabla\mu_C$ and $\nabla\mu_B^{eff} =\nabla\mu_B-\nabla\mu_C $. Substituting Eq.~\eqref{app_Gibbs_Duhem} into Eq.~\eqref{app_jabc} yields,
\begin{equation}\label{app_ja}
\mathbf{\bar J}_A = -\left[\left(1-c_A\right)^2M_A+c_A^2\left(M_B+M_C\right)\right]\nabla\mu_A^{eff}+\left[c_BM_A\left(1-c_A\right)+c_AM_B\left(1-c_B\right)-c_Ac_BM_C\right]\nabla\mu_B^{eff}
\end{equation}
and
\begin{equation}\label{app_jb}
\mathbf{\bar J}_B = -\left[\left(1-c_B\right)^2M_B+c_B^2\left(M_A+M_C\right)\right]\nabla\mu_B^{eff}+\left[c_AM_B\left(1-c_B\right)+c_BM_A\left(1-c_A\right)-c_Ac_BM_C\right]\nabla\mu_A^{eff}.
\end{equation}
We define the effective mobilities as,
\begin{eqnarray}\label{app_mobility}
M_{AA}&=&\left(1-c_A\right)^2M_A+c_A^2\left(M_B+M_C\right), \nonumber\\
M_{BB}&=&\left(1-c_B\right)^2M_B+c_B^2\left(M_A+M_C\right), \; \text{and} \nonumber\\
M_{AB}=M_{BA}&=&\left(1-c_A\right)c_BM_A+c_A\left(1-c_B\right)M_B-c_Ac_BM_C. 
\end{eqnarray}
Following Eq.~\eqref{app_mobility}, Eqs.~\eqref{app_ja} and~\eqref{app_jb} can be written compactly as
\begin{eqnarray}\label{app_ji_compact}
\mathbf{\bar J}_A &=& -M_{AA}\nabla\mu_A^{eff}+M_{AB}\nabla\mu_B^{eff} \; \text{and} \nonumber\\
\mathbf{\bar J}_B &=& -M_{BB}\nabla\mu_B^{eff}+M_{AB}\nabla\mu_A^{eff}.
\end{eqnarray}
We compute $\mu_i^{eff}$ in Eq.~\eqref{app_ji_compact} using the variational derivative~\cite{arfken1999,riley2002} of $\mathcal{F}$ (Eq.~\eqref{eq_ch}),
\begin{equation}\label{app_mu}
\mu_i^{eff}=\frac{\delta \mathcal{F}}{\delta c_i}, \; i =  A, B,
\end{equation}
where 
\begin{equation}\label{app_variational}
\frac{\delta \mathcal{F}}{\delta c_i} = \frac{\partial \mathcal{F}}{\partial c_i}-\nabla \cdot \frac{\partial \mathcal{F}}{\partial \nabla c_i}.
\end{equation}
We obtain the following expressions of $\mu_i^{eff}$  (bulk free energy density $f$ is defined in Eq.~\eqref{eq_bf}),
\begin{eqnarray}\label{app_muab}
\mu_A^{eff}&=&\frac{\partial f}{\partial c_A}-2\left(\kappa_A+\kappa_C\right)\nabla^2c_A - 2\kappa_C\nabla^2c_B\; \text{and} \nonumber\\
\mu_B^{eff}&=&\frac{\partial f}{\partial c_B}-2\left(\kappa_B+\kappa_C\right)\nabla^2c_B-2\kappa_C\nabla^2 c_A,
\end{eqnarray}
where
\begin{eqnarray}\label{app_dfdc}
\frac{\partial f}{\partial c_A}&=&\ln c_A-\ln c_C+\left(\chi_{AB}-\chi_{BC}\right)c_B+\chi_{AC}\left(c_C-c_A\right) \; \text{and} \nonumber\\
\frac{\partial f}{\partial c_B}&=&\ln c_B-\ln c_C+\left(\chi_{AB}-\chi_{AC}\right)c_A+\chi_{BC}\left(c_C-c_B\right).
\end{eqnarray}
Substituting Eqs.~\eqref{app_ji_compact}--\eqref{app_dfdc} into Eq.~\eqref{app_continuity2} leads to the equations of motion given by Eqs.~\eqref{eq_dcadt} and \eqref{eq_dcbdt} (Sec.~\ref{sec_model}). 

In our ternary phase-field model, two constraints need to be satisfied, Eqs.~\eqref{eq_sum1} and~\eqref{app_j0}. Therefore, in this work, even though we set $M_C = 0$ (Eq.~\eqref{app_mobility}) that renders $\mathbf{J}_C = 0$ (Eq.~\eqref{app_mobility1}) to render the C-rich particles immobile, the net mass flux of C due to the exchange of A and B atoms with vacancies is still present and contribute to $\mathbf{\bar{J}}_C$ (Eq.~\eqref{app_ji2}). Mathematically speaking, $\mathbf{J}_C$ can be redundant, but not $\mathbf{\bar{J}}_C$, and hence the $c_C$-field is not frozen (at least in the matrix) in our simulations. Thus, we need to solve for two independent nonlinear diffusion equations for $c_A$ (Eq.~\eqref{eq_dcadt}) and $c_B$ (Eq.~\eqref{eq_dcbdt}) in our phase-field simulations. Also note, to define the effective mobilities in Eq.~\eqref{app_mobility}, both $c_A$ and $c_B$ are necessary.

%---------------------------------------------------------

\section{Ternary Phase Equilibria}\label{sec_appendix2}

In this work, A--rich $\alpha$, B--rich $\beta$, and C--rich $\gamma$ phases constitute our ternary system. During ternary phase equilibrium~\cite{lupis,gaskell2003}, chemical potential of the components ($i$) in the coexisting phases ($j$) becomes equal,
\begin{eqnarray}\label{app_eq1}
\mu_i^\alpha &=& \mu_i^\beta = \mu_i^\gamma, \; i =  A, B, C.
%\mu_B^\alpha &=& \mu_B^\beta = \mu_B^\gamma, \; \text{and} \nonumber\\
%\mu_C^\alpha &=& \mu_C^\beta = \mu_C^\gamma.
\end{eqnarray}
Following Refs.~\cite{lupis,gaskell2003}, $\mu_i$ can be approximated by ($f$ defined in Eq.~\eqref{eq_bf}),
\begin{eqnarray}\label{app_eq2}
\mu_A &=& f-c_B\frac{\partial f}{\partial c_B}-c_C\frac{\partial f}{\partial c_C}, \nonumber\\
\mu_B &=& f+(1-c_B)\frac{\partial f}{\partial c_B}-c_C\frac{\partial f}{\partial c_C}, \; \text{and} \nonumber\\
\mu_C &=& f-c_B\frac{\partial f}{\partial c_B}+(1-c_C)\frac{\partial f}{\partial c_C}.
\end{eqnarray} 
Thus, for example, $\mu_A^\alpha $ and $ \mu_A^\beta $ can be written as,
\begin{eqnarray}\label{app_eq3}
\mu_A^\alpha &=& f^\alpha-c_B^\alpha\frac{\partial f^\alpha}{\partial c_B^\alpha}-c_C^\alpha\frac{\partial f^\alpha}{\partial c_C^\alpha} \; \text{and} \nonumber\\
\mu_A^\beta &=& f^\beta - c_B^\beta\frac{\partial f^\beta}{\partial c_B^\beta}-c_C^\beta\frac{\partial f^\beta}{\partial c_C^\beta}.
\end{eqnarray}
Note that six distinct relationships are given by Eq.~\eqref{app_eq1}. For instance, substituting Eq.~\eqref{app_eq3} in $\mu_A^\alpha = \mu_A^\beta$ yields,
\begin{equation}\label{f1}
\ln c_A^\alpha-\ln c_A^\beta+\left(c_B^\alpha c_C^\alpha - c_B^\beta c_C^\beta\right)\left(\chi_{AB}-\chi_{BC}+\chi_{AC}\right)+\left({c_B^{\alpha}}^2-{c_B^{\beta}}^2\right)\chi_{AB}+\left({c_{C}^{\alpha}}^2-{c_C^{\beta}}^2\right)\chi_{AC} = 0.
\end{equation} 
Although not given here, all other equations in Eq.~\eqref{app_eq1} can be derived~\cite{Ghosh} similarly following:
$\mu_A^\alpha = \mu_A^\gamma$,
$\mu_B^\alpha = \mu_B^\beta$,
$\mu_B^\alpha = \mu_B^\gamma$,
$\mu_C^\alpha = \mu_C^\beta$, and
$\mu_C^\alpha = \mu_C^\gamma$.
These nonlinear equations are then numerically solved with respect to the following constraints,
\begin{equation}
c_A^j + c_B^j + c_C^j = 1, \; j = \alpha, \beta, \gamma,
\end{equation}
to determine the (isothermal) ternary phase equilibrium compositions ($c_i^\alpha$, $c_i^\beta$, $c_i^\gamma$) as a function of $\chi_{AB}$, $\chi_{BC}$, and $\chi_{AC}$ (Sec.~\ref{sec_parameters}).

\bibliography{spinodal}
%--------------------------------------------------
\end{document}